\documentclass[a4paper,11pt]{article}
\pdfoutput=1
\usepackage{jheppub,bm}
\usepackage{graphicx}% Include figure files
\usepackage{dcolumn}% Align table columns on decimal point
\usepackage{amsmath}
\usepackage[utf8]{inputenc}
\usepackage{booktabs}

\newcommand{\pythia}  {{\sc Pythia}}
\newcommand{\pythiaqed}  {{\sc Pythia-qed}}
\newcommand{\dyqt}  {{\sc DYqT}}
\newcommand{\dyres}  {{\sc DYRes}}
\newcommand{\resbos}  {{\sc ResBos}}
\newcommand{\horace}  {{\sc Horace-3.1}}
\newcommand{\horaceo}  {{\sc Horace}}
\newcommand{\wzgrad}   {{\sc WZgrad}}

\newcommand{\rady}   {{\sc Rady}}
\newcommand{\sanc}   {{\sc Sanc}}
\newcommand{\photos}   {{\sc Photos}}
\newcommand{\powheg}   {{\sc Powheg}}
\newcommand{\powhegvtwo}   {{\sc Powheg-v2}}

\newcommand{\herwig} {{\sc Herwig}}
\newcommand{\sherpa} {{\sc Sherpa}}
\newcommand{\dzero}    {{\sc D\O}}
\newcommand{\cdf}    {{\sc Cdf}}
\newcommand{\atlas}    {{\sc Atlas}}
\newcommand{\cms}    {{\sc Cms}}
\newcommand{\qcdqed} {QCD$_{\rm NLOPS}\times$QED$_{\rm PS}$}
\newcommand{\qcdew} { QCD$_{\rm NLOPS}$$\times$EW$_{\rm NLOPS}$}
\newcommand{\smartpap}{p\hskip-7pt\hbox{$^{^{(\!-\!)}}$}}
\newcommand{\oalpha}  {\mbox{${\cal O}(\alpha)$}}
\newcommand{\myref}[1]{(\ref{#1})}

\newcommand{\bea}{\begin{eqnarray}}
\newcommand{\eea}{\end{eqnarray}}
\newcommand{\be}{\begin{equation}}
\newcommand{\ee}{\end{equation}}

\newcommand{\gev}   {\mbox{${\rm GeV}$}}

\newcommand{\mwzero}   {\mbox{$M_{W,0}$}}
\newcommand{\mw}   {\mbox{$M_W$}}
\newcommand{\mh}   {\mbox{$M_H$}}

\newcommand{\mz}   {\mbox{$M_Z$}\,}
\newcommand{\gw}   {\mbox{$\Gamma_{W}$}}
\newcommand{\smallw}{{\scriptscriptstyle W}}
\newcommand{\tw}{\theta_{\smallw}} 

\newcommand{\mtr}  {\mbox{$M_T$}}
\newcommand{\ptl}  {\mbox{$p_T^l$}}

\newcommand{\shat}   {\mbox{$\hat{s}$}} 
\newcommand{\metEq}  {{E}\!\!\!/_T }
\newcommand{\met}    {\mbox{$\metEq$}}
\newcommand{\oaas}{\mbox{${\cal O}(\alpha \alpha_s)$\,}}
\newcommand{\oalphasq}{\mbox{${\cal O}(\alpha^2)~$}}

\newcommand{\oa}{${\cal O}(\alpha)$} 
\newcommand{\rpm}{\raisebox{.2ex}{$\scriptstyle\pm$}}

\begin{document}

\title{Precision Measurement of the $W$--Boson Mass: Theoretical Contributions and Uncertainties}

\author[a]{Carlo Michel Carloni Calame,}
\author[a,1]{Mauro Chiesa \note{   
Present address: Julius-Maximilians-Universit\"at 
W\"urzburg, Institut f\"ur Theoretische Physik und Astrophysik, D-97074 W\"urzburg, Germany.
},}
\author[b]{Homero Martinez,}
\author[b]{\\Guido Montagna,}
\author[a]{Oreste Nicrosini,}
\author[a]{Fulvio Piccinini,}
\author[c]{Alessandro Vicini\,}

\affiliation[a]{INFN, Sezione di Pavia, Via A. Bassi 6, 27100, Pavia, Italy}
\affiliation[b]{Dipartimento di Fisica, Universit\`a di Pavia, and INFN, Sezione di Pavia, \\Via A. Bassi 6, 27100, Pavia, Italy}
\affiliation[c]{Tif lab, Dipartimento di Fisica, Universit\`a di Milano, and INFN, Sezione di Milano, \\Via G. Celoria 16, 20133, Milano, Italy}

\preprint{TIF-UNIMI-2016-10}
\abstract{
We perform a comprehensive analysis of electroweak, QED %contributions 
and mixed QCD-electroweak corrections underlying 
the precise measurement of the $W$-boson mass $M_W$ at hadron colliders. 
By applying a template fitting technique, we detail 
the impact on $M_W$ of next-to-leading order electroweak and QCD corrections, 
multiple photon emission,  lepton pair radiation and factorizable QCD-electroweak 
contributions. As a by-product, we provide an up-to-date estimate of the main theoretical uncertainties 
of perturbative nature. Our results can serve as a guideline 
for the assessment of the theoretical systematics at the Tevatron and LHC 
and allow a more robust precision measurement  of the $W$-boson mass at 
hadron colliders. 
}


\maketitle
\flushbottom

\section{\label{sec:introduction}Introduction}

The high-precision measurement of the $W$-boson mass ($\mw$) 
offers the possibility of a stringent test of the Standard Model (SM) of the electroweak (EW)
and strong interactions.
The recent results by the Tevatron collaborations 
\cdf\ ($\mw=80.387\pm 0.019$ GeV) \cite{Aaltonen:2013vwa}
and 
\dzero\ ($\mw=80.375 \pm 0.023$ GeV) \cite{D0:2013jba,:2014pka,Li:2016jal}
dominate the current world average
($\mw=80.385 \pm 0.015$ GeV) 
\cite{Olive:2016xmw},
which has now an accuracy of $2\cdot 10^{-4}$.
Also the LHC experiments {\atlas}\, and {\cms}\, are planning to measure $\mw$
and the possibility of reaching a final error of 15~MeV or eventually of 10 MeV 
has been discussed~\cite{Buge:2006dv,Besson:2008zs,CMS:2016nnd}~\footnote{Recently, 
the {\atlas}\, collaboration published the first $W$-boson mass measurement 
with a total uncertainty of $19$~MeV~\cite{Aaboud:2017svj}.}.

The SM prediction for $\mw$ has been reevaluated in ref.~\cite{Degrassi:2014sxa}
($\mw=80.357 \pm 0.009 \pm 0.003$ GeV),
including the full 2-loop EW corrections derived from the study of the muon-decay amplitude,
the leading 3-loop QCD effects and some classes of loop corrections to all orders.
The two main sources of error in the theoretical predictions are the parametric uncertainty induced
by the top quark mass value and the effect of missing higher orders, with similar importance in the final result.
The comparison of the experimental world average with the SM prediction,
in a global fit of the EW parameters that includes, among others,
the top and the Higgs boson masses,
is of primary importance to appreciate possible tensions in the SM~\cite{Baak:2014ora}.
Furthermore, in case of a significant discrepancy, 
it might be possible to derive some indirect hints of
physics beyond the SM~\cite{Baak:2013fwa} or put constraints within the Standard Model Effective 
Field Theory framework~\cite{Bjorn:2016zlr}. 
For this reason, it is crucial to have control on all the contributions to the final 
systematic error of the experimental result, both of experimental and theoretical origin.

The $W$-boson mass is measured at hadron colliders from the analysis of the 
distributions of the kinematic variables of the final state leptons 
of the charged current (CC) Drell-Yan (DY) process
($p\smartpap \to l \nu_l+X,\,\, l=e,\mu$). 
The lepton-pair transverse mass, the charged lepton transverse momentum and the missing transverse momentum
distributions have a jacobian enhancement that makes them sensitive to the precise value of $\mw$.
The measurement of the latter is derived from the accurate knowledge of the shape of these
differential cross sections,
via a template fit procedure. 

The templates are computed with Monte Carlo simulations that include higher-order radiative corrections
and that allow to account for the acceptance cuts and for the response of the detector. Each element
of the simulation may affect the basic shape of the distributions and in turn have an impact on the
central value extracted from the fitting procedure.
The whole approach requires $i)$ the discussion of the effects that are available as well as those
that are not included in the preparation of the templates and $ii)$ the propagation onto the $\mw$
determination of the uncertainties affecting each of the elements entering in the Monte Carlo simulation.

The goals of the paper are the following. First, we assess the impact on the $\mw$ determination 
due to different subsets of EW corrections which are presently available in public simulation codes. 
However, since we want to focus on the impact of the EW corrections
on the DY observables relevant to measure $\mw$ at hadron colliders,
we also need to consider their interplay with the QCD description of the process. 
Secondly, we estimate the impact of missing higher-order EW and mixed QCD-EW corrections, 
which are among the sources of theoretical uncertainty in the $\mw$ measurement.

To achieve these goals, we developed an improved version of the EW Monte Carlo program
\horace, to study the impact of EW input scheme variations and simulate the radiation of
light lepton pairs. In the present study we do not consider the effect of 
$\gamma$-induced processes, which is left to a future study. 
Along the lines described in Ref.~\cite{Jezo:2015aia}, we also implemented a new version of the \powhegvtwo\, generator 
(named \powhegvtwo\, {\tt two-rad} in the following) for the simulation of
DY processes in the presence of both QCD and EW corrections.

The paper is organized as follows. After a discussion of the main aspects of the theoretical and experimental framework 
in which the $\mw$ measurement is performed (section~\ref{sec:frame}), we describe in section~\ref{sec:tools} the theoretical and computational 
features of the tools used in our study. In section~\ref{sec:thunc} we briefly discuss the uncertainties 
induced by \oalphasq EW corrections (EW input scheme) and \oaas corrections. Section~\ref{sec:distributions} and section~\ref{sec:numerics} are devoted to the
presentation and discussion of our numerical results. In section~\ref{sec:distributions} we detail the impact of various 
sources of higher-order corrections on the differential cross sections used to extract \mw\ from the data, 
while in section~\ref{sec:numerics} we quantify the \mw\ shifts by the same higher-order contributions. 
Results for both the Tevatron and the LHC are given. 
Our conclusions are drawn in section~\ref{sec:conclusions}.

%%%%%%%%%%%%%%%%%%%%%%%%%%%%%%%%%%%%%%%%%%%%%%%%%%%%%%%%%%%%%%%%%%%%%%%%%%%%%%%
\section{Theoretical and experimental framework}
\label{sec:frame}
\subsection{Existing calculations and simulation codes}
\label{sec:existing}
The DY processes start at Leading Order (LO)
with purely EW amplitudes and receive radiative corrections that are exactly
known up to ${\cal O}(\alpha_s^2)$~\cite{Hamberg:1990np,Anastasiou:2003ds,Anastasiou:2003yy,Melnikov:2006di,Melnikov:2006kv}
in the strong interaction coupling,
and that are implemented, at fully differential level with respect to the leptonic variables, 
in codes like {\sc{FEWZ}}~\cite{Gavin:2012sy}, {\sc{DYNNLO}}~\cite{Catani:2009sm}, 
{\sc{MCFM}}~\cite{Boughezal:2016wmq} and {\sc{SHERPA}}~\cite{Hoeche:2014aia}.
The N$^3$LO threshold corrections for the inclusive cross section and for the rapidity distributions
of the dilepton pair have been presented in refs.~\cite{Ahmed:2014cla,Ahmed:2014uya}. 
The corrections up to ${\cal O}(\alpha)$ \cite{Wackeroth:1996hz,Baur:1998kt,Baur:2001ze,Dittmaier:2001ay} in the EW coupling
are also available and are implemented in different public codes 
like \wzgrad~\cite{Baur:1997wa,Baur:1998kt,Baur:2001ze,Baur:2004ig},
\rady~\cite{Dittmaier:2001ay,Dittmaier:2009cr},
\sanc~\cite{Arbuzov:2005dd,Arbuzov:2007db,Bardin:2012jk},
\horace~\cite{CarloniCalame:2006zq,CarloniCalame:2007cd}. A systematic overview
and comparison among different codes for DY simulations have been presented
in ref.~\cite{Alioli:2016fum}.

Fixed-order results are not sufficient to reach the level of accuracy needed for a precise measurement of $\mw$.
The approximated inclusion of multiple parton initial state  radiation (ISR) to all perturbative orders
is necessary to obtain a sensible description of the lepton-pair transverse momentum distribution and is known up to next-to-next-to-leading logarithmic (NNLL) QCD accuracy \cite{Collins:1984kg,Catani:2000vq} 
with respect to the $L_{\rm QCD} \equiv\log(p_\perp^V/m_V)$ factor, 
where $p_\perp^V$ is the lepton-pair transverse momentum and $m_V$ is the relevant gauge boson mass ($V=W,Z$);
these corrections have been implemented in simulation codes like 
\resbos\, \cite{Balazs:1997xd} or \dyqt/\dyres\, \cite{Bozzi:2010xn,Catani:2015vma}.
The inclusion of multiple photon radiation effects is necessary to 
achieve an accurate description of the leptonic observables,
it is known up to leading-logarithmic (LL) accuracy 
with respect to the $L_{\rm QED}\equiv\log(\hat s/m_l^2)$ enhancement factor, where $\hat s$ is the partonic Mandelstam variable and $m_l$ is the final state lepton mass;
these corrections are available in codes like \photos\ or \horace\
\cite{Barberio:1990ms,Barberio:1993qi,Golonka:2005pn,CarloniCalame:2003ux,CarloniCalame:2005vc}.
The problem of merging fixed-order and all-order results, avoiding double counting, has been separately discussed in QCD \cite{Balazs:1997xd,Bozzi:2008bb,Frixione:2002ik,Nason:2004rx,Hoeche:2014aia,Karlberg:2014qua,Alioli:2015toa} 
and in the EW SM \cite{CarloniCalame:2006zq,Placzek:2003zg}.
QCD and EW results have to be combined together to obtain a realistic description of the DY final states:
general purpose Shower Monte Carlo programs,
like \pythia \cite{Sjostrand:2006za,Sjostrand:2007gs}, 
\herwig \cite{Corcella:2000bw} or \sherpa \cite{Gleisberg:2008ta}, 
include the possibility of multiple photon, gluon and quark emissions
via a combined application of QCD and QED Parton Shower (PS), 
retaining only LL accuracy
in the respective large logarithmic factors.
In the absence of an exact calculation of the first mixed 
\oaas corrections,
several recipes have been devised to include the bulk of the two sets of effects 
including factorizable subleading effects both of QCD and EW origin
\cite{Cao:2004yy,Adam:2008pc,Adam:2008ge,Adam:2010tg,Buttar:2008jx,Balossini:2009sa}, 
and preserving the NLO-QCD and NLO-EW accuracy on the quantities inclusive with respect to additional radiation
\cite{Bernaciak:2012hj,Barze:2012tt,Barze':2013yca}.

\subsection{Classification of theoretical uncertainties}
\label{sec:classification}
The value of $\mw$ is extracted with a template fit technique.
The templates are theoretical distributions of the relevant observables, prepared with Monte Carlo simulation codes and keeping $\mw$ as a free parameter; 
the accuracy of all the elements that enter in these codes has a direct impact on the final accuracy of the $\mw$ determination.
According to the factorized formulation of the hadron level cross section, we can identify two groups of effects as main sources of theoretical uncertainty: 
$i)$ the non-perturbative effects stemming from the the proton description at low transverse momentum scales 
(probability density functions (PDFs), lepton-pair transverse momentum ($p_\perp^W$) modeling, intrinsic transverse momentum of the partons in the proton)
and
$ii)$
the perturbative radiative corrections (QCD, EW, mixed QCD-EW) to the partonic cross section.

\subsubsection{Non-perturbative effects: PDFs and $p_\perp^W$ modeling}
\label{sec:nonpert}
The knowledge of the structure of the proton in terms of elementary constituents (quarks, gluons, photons)
is affected by different uncertainties stemming at low-momentum scales, where QCD non-perturbative effects are important.

The proton collinear PDFs suffer of the errors of the experimental data from which they are extracted; these uncertainties are represented by the PDF collaborations with sets of replicas that have to be used to propagate the PDF error to the observables under study. These replicas are parametrizations of the proton structure equally compatible with the data, with a given confidence level; they are all on the same footing for the preparation of the templates used to fit $\mw$.
The \cdf\, and \dzero\, collaborations have used the PDF set CTEQ6M to estimate the corresponding uncertainty on $\mw$ and reported respectively a PDF contribution of 10 MeV \cite{Aaltonen:2013vwa} and of 11 MeV \cite{D0:2013jba} to the systematic error.
An extensive survey on the PDF uncertainties on $\mw$, considering different sets of global PDFs whose results are then combined according to the PDF4LHC recipe \cite{Alekhin:2011sk}, has been performed at generator level (i.e. without including the effects of the detector response) in refs. \cite{Bozzi:2011ww,Bozzi:2015hha,Bozzi:2015zja,Quackenbush:2015yra,Bodek:2016olg}.

The prediction of the lepton-pair transverse momentum distribution requires, in the limit $p_\perp^W\to 0$, the resummation to all orders of perturbative corrections which are logarithmically divergent. 
In the region of small $p_\perp^W$ there might be non-perturbative contributions to the distributions, on top of the perturbative component obtained via the resummation procedure.
In ref.~\cite{Konychev:2005iy} the size and the universality of these effects have been discussed in detail. 
In codes like \resbos\, or \dyqt\, a general parametrization, in the resummation formalism, allows to account for these non-perturbative contributions,
whose size is intertwined with the logarithmic accuracy of the $p_\perp^W$ resummation.
In the PS framework there are simple models to generate an intrinsic parton transverse momentum, with a typical size that is related to the LL accuracy of the PS resummation.
The presence of heavy flavors inside the proton has an impact on the $p_\perp^W$
distributions, because of the different spectrum, with respect to the light quarks, induced by the heavy quarks PDFs.

The \cdf\, and \dzero\, experiments based their description of the lepton-pair transverse momentum distribution on the code \resbos, which has NNLL-QCD accuracy in the $p_\perp^W$ resummation and a partial inclusion of the full set of NNLO-QCD corrections;
for a fixed choice of the PDF set, factorization and renormalization scales,
they estimated the impact on $\mw$ of the $p_\perp^W$ modeling 
by varying the coefficients of the non-perturbative functions present
in the resummed cross section; \cdf\, found this effect to be of 5~MeV
(according to table~XIV of ref.~\cite{Aaltonen:2013vwa}) while \dzero\, 
found it to range between 2 and 5~MeV (according to table~6 of
ref.~\cite{D0:2013jba}), depending on the observable.

A global constraint on all these different sources of uncertainty comes from the study of the lepton-pair transverse-momentum distribution in the case of Neutral Current (NC) DY, where the gauge boson kinematics can be fully reconstructed.
A detailed study of these uncertainties is beyond the scope of this paper.

\subsubsection{EW and QED uncertainties: the Tevatron approach}
\label{sec:unctev}
Both Tevatron experiments, \cdf\, and \dzero,
use the \resbos\, generator for the treatment of QCD radiation
and simulate the QED corrections to $W/Z$ decay with \photos.
As described in refs.~\cite{Aaltonen:2012bp,Abazov:2012bv,Group:2012gb}, 
the EW uncertainty is derived by performing cross-checks of the \photos\ predictions with those  of \horace\, in \cdf\, and with those of \wzgrad\,
in \dzero.
The basic idea of the procedure described in those papers
is that an estimate of the EW 
systematics can be derived by comparing independent tools which are 
rather different in their theoretical contents. 
A detailed example of 
this procedure can be found in ref.~\cite{Kotwal:2015bfa}. 
Actually, \photos\, is a process-independent tool ~\cite{Barberio:1993qi,Barberio:1990ms,Davidson:2010ew} 
used at the Tevatron (as well as at the LHC) 
to simulate multiple photon final-state radiation (FSR) 
in the LL approximation,
with weights to correct to the QED matrix element calculation ~\cite{Golonka:2005pn}. 
Indeed, FSR  is the largely dominant contribution to the EW corrections 
to $W/Z$ production in the resonance region, as shown in  
many calculations and phenomenological studies~\cite{CarloniCalame:2007cd,CarloniCalame:2006zq,Baur:2001ze,Baur:1998kt,Dittmaier:2001ay,Arbuzov:2005dd,Arbuzov:2007db}. 
However, pure weak corrections are neglected in \photos. 
On the other hand, 
the published versions of the  \wzgrad\, codes contain the full NLO EW corrections, without including contributions of photon radiation beyond ${\cal O}(\alpha)$, 
while \horace\, is based on the matching of exact \oalpha ~EW corrections with multiple photon emission. 
Consequently, according to the experimental strategy used by 
both the Tevatron collaborations, 
the main component of uncertainty in the modeling of EW/QED processes 
comes from the missing  inclusion of NLO EW corrections.

The present Tevatron estimate of the EW uncertainty can not be considered, 
strictly speaking, an evaluation of the actual theoretical uncertainty but rather an assessment of the theoretical (or physical) precision of the main tool, {\it i.e.} \photos, used in the experimental analysis. 
Indeed, a more reliable estimate of the theoretical uncertainty should
be associated, by definition, to the best available calculations, 
as induced by the most important missing higher--order contributions.
From this point of view, 
a reassessment of the current status of EW predictions and associated QCD-EW theory results, for DY processes, 
provides a robust framework to put the Tevatron error estimate on firmer theoretical grounds and 
cross-check it according to an independent procedure. 
Furthermore, our results can provide a guideline for an assessment of the theoretical uncertainties in the \mw\ measurement at the LHC.

\subsubsection{Mixed QCD-EW corrections}
\label{sec:uncmix}
At NNLO accuracy level, there are three groups of perturbative terms: 
the ${\cal O}(\alpha_s^2)$, the ${\cal O}(\alpha^2)$ and the \oaas contributions.
The ${\cal O}(\alpha_s^2)$ terms are completely known, 
as already discussed in section~\ref{sec:existing}.
Among the ${\cal O}(\alpha^2)$ terms,
the discussion of the full two-loop amplitudes renormalization has been presented in refs.
\cite{Degrassi:2003rw,Actis:2006ra,Actis:2006rb,Actis:2006rc};
besides the latter,
two-photon and lepton-pair virtual and real emission, discussed 
in section~\ref{sec:horacepairs}, are known and yield a contribution which is classified as dominant in an expansion in powers of the enhancement factor $L_{\rm QED}$.
The mixed \oaas contributions are a potential source 
of relevant theoretical uncertainty and, as a consequence, they deserve particular attention. 
For this reason, recent activities of different groups focused on the calculation 
of partial contributions of the full \oaas perturbative 
corrections to the charged DY cross section and distributions. 
In refs.\cite{Czarnecki:1996ei,Kara:2013dua} the \oaas corrections to the decays
of $Z$ and $W$ bosons were presented, whereas  
the mixed two-loop corrections to the $Z$ boson production form factors 
have been computed in ref.\cite{Kotikov:2007vr}.
Mixed QCD-QED virtual correction to lepton-pair production
have been presented in \cite{Kilgore:2011pa},
while the full set of Master Integrals relevant for the evaluation of the complete set of QCD-EW virtual corrections has been presented in
ref.\cite{Bonciani:2016ypc}.
The double-real contribution to the total cross section for the on-shell single gauge boson production has been presented in ref.\cite{Bonciani:2016wya}, while the subtraction of initial state collinear singularities can now be accomplished at \oaas with the Altarelli-Parisi splitting functions presented in ref.\cite{deFlorian:2015ujt}.
Additional building blocks are the calculation of the 
complete NLO EW corrections to $V + $~jet production, 
including the vector boson decay into leptons \cite{Denner:2009gj,Denner:2011vu}  
and the calculation of NLO QCD corrections to $V+\gamma$
\cite{Barze:2014zba,Denner:2014bna,Denner:2015fca}.

A systematic approach to the calculation of mixed QCD and EW corrections 
to DY processes has been presented recently in refs.~\cite{Dittmaier:2014qza,Dittmaier:2014koa,Dittmaier:2015rxo,Dittmaier:2016egk},
adopting the pole approximation, 
which has been proven to give reliable predictions at NLO accuracy for observables dominated by a resonant $W$ boson~\cite{Wackeroth:1996hz,Dittmaier:2001ay}. 
In this approximation the corrections split naturally into 
factorizable corrections to $W$ production and decay 
(for which the results of on-shell production and decay can be used as building blocks) 
and non-factorizable corrections, 
which consist of virtual and real soft photons $(E_\gamma \ll \Gamma_W)$ 
connecting the $W$ production and decay stages 
as well as soft photons connecting a final state quark and the lepton. 
The latter ones, which are not enhanced by collinear logarithms, 
have been calculated in ref.~\cite{Dittmaier:2014qza} and 
found to be numerically well below the 0.1\% level 
for the lepton-pair transverse mass as well as 
for the lepton transverse momentum distribution. 
Additionally, these corrections are flat over the entire range of the considered observables. 
As a consequence the non-factorizable \oaas contributions have a 
negligible impact on the $\mw$ determination. 

The factorizable \oaas\ corrections 
comprise three main classes of contributions: 
``initial-initial'', 
consisting of virtual and real parton and photon insertions in the initial state; 
``final-final'', 
consisting of virtual contributions from two-loop counterterms; 
``initial-final'', consisting of NLO QCD (virtual and real) corrections to 
the $W$ production process and NLO EW (virtual and real) corrections to the $W$ decay. The latter 
class, {\em i.e.} ``initial-final'', is expected to give the bulk of the 
total factorizable corrections. In fact to this class contribute initial state logarithms 
of QCD origin and final state collinear logarithms of QED origin. 
In refs.~\cite{Dittmaier:2014qza,Dittmaier:2015rxo,Dittmaier:2016egk} the calculation of the 
``initial-final'' \oaas~\footnote{Also the ``final-final'' corrections
have been calculated and shown to be numerically negligible.} 
corrections is detailed in all its components: double-virtual, real$\times$virtual and 
double-real contributions. Only the latter are not completely factorized into a product 
of NLO QCD and EW corrections. The factorized ``initial-final'' corrections, 
computed in pole approximation in a full calculation,
are compared with the ansatz of complete factorization of the NLO corrections 
$\delta_{\alpha_s}^\prime \delta_\alpha$~\footnote{$\delta_\alpha$ is the relative 
NLO EW correction w.r.t. the LO cross section and  $\delta_{\alpha_s}^\prime$ 
is the relative NLO QCD correction w.r.t. to the LO contribution calculated 
with NLO parton distribution functions.},
computed separately and then multiplied,
for the lepton-pair transverse mass and lepton transverse momentum distributions. 
In the case of the transverse mass a good level of agreement 
between the complete calculation and the factorized ansatz is found. 
Instead, in the case of the lepton transverse momentum distribution
the factorized ansatz differs from the full result obtained in pole approximation.

In section \ref{sec:powheg} we describe the factorized ansatz for the combination of QCD and EW effects
on which the formulation of \powheg\, is based;
in section \ref{sec:comparison_with_fixed_order} 
we compare the \powheg\, results
with those of refs.
\cite{Dittmaier:2014qza,Dittmaier:2014koa,Dittmaier:2015rxo,Dittmaier:2016egk}
concerning the impact on the $\mw$ determination
of factorizable \oaas\, corrections.

\section{\label{sec:tools}Theoretical tools}
\subsection{The \horace\ formulation}
\label{sec:horace}
The \horace\, event generator merges
the exact ${\cal O}(\alpha)$ corrections to the CC and NC DY processes 
with an all order QED-PS, with photons being radiated from all the electrically charged scattering particles.
It reaches NLO-EW accuracy in the description of the observables inclusive over QED radiation and it includes the effects of the all order resummation of the final state collinear mass logarithms, with LL accuracy.
The double counting between the exact matrix elements and the PS algorithm is avoided, relying on the following formula for the event generation:
\bea
\!\! d\sigma^{\infty} \!\! =\!
F_{SV}~\Pi(Q^2,\varepsilon)
\!\sum_{n=0}^\infty \frac{1}{n!}
\left( \prod_{i=0}^n F_{H,i}\right)
\! |{\cal M}_{n,LL}|^2
d\Phi_n
\label{eq:matchedinfty}
\eea
The basic structure in equation \myref{eq:matchedinfty} 
for the fully differential cross section
is the sum 
over all photon multiplicities that accompany the hard scattering process (in this notation $n=0$). The emission of $n$ photons is described by LL-accurate matrix elements ${\cal M}_{n,LL}$, with phase space $d\Phi_n$; the unitarity of the process is guaranteed by the Sudakov form factor $\Pi(Q^2,\varepsilon)$, with $Q^2$ the hard scale of the process and $\varepsilon$ the photon-energy detection threshold.
The effect of the exact ${\cal O}(\alpha)$ matrix elements
is given by the correction factors $F_{SV}$ and $F_{H,i}$;
the latter are by construction IR-finite, since the full IR divergent structure is already present in the all orders LL formulation.
The virtual corrections are given by the overall factor $F_{SV}=1+\delta_{SV}$,
where $\delta_{SV}$ is a term of ${\cal O}(\alpha)$, dependent on the LO kinematical invariants;
it includes the effect of 
the renormalization scheme-dependent terms,
of all the virtual diagrams and in particular of the loops where weak massive bosons are exchanged, whose effect becomes sizeable at large partonic center-of-mass energies.
The one-photon real matrix element corrections $F_{H,i}=1+\delta_{H,i}$,
where $\delta_{H,i}$ is a term of ${\cal O}(\alpha)$,
are applied to each photon emission labelled by the index $i$ and are evaluated with an effective single-photon kinematics configuration.
The presence of the correction factors $F_{H,i}$ for all emissions does not spoil the LL accuracy of the resummation via PS, which is important in the description of soft and/or collinear QED radiation.
The exact NLO-EW accuracy is preserved as it can be checked by expanding equation \myref{eq:matchedinfty} up to ${\cal O}(\alpha)$.
There are several terms of ${\cal O}(\alpha^2)$ and higher which are introduced because of the factorized structure of equation \myref{eq:matchedinfty}; their effect is beyond the accuracy of the calculation; 
we recognize 
those generated by the product of $F_{SV}$ with  all photon multiplicities and 
those due to the product of one $F_{H,i}$ factor with the description of further real photon emissions.

The public version of the \horace\, code allows to generate events in various approximations
that account for the effect of different gauge invariant subsets of corrections:
$1)$ one FSR photon in QED-LL approximation;
$2)$ multiple FSR photon radiation in QED-LL approximation;
$3)$ exact ${\cal O}(\alpha)$ NLO-EW corrections;
$4)$ exact ${\cal O}(\alpha)$ NLO-EW corrections matched with 
   multiple-photon radiation from all the charged legs of the process in
   QED-LL approximation.
The cases $1)$ and $2)$ are obtained from eq. \myref{eq:matchedinfty}
by setting $F_{SV}=F_{H,i}=1$ and by a consistent description only of final state radiation\footnote{We remind that for the CC DY process the treatment of 
FS QED--like corrections poses problems of gauge invariance. When referring to FS radiation, 
it is understood that the contribution of both lepton and internal $W$ photon emission is 
consistently included in the calculations without violating gauge invariance. For the 
NC DY process this difficulty is absent, FS QED contributions being a gauge--invariant subset of the 
full EW correction.}.
We stress that the outcome of points $1)$, $2)$ and $4)$ depends on the details
of the \horace\, implementation of multiple photon radiation
and on the matching scheme adopted to merge fixed- and all-order results.
Alternative results, that share the same nominal accuracy and differ by subleading higher-order terms, can be found in \photos \cite{Golonka:2005pn}
for approximations $1)$ and $2)$ or in the EW part of \powhegvtwo \cite{Barze:2012tt} for approximation $4)$.
A new subset of ${\cal O}(\alpha^2)$ corrections, namely those due to the emission of an additional lepton pair, are now implemented in \horace\, and will be described in section \ref{sec:horacepairs}.
The effects induced by a change of the renormalization scheme,
which is an option present in the code,
will be discussed instead in section \ref{sec:inputscheme}.

\subsection{Lepton-pair corrections in \horace}
\label{sec:horacepairs}

In the theoretical tools used at the Tevatron and, in particular, in the 
publicly available version of the \horace\, generator, 
the contribution due to the emission of photons converting into lepton 
pairs is not accounted for. The corresponding correction, of the 
order $\alpha^2 L^2$, 
is {\it a priori} at the same level as the two--photon contribution considered in the experimental 
analysis. Therefore, it is important to understand how the $\mw$ measurement 
is influenced by this effect. To this end
we have introduced the possibility of accounting for the emission of
extra lepton pairs in the \horace\, PS mode describing QED FSR.
The procedure is based on the Structure Function (SF) approach \cite{Kuraev:1985hb,Nicrosini:1986sm,Arbuzov:2010zzb} and 
closely follows the recipe described in ref.~\cite{Skrzypek:1992vk}, used in
the 1990s to calculate initial-state pair corrections to the 
$Z$ line shape parameters~\cite{Jadach:1992aa}. 
In its standard collinear formulation, the SF method 
provides the probability of finding a fermion
inside a parent fermion whose energy has been reduced by a factor $x$ 
because of the emission, at a given energy scale $Q$,
of (strictly collinear) QED ``partons". The energy loss is due to photon radiation 
and/or the emission of an additional light--fermion pair.
Accordingly, the SF can be organized
into a Non--Singlet (NS) $D^{NS}(x,Q^2)$ contribution (if the emitting 
fermion arrives at an annihilation or detection point without passing through a photon line) 
and a Singlet (S) $D^{S}(x,Q^2)$ contribution (if the fermion transforms into a 
fermion passing 
through a photon line). 

In the SF language, the probability of radiating a fermion
pair can be expressed using a running electromagnetic coupling constant, whose value is
related to the number of emitted flavors, {\it i.e.} \cite{Skrzypek:1992vk}
\be
\alpha(s) = 
\left\{
\begin{array}{ll}
{\alpha}/\left({1-\frac{\alpha}{3\pi}\ln\frac{s}{m_e^2}}\right)
& \qquad \qquad {\rm  electrons~only} \\
{\alpha}/\left({1-\frac{\alpha}{3\pi}\ln\frac{s}{m_e^2}
-\theta(s-m_\mu^2) \frac{\alpha}{3\pi}\ln\frac{s}{m_\mu^2}} \right)
& \qquad \qquad {\rm  electrons+muons} 
\end{array}
\right.
\label{eq:alfas}
\ee

The explicit analytical expressions of the QED SF are usually given
in powers of the expansion parameter $\beta_f\equiv2\frac{\alpha}{\pi}\log\left(\frac{s}{m_f^2}
\right)$, where $f$ indicates the flavor of the radiating particle
(not to be confused with the flavor of the additional emitted pair) and $s$ the squared energy scale of the 
  process under consideration.
When using a running coupling constant, the definition of $\beta_f(s)$
is
\be
\beta_f(s)=
\int_{m_f^2}^s \frac{ds^\prime}{s^\prime}~
\frac{2\alpha(s^\prime)}{\pi}
\label{eq:betart}
\ee
For the NS contribution, the effect of a light fermion pair emission
from an emitting particle of flavor $f$
can be simply taken into account 
 with the replacement
$\beta_f\to\beta_f(s)$ 
in the analytical expression of the SF $D^{NS}(x,Q^2)$~\cite{Skrzypek:1992vk,Arbuzov:2010zzb}.

Therefore, for electrons or muons coming from the 
$W$ decay and emitting electron and muon pairs, the formulae of
interest, as obtained from eq.~\myref{eq:betart}, are
\be
\beta_f(s) =
\left\{
\begin{array}{ll}
2\frac{\alpha}{\pi}\log\frac{s}{m_f^2}
& ~~~~~ f=e,\mu {\rm ~~~no~pair~emission}\\
6\left( 
\log\left(1-\frac{\alpha}{3\pi}\log\frac{m_f^2}{m_e^2}\right) 
-\log\left(1-\frac{\alpha}{3\pi}\log\frac{s}{m_e^2}\right) 
\right)
& ~~~~~ f=e,\mu ~~~ e^+e^- {\rm emission}\\
-3\log\left(
\left(1-\frac{\alpha}{3\pi}\log\frac{s}{m_e^2}\right)^2
-\left(\frac{\alpha}{3\pi} \right)^2
\log^2\left(\frac{s}{m_\mu^2}\right)
\right)
& ~~~~~ f=e \quad ~~~ e^+e^- {\rm ~and~} \mu^+\mu^- {\rm emission}\\
3\log
\frac{
1-\frac{\alpha}{3\pi}\log\frac{s}{m_e^2}
  +\frac{\alpha}{3\pi}\log\frac{s}{m_\mu^2}
}
{
1-\frac{\alpha}{3\pi}\log\frac{s}{m_e^2}
  -\frac{\alpha}{3\pi}\log\frac{s}{m_\mu^2}
}
& ~~~~~ f=\mu \quad ~~~ e^+e^- {\rm ~and~} \mu^+\mu^- {\rm emission}\\
\end{array}
\right.
\label{eq-betarunning}
\ee

In the PS algorithm implemented in \horace\,
these contributions have been included by means of an appropriate
modification of the Sudakov form factor and the calculation of
the fractions of the emitted electron and muon pairs, 
as detailed in appendix~\ref{sec:lpa}.

In addition to the NS contribution, one must also include 
the singlet contribution to the QED SF, as given {\it e.g.} 
in ref.~\cite{Skrzypek:1992vk}. 
However, we checked that the numerical impact of 
$D^{S}(x,Q^2)$ on the relevant observables is negligible 
at the precision  level of interest here, the singlet SF
being numerically suppressed with respect to $D^{NS}(x,Q^2)$ 
for any $x$ value and vanishing in the most relevant 
phenomenological region, {\it i.e.} in the infrared 
limit $x \to 1$.

\begin{figure}[hbtp]
\begin{center}
\includegraphics[height=55mm]{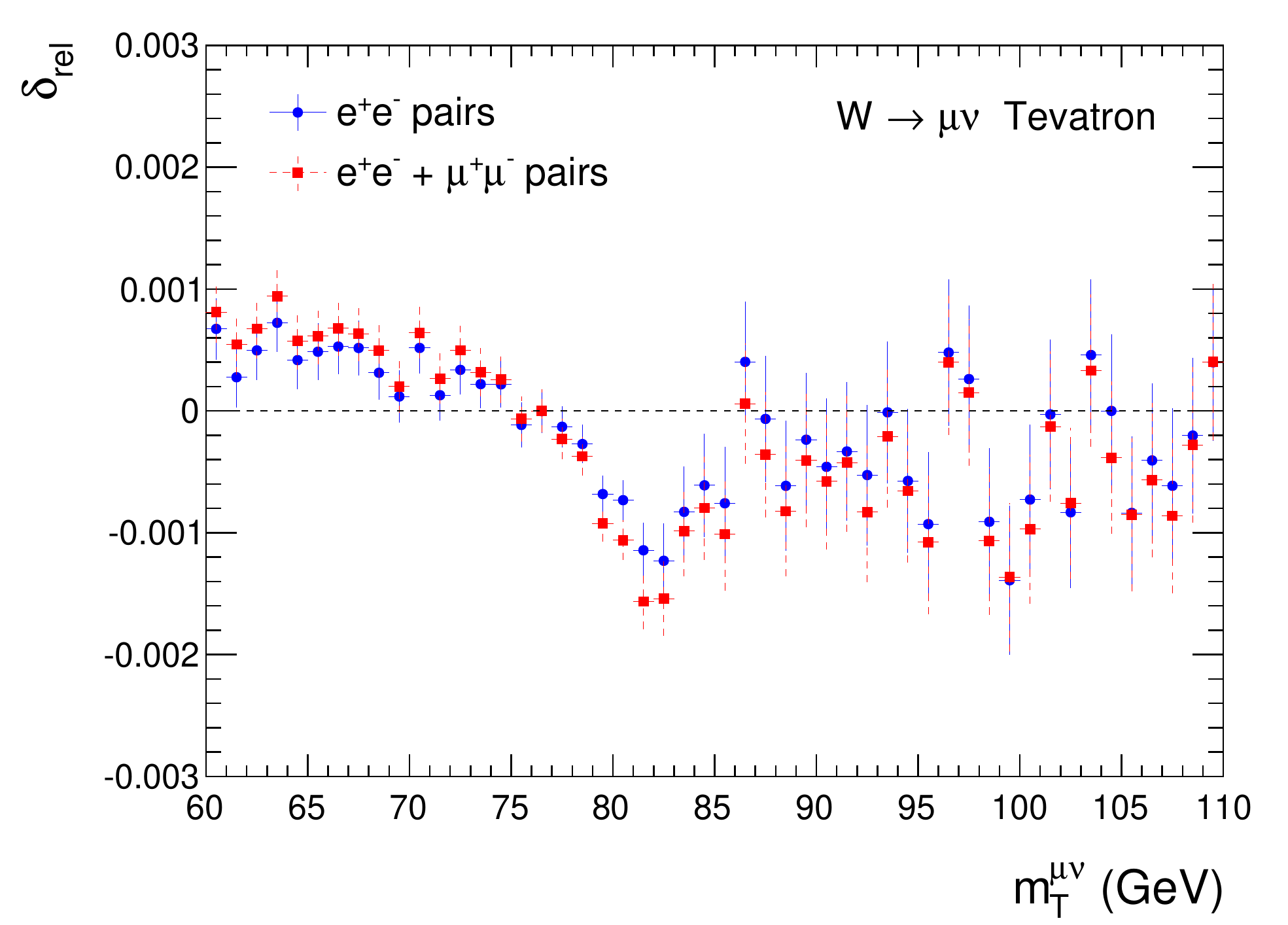}~~
\includegraphics[height=55mm]{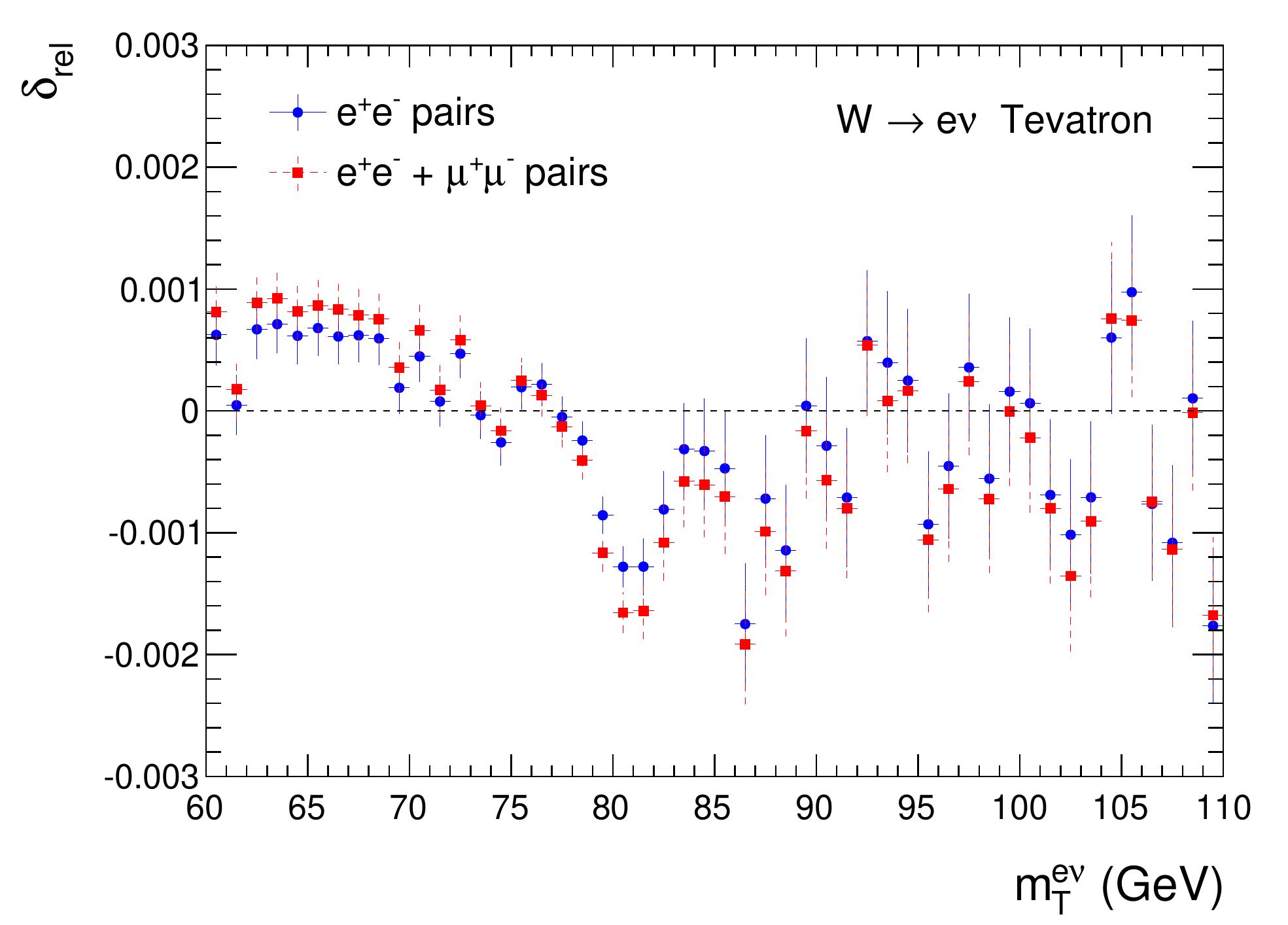}
\caption{\small Relative effect due to lepton-pair corrections
on the $W$ transverse mass distribution, for $W \to \mu \nu$  (left plot) 
and $W \to e \nu$ (right plot) decays at the Tevatron 
($\sqrt{s} =$ 1.96~TeV). The plots show the 
relative difference between 
the \horace\, predictions for multiple FSR with and without pair emission.
\label{fig:pairs}
} 
\end{center}
\end{figure}
The numerical impact on the $W$ transverse mass distribution due to 
lepton-pair emission 
in $W \to e \nu$  and $W \to \mu \nu$ decays at the Tevatron energy ($\sqrt{s} =$ 1.96~TeV) is shown in figure~\ref{fig:pairs}. 
The relative contribution shown in figure~\ref{fig:pairs} is computed 
in terms of \horace\ predictions for the combined multiple photon and pair radiation normalized 
to the results for multiple photon radiation only. As can be seen, 
the correction is largely dominated by the contribution of 
the lightest radiated particles, {\it i.e.} by electron pair emission, which is 
a direct consequence of eq.~\myref{eq-betarunning}. Around the 
Jacobian peak, the pair correction amounts to about 
$0.1-0.2\ $\% for both decay channels and 
modifies the shape of the transverse mass distribution, 
similarly to the effect introduced by photon emission \cite{Baur:1998kt,Dittmaier:2001ay,
CarloniCalame:2006zq,CarloniCalame:2004qw}.

%%%%%%%%%%%%%%%%%%%%%%%%%%%%%%%%%%%%%%%%%%%%%%%%%%%%%%%%%%%%%%%%%%%%%%%%%%%%%
\subsection{\powheg\, with QCD and EW corrections}
\label{sec:powheg}
The implementation of the CC DY process in \powheg\, is documented in ref.~\cite{Alioli:2008gx}, at NLO QCD accuracy. 
The extension to include both NLO QCD and NLO EW corrections 
for this process in \powheg\ is documented in
refs.~\cite{Barze:2012tt,Barze':2013yca}~\footnote{An independent implementation of EW corrections
  to the CC DY process is presented in ref.~\cite{Bernaciak:2012hj}.}.
In this implementation, the overall cross section has 
NLO (QCD+EW) accuracy, and the real radiation can be of QCD as well as QED origin.  
According to the \powheg\, method, the cross section for a given process is written as:
\begin{multline}
d\sigma = \sum_{f_b} \bar{B}^{f_b}(\mathbf{\Phi}_n) \, d\mathbf{\Phi}_n \Biggl\{ \Delta^{f_b} 
(\mathbf{\Phi}_n,p_T^{min}) \\ 
+ \sum_{\alpha_r \in \{ \alpha_r | f_b \} } \frac{ \left[ d\Phi_{rad} \, \theta(k_T - p_T^{min}) 
\, \Delta^{f_b}(\mathbf{\Phi}_n, k_T) \, R(\mathbf{\Phi}_{n+1}) 
\right]_{\alpha_r}^{\bar{\mathbf{\Phi}}_n^{\alpha_r} = \mathbf{\Phi}_n} }{ B^{f_b} 
(\mathbf{\Phi}_n)} \Biggr\}\, .
\label{eq:powheg}
\end{multline}
The function $\bar{B}^{f_b}$ gives the NLO (QCD+EW) inclusive cross section, 
and the term between curly brackets controls the hardest emission 
(for more details on the notation, see ref.~\cite{Frixione:2007vw}). 
The inclusion of NLO EW corrections, 
with respect to the version including only QCD corrections, 
amounts to a modification of  $\bar{B}^{f_b}$ in order to include the 
virtual EW and real QED contributions, 
and the addition of subtraction counterterms and collinear remnants corresponding to the new singular regions, i.e. the ones associated  with the 
emission of a soft/collinear photon by a hard scattering quark or a soft photon by the final state lepton. 
It is worth reminding that in refs.~\cite{Barze:2012tt,Barze':2013yca}
the final state leptons have been treated with full mass dependence, 
in order to deal in a proper way with all event selections 
of experimental interest: bare as well as dressed leptons. 
In this formulation, the photons collinear to leptons give rise to logarithmic but not singular terms and they do not need subtraction. 
The Sudakov form factor $\Delta^{f_b} (\mathbf{\Phi}_n,k_T)$ is the product of individual factors defined for each singular region, 
so it has been modified as well 
in order to take care of the additional singular regions.

The \powheg\, algorithm generates the hardest emission of the event 
by probing in sequence all the singular regions described by the various Sudakov form factors 
and eventually choosing the radiated parton with largest transverse momentum. 
In this sense, the generation of a photon, either from the initial 
state or from the final state by \powheg\,  
is in competition with the generation of a parton. 
Notice that expanding eq.~\myref{eq:powheg}, 
on top of \oa\ and ${\cal O}(\alpha_s)$ corrections for the cross section, also higher order terms, 
in particular of ${\cal O}(\alpha_s^2)$ and of \oaas, 
appear and affect the description of differential distributions. 
This is described in more detail in section~\ref{sec:oalphaalphascorrections}. 

The matching of the generated events with the PS,
reaching the so called NLOPS accuracy, 
is naturally done using a $p_T$ ordered PS: 
the emissions start down from the scale set by \powheg\, 
($p_T$ of the hardest radiated particle). 
This is automatically implemented for the QCD PS when using \pythia. 
Here there is a subtlety: 
the default definition of hardness of the PS emissions is usually kinematically 
different from the way the variable ``scalup'' is calculated in \powheg\, at the level of the Les Houches event file generation 
(the different definitions agree in the limit of collinear radiation). 
In our simulations we use the version {{\sc v2 }} of the {{\sc Powheg-Box}}, 
abbreviated in the following \powhegvtwo, 
where we provide the interface to {{\sc Pythia8}\,} and {{\sc Photos++}}. 
In this interface we allow for QCD and QED radiation of partons or photons up to the kinematical limit 
and then we veto the radiation when one PS radiated parton or photon does not respect the limit imposed by ``scalup'', 
calculated according to the \powheg\, definition. 
The same strategy is adopted for the QED radiation from the resonance, 
either if it is treated with {{\sc Pythia8}\,} 
(where, by default, the starting scale is the resonance mass) 
or if it is treated by {{\sc Photos++}\,}, 
which does not generate $p_T$ ordered radiation. 

In order to perform our studies,
we added a switch to turn on and off the NLO EW corrections directly from the input file, 
thus allowing consistent and tuned estimates of the EW effects. 
We also included an interface to {{\sc Photos++}\,}, 
which takes as input the \powheg\, event file; 
and then writes a second LHE event file 
containing also the photons generated by {{\sc Photos++}\,},
as a preliminary step before the ISR radiation described by the Parton Shower. 
This second file is written in standard {\tt .lhe} format and can be read 
by any other PS Monte Carlo for additional QCD/QED radiation 
from partons. 
When using this interface to {{\sc Photos++}\,}, 
the QED radiation from $W$ and $Z$ resonances should be switched off in the PS Monte Carlo, to avoid double counting.  

An important remark on the lepton event selection is in order. 
By using fully massive kinematics for leptons, 
the generator can handle, in principle, any event selection, including,
in particular, bare leptons, i.e. the lepton momentum is never recombined with any photon.
However, at the generation stage, 
the separation between radiative events, with resolved photon radiation, 
and elastic events, with unresolved photon radiation, 
is done through the relative transverse momentum between lepton and photon. This means that, in principle, the lepton can never be considered bare 
(for this it would be necessary to set the resolution parameter
{\tt kt2minqed}\footnote{
The parameter {\tt kt2minqed} sets in \powhegvtwo\ the lowest possible value of the transverse momentum of an emitted real photon.
}
to zero, which produces an infrared divergence). 
In practice, since the peak of the distribution of the relative momentum between lepton and photon is of the order of $m_\ell$,
the lepton can be considered as a bare lepton if the condition {\tt kt2minqed} $\ll m_\ell^2$ is fulfilled. 
While for muons this is the case for the default value
{\tt kt2minqed}$ = 10^{-6}$~GeV$^2$, 
for electrons it would be required to set the scale to much lower values. 
Since at the LHC bare leptons are experimentally well defined only for muons,
the results obtained in this study with \powheg\, are presented accordingly 
only for bare muons and dressed electrons. 
We observe that the default value of {\tt kt2minqed} is consistent with the definition of dressed electrons.

\subsubsection{\powhegvtwo\, {\tt two-rad} improved version for DY processes}
\label{sec:powheg-improved}
The above brief description of the \powheg\, code 
for the CC DY process applies to the library 
{{\sc W\_ew-BMNNP}\,} svn revision {{\sc 3369}\,} 
(the same applies to {{\sc Z\_ew-BMNNPV}\,} svn revision {{\sc 3370}\,} 
for the NC DY process). 
In the treatment of both  QCD and ISR/FSR QED radiations there is a potential problem at the level of event generation, 
which can become phenomenologically important for some exclusive observables.

In fact, the largest transverse momentum of a coloured parton or photon, extracted by means of the Sudakov form factor, 
sets the maximum scale of the (QCD and QED) PS radiation. 
On average, the scale of QCD radiation, 
obtained by the inversion of the ISR Sudakov form factor,
is larger than the one of QED FSR, 
obtained by the inversion of the QED FSR Sudakov form factor. 
As a consequence, the \powheg\, first radiation is typically a QCD parton 
which sets the scale for both QCD and QED PSs.
This behaviour entails a double counting between QED radiation from matrix element
and PS, which can become severe for observables particularly sensitive 
to final state QED radiation, as is the case under study in the present work 
(quantitative estimates of the impact on $\mw$ of the effect will be given 
in section.~\ref{sec:v2vstworad}). Additionally, also spurious \oaas\ corrections can
be introduced. 

A way out of this problem is provided by the general treatment of the NLOPS matching of ref.~\cite{Jezo:2015aia} 
in the presence of radiation from final state resonances. 
According to this approach, the competition between QCD and QED radiation is present only in the initial state. 
The QED radiation from the $W$ resonance is treated separately and, 
in particular, the hardest photon scale, 
obtained through the inversion of the FSR QED Sudakov form factor, 
is kept as input to the QED PS from the resonance. 
Moreover, the events generated by the current release of {{\sc Powheg-Box-v2}} 
({{\sc W\_ew-BMNNP}\,} svn revision 3375 and {{\sc Z\_ew-BMNNPV}\,} 
svn revision 3376) can contain up to two radiated particles 
(one ISR parton/photon and one FSR photon) 
and the information about the two ISR and FSR scales. 
For this reason the matching with the QCD and QED PSs 
has been modified with respect to the original version described in section \ref{sec:powheg}. 
We have modified the {{\sc Powheg-v2/W\_ew-BMNNP}\,} code 
according to the approach of ref.~\cite{Jezo:2015aia}~\footnote{
A general {{\sc Powheg-Box-Res}\,} version,
able to treat in an automatic way arbitrary processes with resonances, is under development. 
It has already been applied to the simulation of $t$-channel single-top production~\cite{Jezo:2015aia}
and $p p \to \ell^+ \nu_\ell \ell^- {\bar \nu_\ell} b \bar b$~\cite{Jezo:2016ujg}. 
For the DY case we simply modify the {{\sc v2}\,} process libraries, since other subtleties related
to the presence of resonance and additional particles in the final state do not affect the
DY processes. In any case, the DY process libraries
will be also available under the {{\sc Powheg-Box-Res}\,}
version.}~\footnote{Similar considerations hold also
for the {{\sc Z\_ew-BMNNPV}} package.
}; we dub {\tt two-rad} the results obtained with this improved version.

\begin{figure}[!hbt]
\begin{center}
\includegraphics[height=55mm]{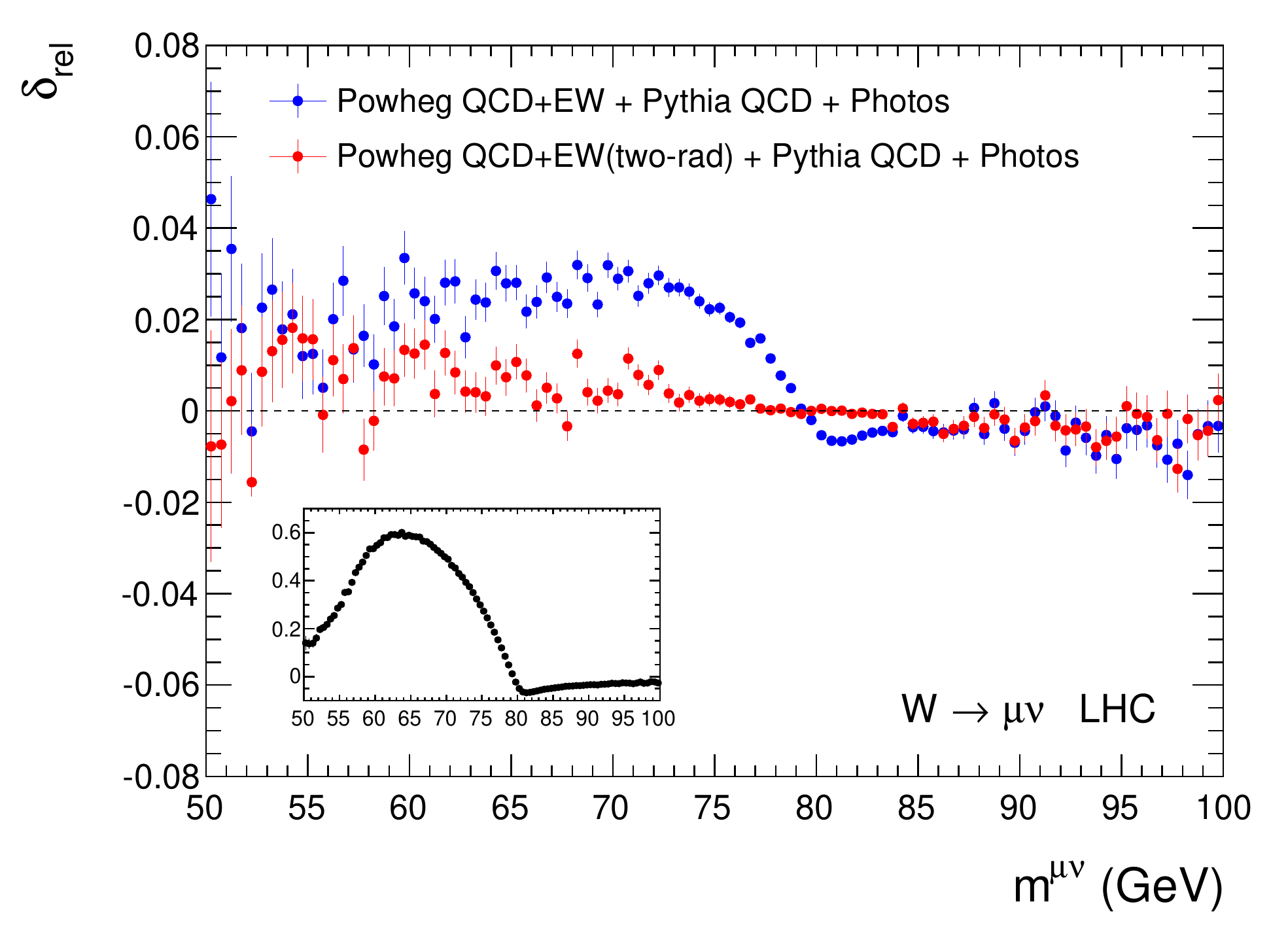}
\caption{\small 
Relative difference, for the $\mu^+ \nu$ invariant mass distribution,
normalized to the prediction of \powhegvtwo\,
with NLO QCD corrections interfaced to the \pythia\, QCD PS, 
of two different implementations of EW corrections: 
predictions of \powhegvtwo\, {\tt two-rad} with NLO QCD+EW corrections
(red dots), 
and of the old version of \powhegvtwo\, with NLO QCD+EW corrections 
(blue dots). 
Both codes are interfaced to the \pythia\, QCD PS and to \photos.  
For reference, the EW corrections normalized to the LO are reported in the insets. 
\label{fig:mixed-invm-lhc-mu}
} 
\end{center}
\end{figure}

\begin{figure}[!hbt]
\begin{center}
\includegraphics[height=55mm]{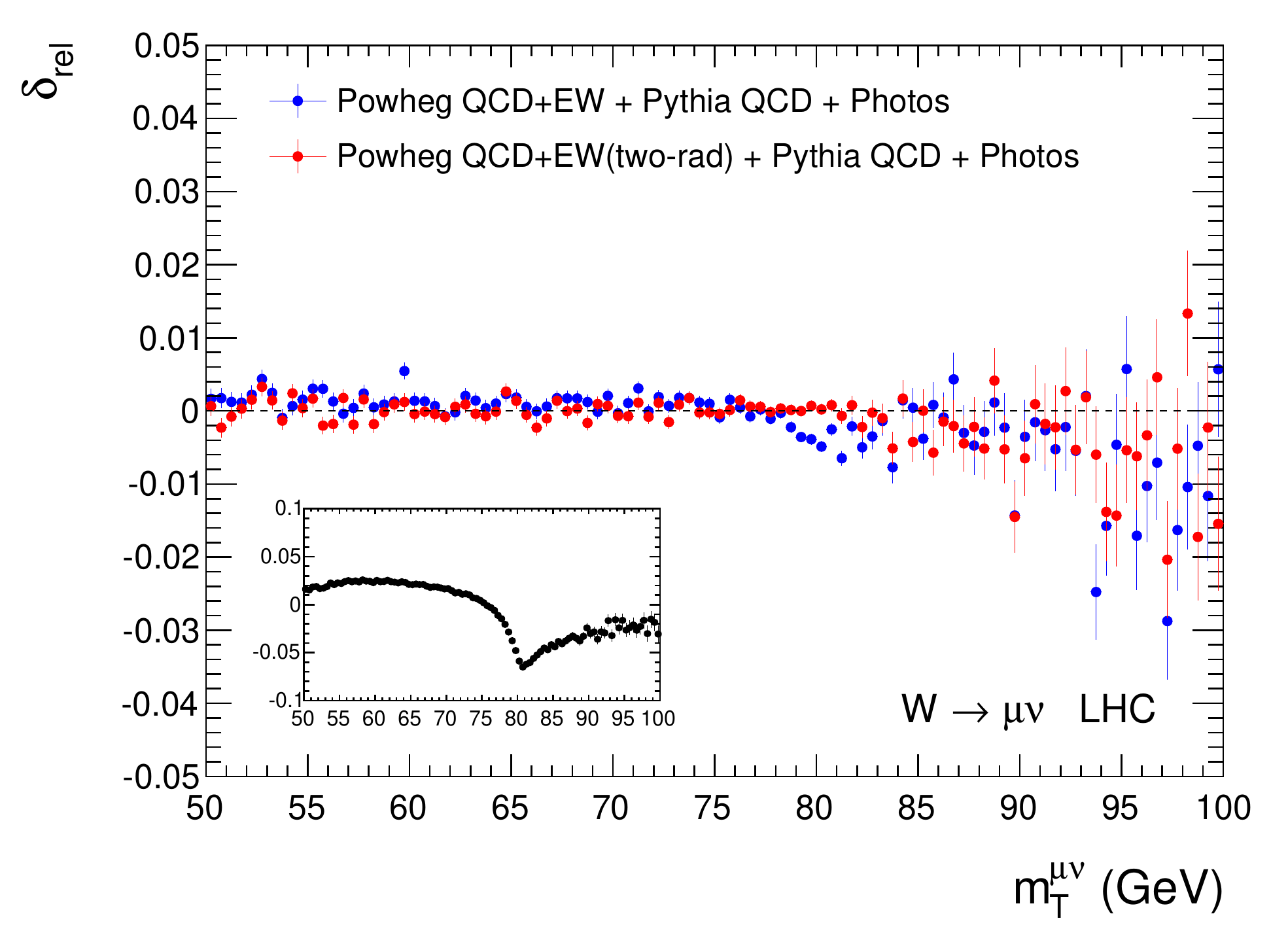}~~
\includegraphics[height=55mm]{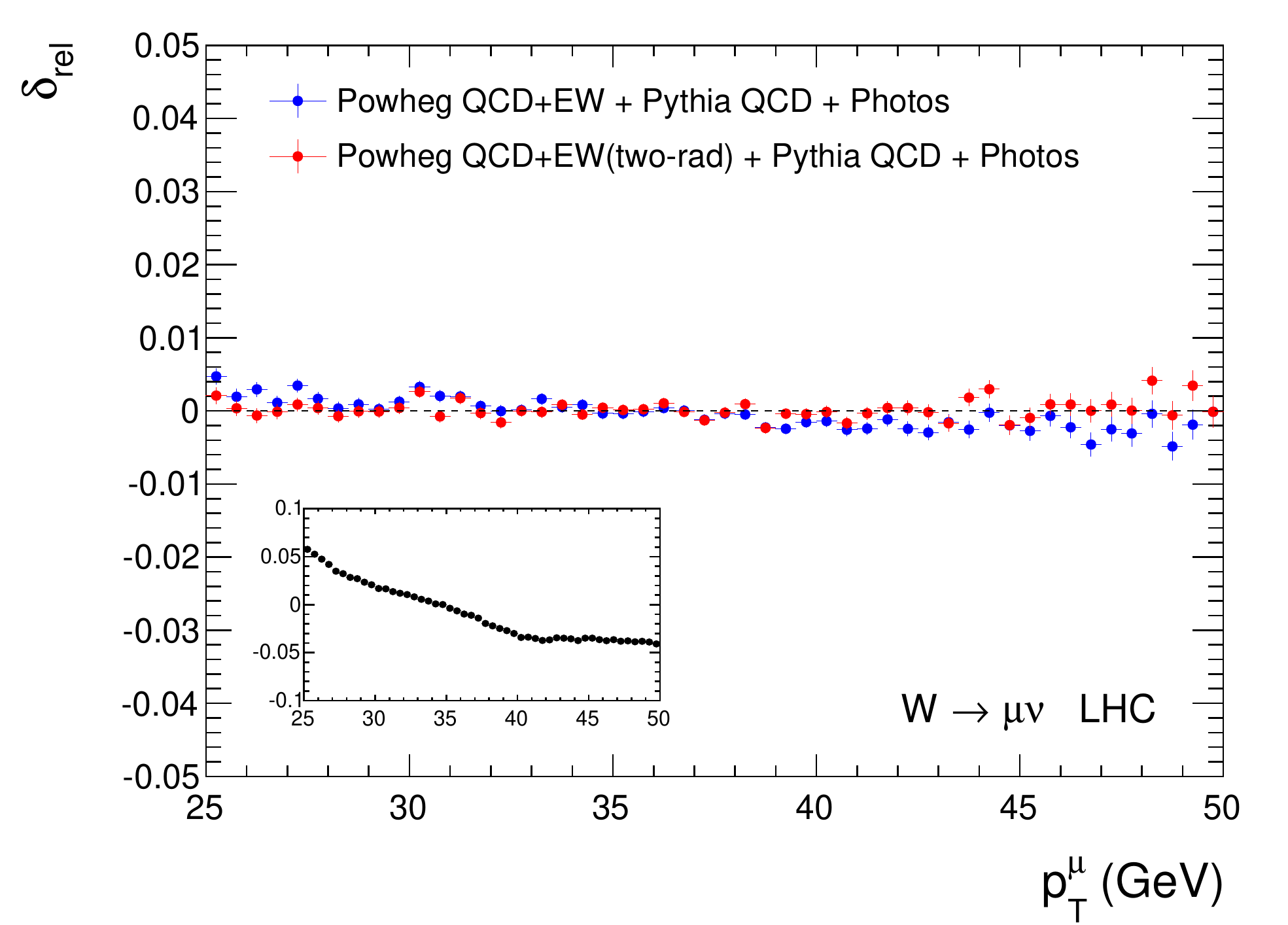}
\caption{\small Same as in figure~\ref{fig:mixed-invm-lhc-mu} for the 
lepton-pair transverse mass (left plot) and
for the lepton transverse momentum (right plot).
\label{fig:mixed-mt-pt-lhc-mu}
} 
\end{center}
\end{figure}

In figure~\ref{fig:mixed-invm-lhc-mu} we show the lepton-neutrino invariant mass,
which is the most sensitive observable to different treatments of QED radiation,
since the relative QED correction can become very large for this distribution, 
up to 60\% of the lowest order prediction,
as it can be seen from the inset of the figure.
The blue dots represent the relative difference between the predictions of
\powhegvtwo\, with NLO QCD$+$EW corrections interfaced to \pythia$+$\photos\, 
and the predictions of 
\powhegvtwo\, with only NLO QCD corrections interfaced to \pythia$+$\photos. 
As it can be seen, 
the differences start around the $W$ mass peak and reach the maximum ($\sim 2$\%) at $M(\mu^+ \nu) \simeq 70$~GeV. 
The red dots show the corresponding difference for the {\tt two-rad}
upgraded version of \powhegvtwo. 
With respect to 
the predictions of
\powhegvtwo\, with only NLO QCD corrections interfaced to \pythia$+$\photos,
there is a moderate slope but the relative effects are well below the \% level in the region around the $W$ mass peak, 
where the bulk of the cross section is concentrated.
The same relative difference is investigated in figure~\ref{fig:mixed-mt-pt-lhc-mu} for the $W$ transverse
mass (left plot) and for the lepton transverse momentum (right plot). 
Since the QED corrections to these two observables are smaller w.r.t. the ones for the lepton-pair invariant mass, 
the differences between \powhegvtwo\, standard version  and the {\tt two-rad} upgraded one are much smaller. 
However we can note a difference reaching the $0.5$\% level on the
transverse mass around the jacobian peak and a mild slope in the lepton transverse momentum distribution. 
These different shapes are at the origin of the differences
of the shifts $\Delta \mw$ 
discussed in section \ref{sec:v2vstworad}.

\section{\label{sec:thunc}Perturbative uncertainties}

%%%%%%%%%%%%%%%%%%%%%%%%%%%%%%%%%%%%%%%%%%%%%%%%%%%%%%%%%%%%%%%%%%%%%%
\subsection{EW input scheme}
\label{sec:inputscheme}

The evaluation of the NLO-EW corrections requires 
a renormalization procedure 
to define the couplings and the masses which appear in the scattering amplitudes;
the renormalized parameters are then expressed in terms of physical observables. 
The couplings of the gauge sector of the EW SM, 
namely the $SU(2)_L$ and $U(1)_Y$ gauge couplings $(g,g')$, 
the vacuum expectation value $v$ of the Higgs doublet and 
the quartic coupling $\lambda$ of the Higgs scalar potential,
can be expressed in terms of
different possible sets of measured quantities.
The difference between two input choices can be regarded as a finite renormalization; as such, it yields terms of higher perturbative order, with respect to the NLO-EW accuracy of the calculation, 
namely terms of ${\cal O}(\alpha^2)$.

The input parameter choices affect not only the overall normalization of the cross section, but also the precise shape of the differential distributions and eventually the $\mw$ determination.
The size of these effects depends on the precise formulation of the calculation, whether it is a purely fixed order analysis or whether it involves the matching of exact matrix elements with a Parton Shower.

We consider in this section three input schemes:
the so called $\alpha_0$ scheme and two variants of the so called $G_\mu$ scheme.
The $\alpha_0$ scheme is based on the use, as input parameters,
of $(\alpha(0),\mw,\mz,\mh)$, 
which are respectively the Thomson value of the fine structure constant and the $W$, $Z$ and Higgs boson masses; 
the renormalization is performed in the on--shell scheme, computing the
counterterms associated to the electric charge, the $W$ and $Z$ boson masses.
As shown in the literature \cite{Dittmaier:2001ay,Arbuzov:2005dd}, 
this scheme turns out to be rather unnatural for the description of the CC DY 
process, because it maximizes the contribution of NLO EW corrections 
through the fermion--loop contribution to the $W$ self--energy. 
The latter can be reabsorbed in the measured value of the muon decay constant $G_\mu$ and 
the input parameter scheme defined in terms of the 
quantities ($G_\mu$, $M_W$, $M_Z$, $\mh$) is known as $G_\mu$ scheme, 
typically used in the literature for the calculation of the radiative
corrections to the CC DY process.
However, the $\alpha_0$ 
option, in spite of its drawbacks, can not be left out {\it a priori}, being 
a fully consistent and predictive theoretical scheme.

The choice of the EW input scheme affects
the computation of the observables of the CC DY process in the presence of 
exact \oa\ corrections. 
To introduce the $G_\mu$ scheme, it is worth noting that 
the EW tree--level amplitude is proportional to the $SU(2)$ gauge coupling
$g^2$ rather than to $\alpha_0$.
Therefore it is possible to exploit the well known relation between
the muon decay constant $G_\mu$ and the radiative corrections to the
muon decay amplitude represented by the $\Delta r$ parameter to write
\be
\frac{G_\mu}{\sqrt{2}}=\frac{g^2}{8 M_W^2}\left(1+\Delta r  \right) \, .
\label{mudecay}
\ee
For later convenience, we introduce the effective electromagnetic
coupling $\alpha_{\mu}$, at tree--level and in one--loop approximation, 
according to the following definitions:
\bea
\alpha_{\mu}^{tree} &\equiv& \frac{\sqrt{2}}{\pi} G_\mu M_W^2 \sin^2\tw \, ,
\label{amutree}\\
\alpha_{\mu}^{1l} &\equiv& \frac{\sqrt{2}}{\pi} G_\mu M_W^2 \sin^2\tw 
\left(1-\Delta r \right) \, ,
\label{amu1l}
\eea
where $\sin^2\tw = 1 - M_W^2/M_Z^2$ is the squared sine of the on-shell weak mixing angle. In the $G_\mu$ scheme, the Born cross section $\sigma_{0}$ is proportional to $(\alpha^{tree}_\mu)^2$.
When performing a complete \oa\ calculation in this scheme, 
it is necessary to add a
finite countertem $(-2 \sigma_{0}\Delta r)$ to the virtual corrections
to avoid a double counting of the Born contribution,
expressed in terms of $G_\mu$, already present in the diagrammatic results.

We write schematically the \oa\ cross section in the 
$\alpha_0$ and $G_\mu$ schemes,
making explicit the dependence on the couplings 
at the different perturbative orders
($\sigma_{SV}$ and $\sigma_H$ label the soft+virtual and the real hard emission contributions)
and distinguishing two possible options for the $G_\mu$ scheme. 
We have in principle the three following alternatives, which
differ by ${\cal  O}(\alpha^2)$ corrections:
\bea
\alpha_0~~ :&& ~~~ \sigma = \alpha_0^2 \sigma_0 + \alpha_0^3 (\sigma_{SV}+\sigma_H)\, , \label{scha0}  \\
G_\mu~ I~ :&&
~~~ \sigma = (\alpha_\mu^{tree})^2 \sigma_0 + (\alpha_\mu^{tree})^2 \alpha_0
 (\sigma_{SV}+\sigma_H) -2 \Delta r (\alpha_\mu^{tree})^2 \sigma_0\, , \label{schgmu1} \\
G_\mu~ II~ :&&
~~~ \sigma = (\alpha_\mu^{1l})^2 \sigma_0 + (\alpha_\mu^{1l})^2 \alpha_0
 (\sigma_{SV}+\sigma_H) \, . \label{schgmu2}
\eea
We introduce the idea of sharing
i.e. the relative percentage of 0-- and 1--photon contributions
with the real-photon energy greater than a certain threshold.
The 0--photon subset receives contributions from the Born cross section
and from the soft+virtual \oa\ corrections; the latter contain in particular the renormalization terms.
As a consequence of eqs.~(\ref{scha0}-\ref{schgmu2}),
we show in table \ref{tab:schemes}
the expression of the correction factors $F_{SV}$,
introduced in section \ref{sec:horace}.
\begin{table}
$$
\begin{array}{l|l|c}
{\rm Scheme} & \quad\quad F_{SV} & {\rm Couplings~of}~ |{\cal M}_{n,LL}|^2 \\
\hline
\alpha_0& 
~~~1+\alpha_0  \frac{\sigma_{SV}-\sigma_{SV}^{LL}}{\sigma_0}\label{fsva0} & 
~~~\alpha_0^2 \alpha_0^n\\
G_\mu~I& 
~~~1+\alpha_0 \frac{\sigma_{SV}-\sigma_{SV}^{LL}}{\sigma_0} -2\Delta r \label{fsvgmu1} & 
~~~(\alpha_\mu^{tree})^2 \alpha_0^n \\
G_\mu~II& 
~~~1+\alpha_0 \frac{\sigma_{SV}-\sigma_{SV}^{LL}}{\sigma_0}~~&
~~~(\alpha_\mu^{1l})^2 \alpha_0^n \label{fsvgmu2}
\end{array}
$$
\caption{\label{tab:schemes} Comparison of different renormalization input schemes: structure of the $F_{SV}$ soft+virtual correction factor and proportionality factor of the matrix element describing the emission of $n$ real photons.}
\end{table}
It can be seen that the correction factors $F_{SV}$ in the $\alpha_0$
and in the $G_\mu~II$ schemes are the same, while the factor in the $G_\mu~I$ scheme is different.
Concerning the squared matrix elements $|{\cal M}_{n,LL}|^2$ according to the three options, 
we show in table~\ref{tab:schemes} their dependence on the coupling constant;
the correction factors $F_{H,i}$ of eq.~\myref{eq:matchedinfty}
are equal in the three schemes since the same proportionality is
present in the exact squared matrix elements $|{\cal M}_{n}|^2$.
In summary, the EW input schemes described above yield a different sharing of 0-- and 1--photon events, 
which in turn can imply a different distortion of
the distributions used to extract the $W$ boson mass. 
In particular, the $\alpha_0$ and the $G_\mu~II$ schemes have the same sharing,
despite of the different normalization.

Now we consider the matching of NLO-EW results with a QED PS,
as described in  eq.~\myref{eq:matchedinfty}.
It is worth noticing that the sharing of the different photon multiplicities is the same in the three schemes discussed above, 
as can be deduced from the facts that the $F_{H,i}$ factors are the same 
and $F_{SV}$ is factorized.
As a consequence, 
we expect that the sensitivity of the matched cross section as 
given by eq.~\myref{eq:matchedinfty} to the
input scheme choice is reduced w.r.t. the pure \oa\ prediction.
We stress that the $F_{SV}$ factor is not constant with respect to the kinematical invariants, but it has a mild dependence on them and thus it can still modify the shape of the distributions.

While eq.~\myref{eq:matchedinfty} describes the structure of a purely EW event generator, it is interesting to consider how the input parameter choices affect the predictions of \powheg, whose formulation is shown in 
eq.~\myref{eq:powheg}, where QCD and EW corrections are mixed.
The EW virtual corrections and all the terms associated with the renormalization are included in the factor $\bar{B} (\mathbf{\Phi}_n)$.
Similarly to the purely EW case, 
this factor has a mild dependence on the event kinematics
and
it rescales in the same way all the real parton multiplicities.

%%%%%%%%%%%%%%%%%%%%%%%%%%%%%%%%%%%%%%%%%%%%%%%%%%%%%%%%%%%%%%%%%%%%%%
\subsection{Mixed ${\cal O}(\alpha\alpha_s)$ corrections }
\label{sec:oalphaalphascorrections}

Due to the factorization properties of the IR soft/collinear 
singularities of QCD and QED origin, 
the available generators, 
used to extract $\mw$ by fitting the experimental data, 
effectively include the leading structures of the factorized mixed QCD-EW corrections. 
It is therefore important to investigate 
the role of the \oaas terms included in these generators
and to attempt an estimate of the impact on $\mw$ of the residual 
\oaas corrections which are not available in the codes.

The distributions predicted by the code adopted in the Tevatron analysis, 
{\em i.e.} \resbos + \photos, 
include the effects, in a factorized form, of initial state QCD corrections and of final state QED corrections.
In the present study we consider a similar combination, 
which is obtained in \powhegvtwo\, code with NLO (QCD+EW) corrections, 
by switching off NLO EW corrections
and by including QED-LL final-state corrections 
to all orders by means of \photos\, or \pythia 8
(for the latter code we dub the corresponding routines \pythiaqed,
to distinguish them from the QCD ISR PS);
this combination includes terms of order 
\begin{equation}
\alpha_s \alpha \left( c_2 L_{\rm QCD}^2 
                     + c_1 L_{\rm QCD} 
                     + c_0 \right) 
                \left( c_{11} L_{\rm QED} l_{\rm QED}
                     + c_{10} L_{\rm QED}
                     + c_{01} l_{\rm QED}
                \right)\, ,
\label{eq:alphas-alpha-resbos}
\end{equation}
where $L_{\rm QCD}$ stands for the logarithm of the scale of the process $Q^2$ 
over the square of the dimensionful observable under study, 
$L_{\rm QED} = \log \left( Q^2 / m_l^2\right)$ 
($m_l$ being the mass of the final state charged lepton) and $l $ is the 
log of soft infrared origin, effectively generated by the applied cuts. 

If, on the other hand, we consider the code \powhegvtwo\, 
with the NLO EW corrections turned on 
and QED-LL final-state  corrections accounted for to all orders 
by means of \photos\, (or \pythiaqed), 
the included \oaas terms have the form 
\begin{equation}
\alpha_s \alpha \left( c_2 L_{\rm QCD}^2 
                     + c_1 L_{\rm QCD} 
                     + c_0 \right) 
                \left( c_{11} L_{\rm QED} l_{\rm QED}
                     + c_{10} L_{\rm QED}
                     + c_{01} l_{\rm QED}
                     + c_{00}
                \right)\, ,
\label{eq:alphas-alpha-powhegqcdew}
\end{equation}
With respect to eq.~\myref{eq:alphas-alpha-resbos}, 
eq.~\myref{eq:alphas-alpha-powhegqcdew} contains in addition the 
term 
$$
\alpha_s \alpha\, c_{00} \left( c_2 L_{\rm QCD}^2 
                     + c_1 L_{\rm QCD} 
                     + c_0 \right). 
$$
This term is available in \powhegvtwo\, as a consequence of the factorized structure of eq.~\myref{eq:powheg} and reproduces correctly a subset of 
\oaas\ in the limit of collinear QCD radiation.
Its inclusion represents a possible improvement of the simulation tools used in the $\mw$ studies, although the \oaas\ accuracy can not be claimed 
because the complete set of the exact matrix elements with this perturbative accuracy is not available.
On the other hand, this term is missing in the Tevatron analysis 
and should thus be treated as a source of theoretical uncertainty affecting the Tevatron $\mw$ determination; we investigate this point in the following sections.

\section{\label{sec:distributions}Impact of radiative corrections on the kinematical distributions}

\begin{figure}[hbtp]
\begin{center}
\includegraphics[height=55mm]{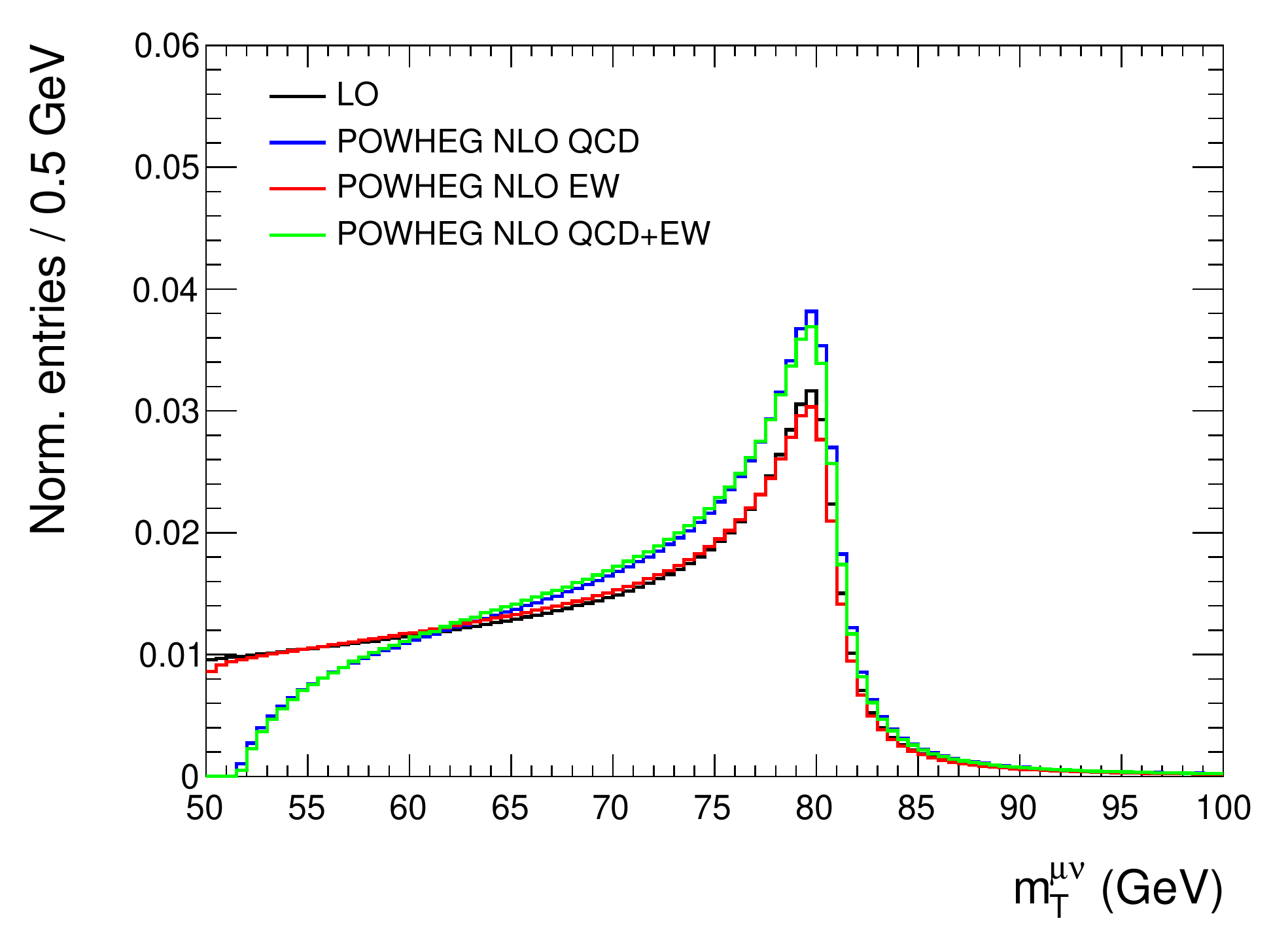}~~
\includegraphics[height=55mm]{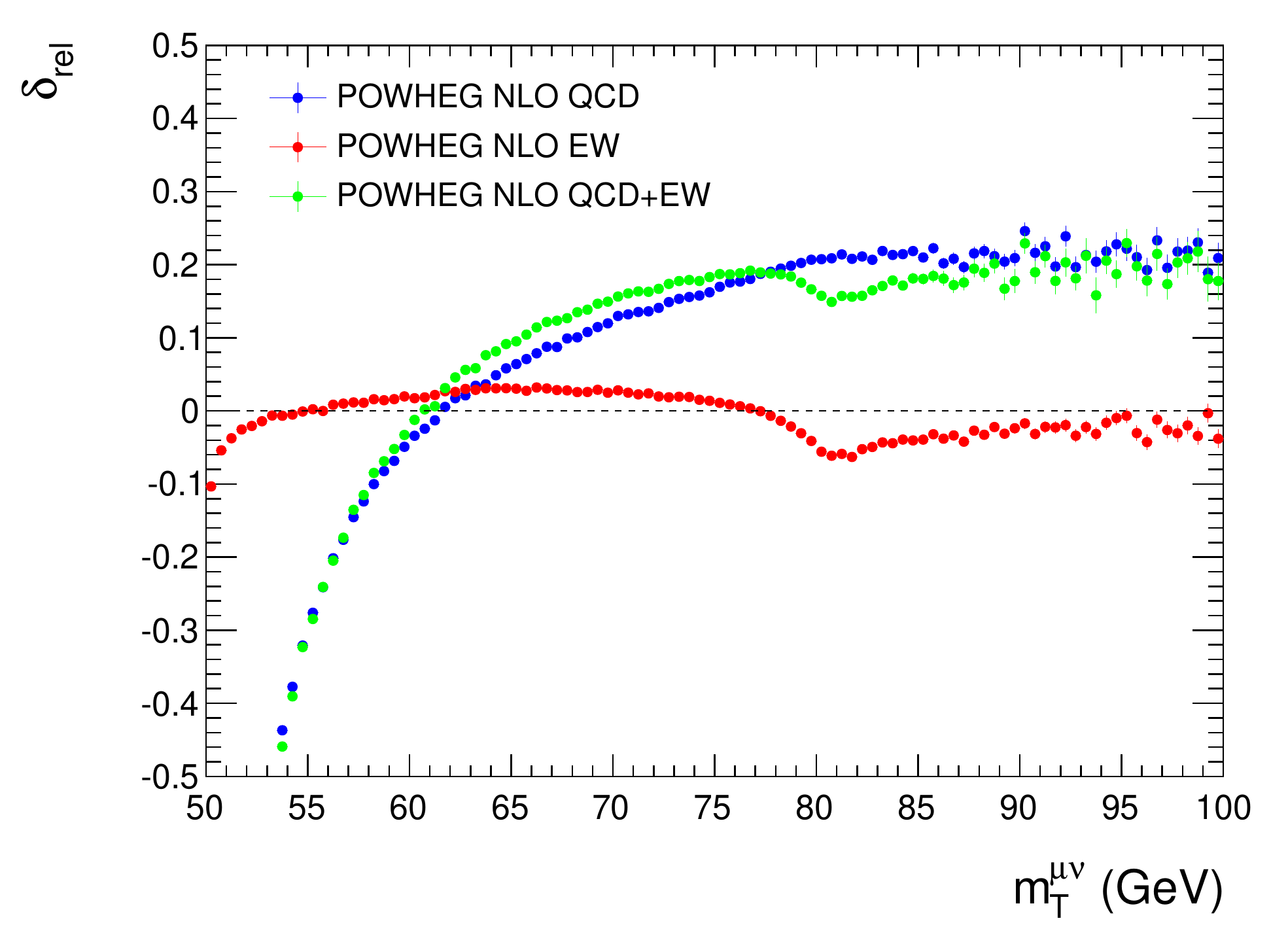}
\caption{\small 
Fixed-order predictions for the transverse mass distribution,
in the case of $W^+$ production with muons in the final state at LHC 14 TeV
and acceptance cuts as in table~\ref{tab:parameters_comparison_with_fixed_order}.
We show different perturbative approximations,
including only NLO QCD, only NLO EW and the sum of the two sets of corrections.
In the left plot we show the shape of the distributions and in the right plot the relative effect of the radiative corrections, normalized to the LO prediction.
\label{fig:nlo}
} 
\end{center}
\end{figure}

In order to set the stage of the discussion,
we present in figure~\ref{fig:nlo} the impact of exact fixed-order corrections
to the lepton-pair transverse mass distribution, with muons in the final state,
at the LHC with $\sqrt{s} = 14$~TeV, in the case of $W^+$ production.
We consider NLO QCD, NLO EW effects and the sum of the two sets of corrections and we show their relative impact
normalized to the LO prediction. We observe the negative impact of EW corrections at the jacobian peak of the distribution and the monotonic increase due to QCD effects.
When summing NLO QCD+EW corrections, we obtain a partial cancellation 
of the radiative effect at the jacobian peak.

In the following sections we study the impact of higher-order radiative corrections on the kinematical distributions relevant for the $\mw$ determination.
In particular we focus on:
{\it i)} light lepton pairs corrections; {\it ii)} the modeling of QED radiation; 
{\it iii)} the influence of QCD contributions and their interplay with purely QED FSR effects 
and 
{\it iv)} mixed \oaas corrections beyond the approximation 
that combines QCD with purely QED FSR corrections.

%%%%%%%%%%%%%%%%%%%%%%%%%%%%%%%%%%%%%%%%%%%%%%%%%%%%%%%%%%%%%%%%%%%%%%%%%
\subsection{Light lepton pairs radiation}
\label{sec:distr:pair}
\begin{figure}[hbtp]
\begin{center}
\includegraphics[height=55mm]{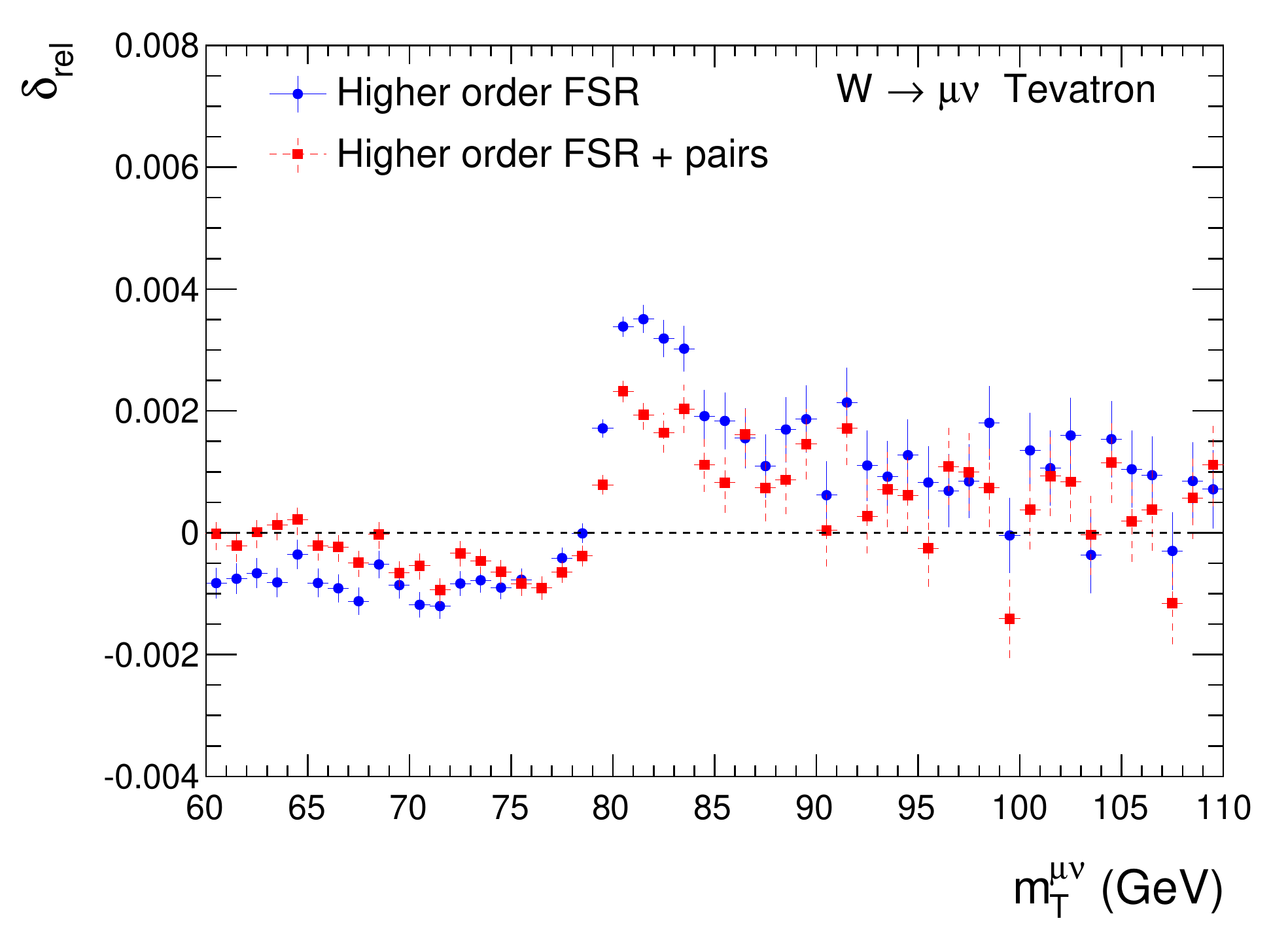}~~
\includegraphics[height=55mm]{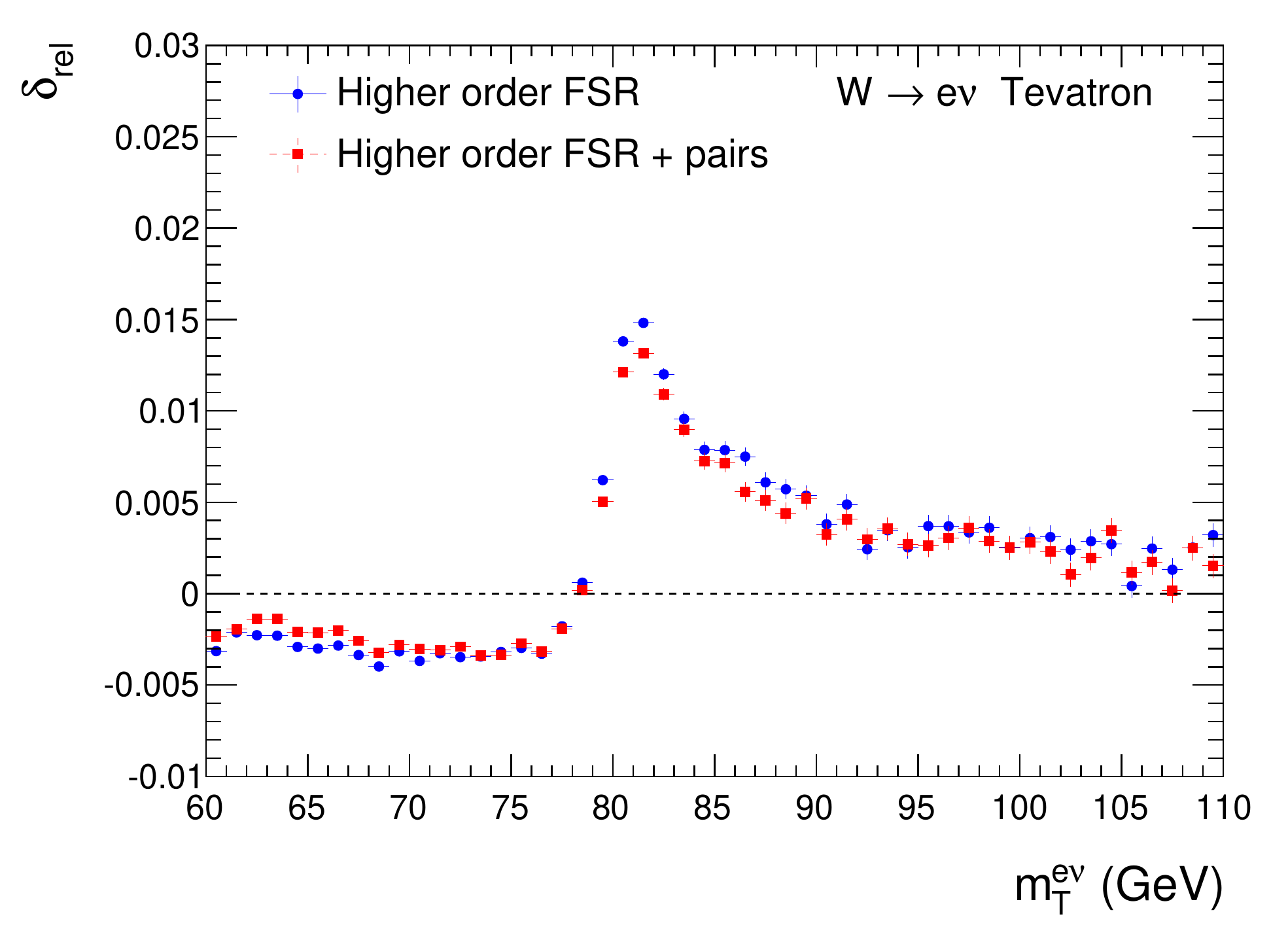}
\caption{\small Relative contribution of higher-order FSR and higher-order FSR plus pair radiation normalized to one-photon emission, 
for $W^+$ decay into muons (left) and into bare electrons (right). 
Predictions from \horace\, at Tevatron energy, with the acceptance cuts of table~\ref{tab:parameters_qed}.
\label{fig:pairsvsho}
} 
\end{center}
\end{figure}

In figure~\ref{fig:pairsvsho} we study the effect of two 
sets of corrections that start at ${\cal O}(\alpha^2)$: 
we show the contribution of additional light lepton pairs radiation 
in comparison with the contribution of multiple FSR beyond \oa,
both normalized to the distribution computed strictly with one FSR photon emission. 
The relative effects are calculated 
in the case of the decays $W \to \mu \nu$ and $W \to e \nu$ 
(where $e$ is a bare electron) 
at the Tevatron.
We observe that higher-order FSR yields at the jacobian peak an increase of the transverse mass distribution, ranging from 0.35\% for muons to 1.5\% for bare electrons,
at variance with the \oa\ effect which is instead negative.
The additional ${\cal O}(\alpha^2)$ contribution due to the emission of light pairs is instead negative, at the level of -0.15\%, to a large extent independent of the radiating lepton.
As detailed in appendix \ref{sec:lpa}, 
this behavior is simply a consequence of the dependence of the expansion parameter $\beta_f (s)$ 
given in eq.~\myref{eq-betarunning} on $m_f$, 
where $m_f$ is the mass of the radiating particle of flavor $f$.
Notice that
in the case of dressed electrons, 
for which accompanying photons are recombined with the electrons within a given angular resolution, 
the contribution of FSR will diminish while the 
correction due to undetected lepton-pair radiation will be left unchanged. Indeed, 
lepton pairs, at variance with respect to photons, can not be recombined 
with the emitting particle, because of 
physical and experimental reasons.

\subsection{QED FSR modeling}
\label{sec:distr:fsr}

The description of multiple QED FSR can be achieved with different tools, like \horaceo, \photos\, or \pythiaqed. The $\mw$ determination is very sensitive to the details of QED FSR description, so that a comparison of the predictions of these three tools and the evaluation of the corresponding impact on $\mw$ are in order.
\begin{figure}[!hbt]
\begin{center}
\includegraphics[height=55mm]{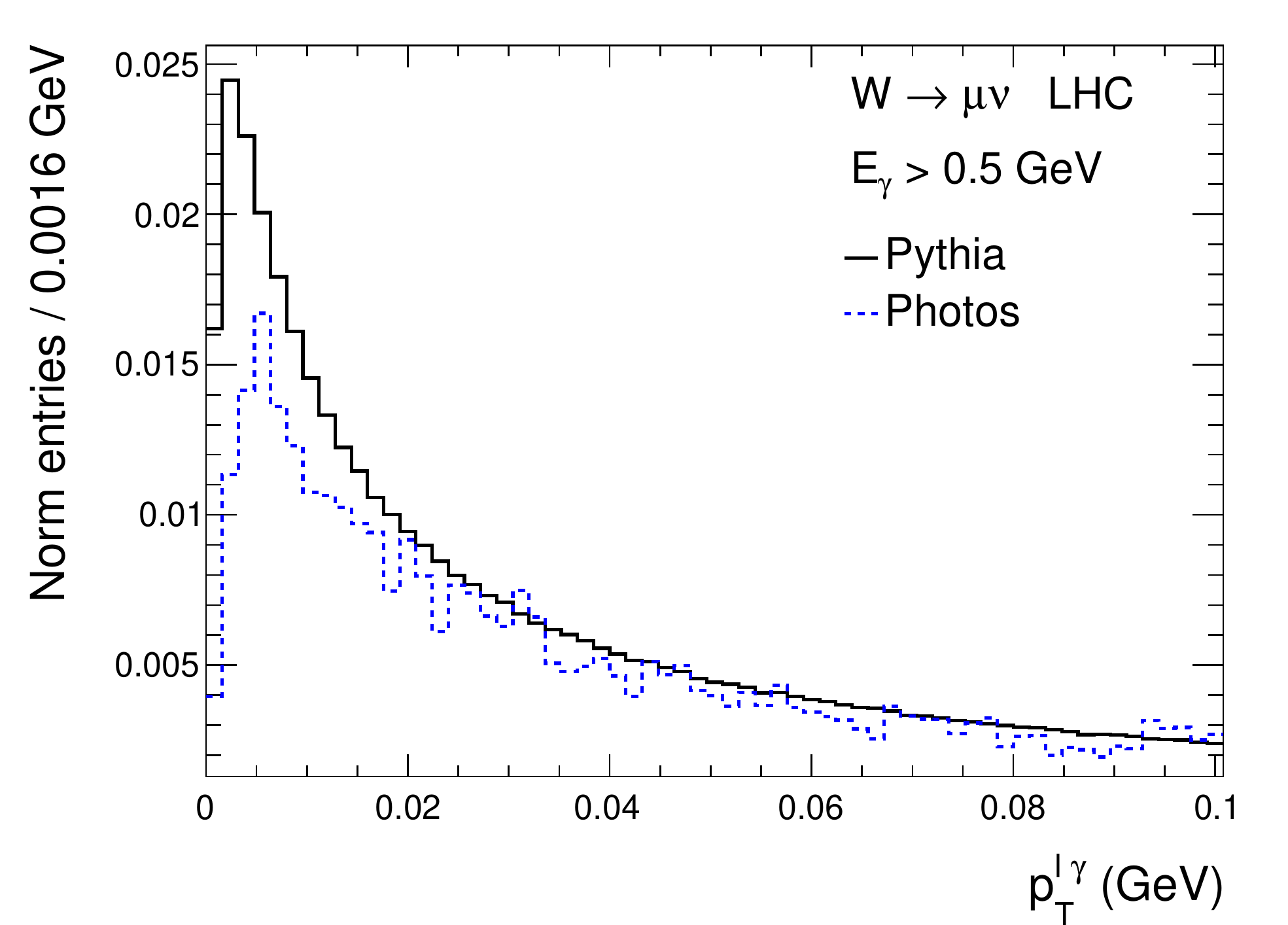}~~
\includegraphics[height=55mm]{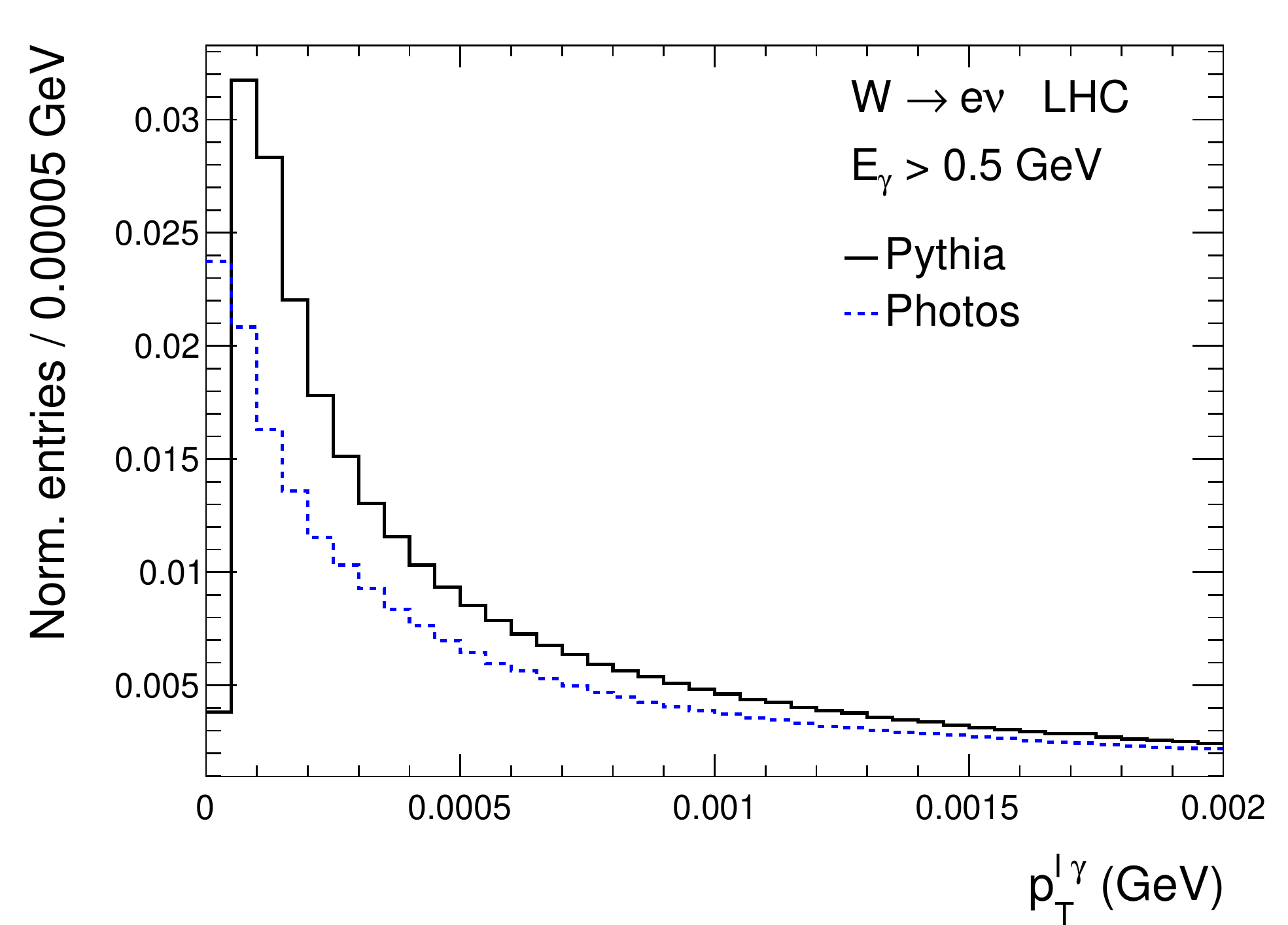}\\
\includegraphics[height=55mm]{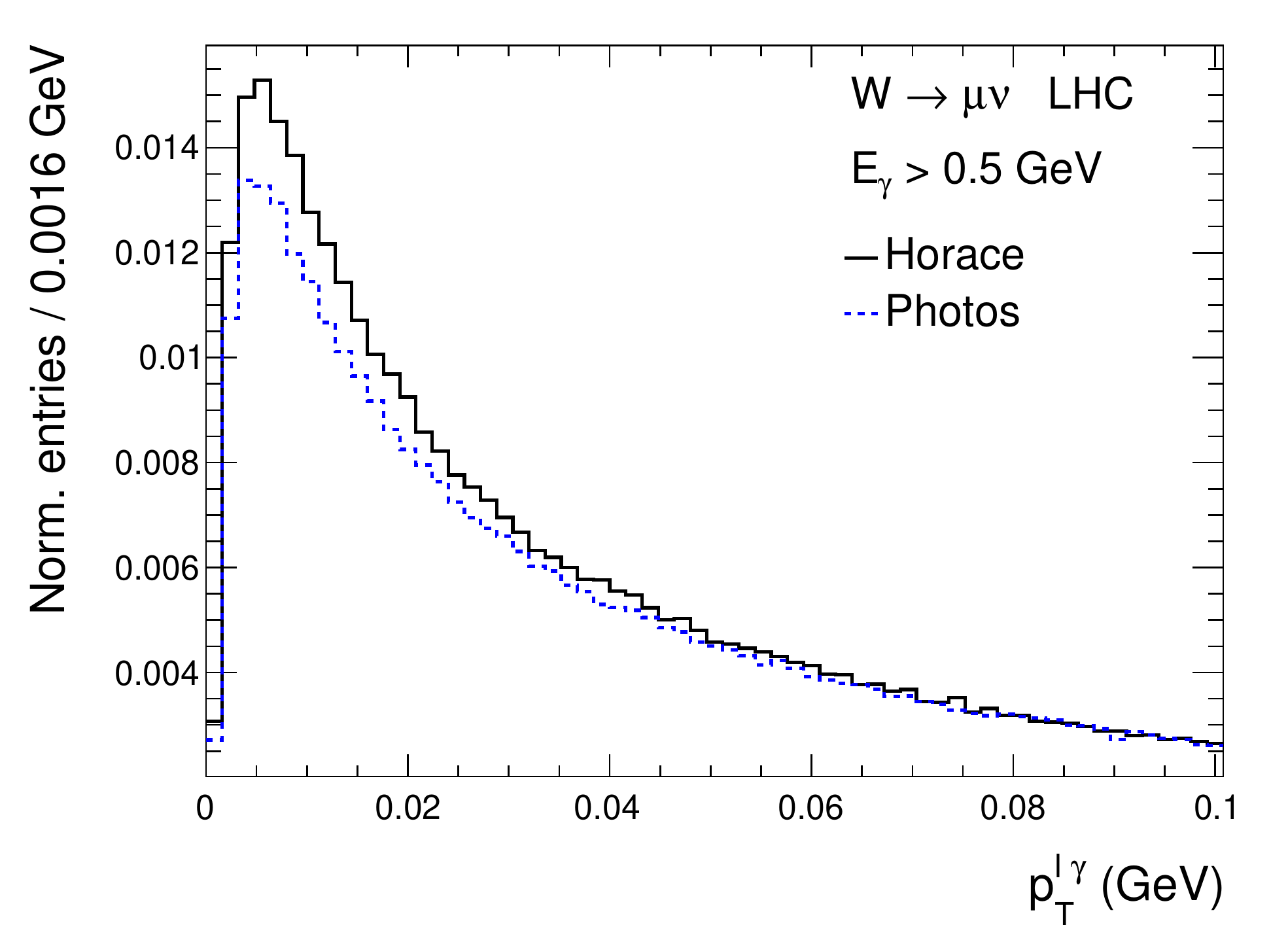}~~
\includegraphics[height=55mm]{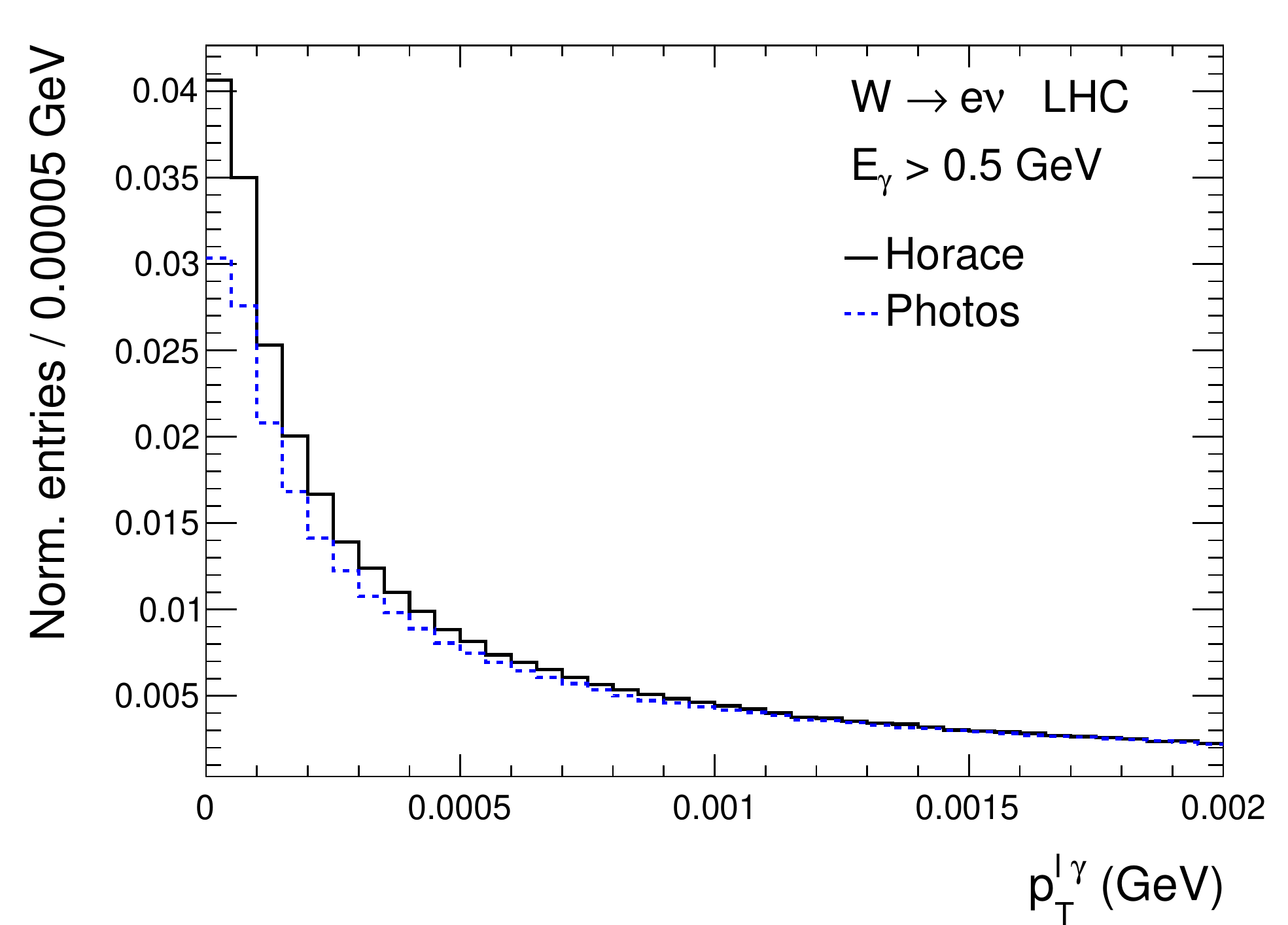}\\
\caption{\small 
Shape of the distribution of the relative lepton-photon transverse momentum $p_T^{\ell\gamma}$
for the decays $W^+ \to \mu^+ \nu$ (left plots) 
and $W^+ \to e^+ \nu$ (right plots) at the LHC 14 TeV, with acceptance cuts as in table~\ref{tab:parameters_lhc}. 
In the upper plots we show the 
comparison of the results obtained interfacing \powhegvtwo\, NLO QCD results with \pythiaqed\,
and \photos.
In the lower plots we show the
comparison of the results obtained with \horaceo\, including QED FSR effects
and those obtained interfacing \horaceo\, LO with \photos,
in both cases without QCD corrections.
\label{fig:qed}
} 
\end{center}
\end{figure}
We analyze the predictions obtained with the three codes 
for the energy and angular distributions of the emitted radiation in 
$W \to \mu \nu$ and $W \to e \nu$ decays. 
We find that the predictions for the photon energy spectrum
are in good agreement, whereas the angular distributions show some discrepancies.

For example, figure~\ref{fig:qed}
shows the comparison of the \photos, \pythia\, and \horaceo\, predictions 
for the relative lepton-photon transverse momentum
$p_T^{\ell\gamma}$, $\ell = \mu, e$, distributions at the LHC. 
We impose $E_{\gamma}^{\rm min} = 0.5$~GeV and 
we consider the events with strictly only one-photon in the final state.
In the upper plots of figure~\ref{fig:qed},
we focus on \photos\, and \pythiaqed, 
which provide for this observable different predictions, 
with a dependence of the size of the discrepancy 
on the flavor of the final state bare lepton.
This discrepancy disappears in the case of dressed electrons. 
On the other hand, as it is shown in the lower plots of figure~\ref{fig:qed},
\photos\, and \horaceo\, results are in much closer agreement.

The observed discrepancy can be ascribed to the theoretical model implemented in \pythiaqed, 
which simulates the angular degree of freedom of QED radiation 
according to the distribution $d p_T / p_T$, $p_T \equiv p_T^{\ell\gamma}$. 
On the contrary, 
\photos\, and \horaceo\, describe the angular variable using a similar model, dictated by the eikonal approximation 
$d \sigma / d \cos \theta \propto 1 / (1 - \beta \cos \theta)$, 
where $\cos\theta \equiv \cos\theta_{\ell \gamma} $ and $\beta$ is the lepton velocity.
All the three codes have LL accuracy with respect to $L_{\rm QED}$, but differ at the level of subleading terms, with different approximations of the exact matrix element for one photon emission.
The different formulation allows to explain the flavor dependence of the discrepancy, with an enhancement in the case of bare electrons where 
$L_{\rm QED}$ has a larger value. 
In the case of dressed electrons, the recombination procedure integrates over the phase-space region where the discrepancy stems from, letting the latter vanish.
%%%%%%%%%%%%%%%%%%%%%%%%%%%%%%%%%%%%%%%%%%%%%%%%%%%%%%%%%%%%%%%%%%%%%%%%%%
\subsection{QCD corrections and QCD-QED interplay}
\label{sec:distr:qcd}
Fixed- and all-orders QCD corrections have a fundamental role in the description of the observables relevant for the $\mw$ determination. 
NLO QCD corrections yield a large positive increase 
of ${\cal O}(20\%)$ of the cross section and 
QCD multiple parton ISR is crucial for a realistic prediction of the shape of
kinematical distributions.
The lepton transverse momentum receives large QCD corrections, 
because of the presence of a logarithmic enhancement factor associated to the initial state collinear singularities; the latter require the resummation to all orders of this class of corrections and yield a sizeable effect on this observable.
The lepton-pair transverse mass distribution instead
is mildly affected by perturbative QCD contributions,
because of a more systematic cancellation of these large logarithmic corrections.

\begin{figure}[hbt]
\begin{center}
\includegraphics[height=55mm]{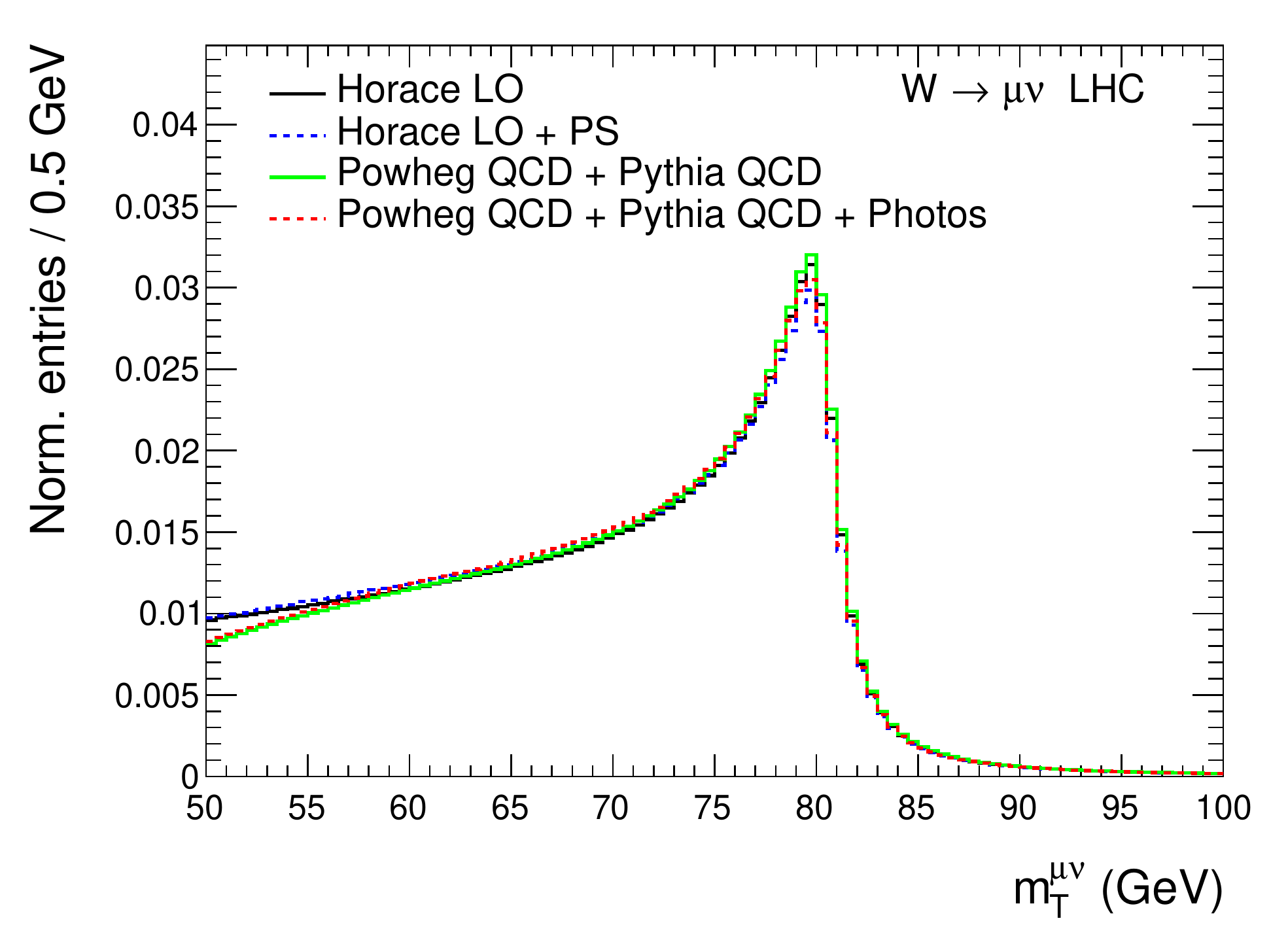}~~
\includegraphics[height=55mm]{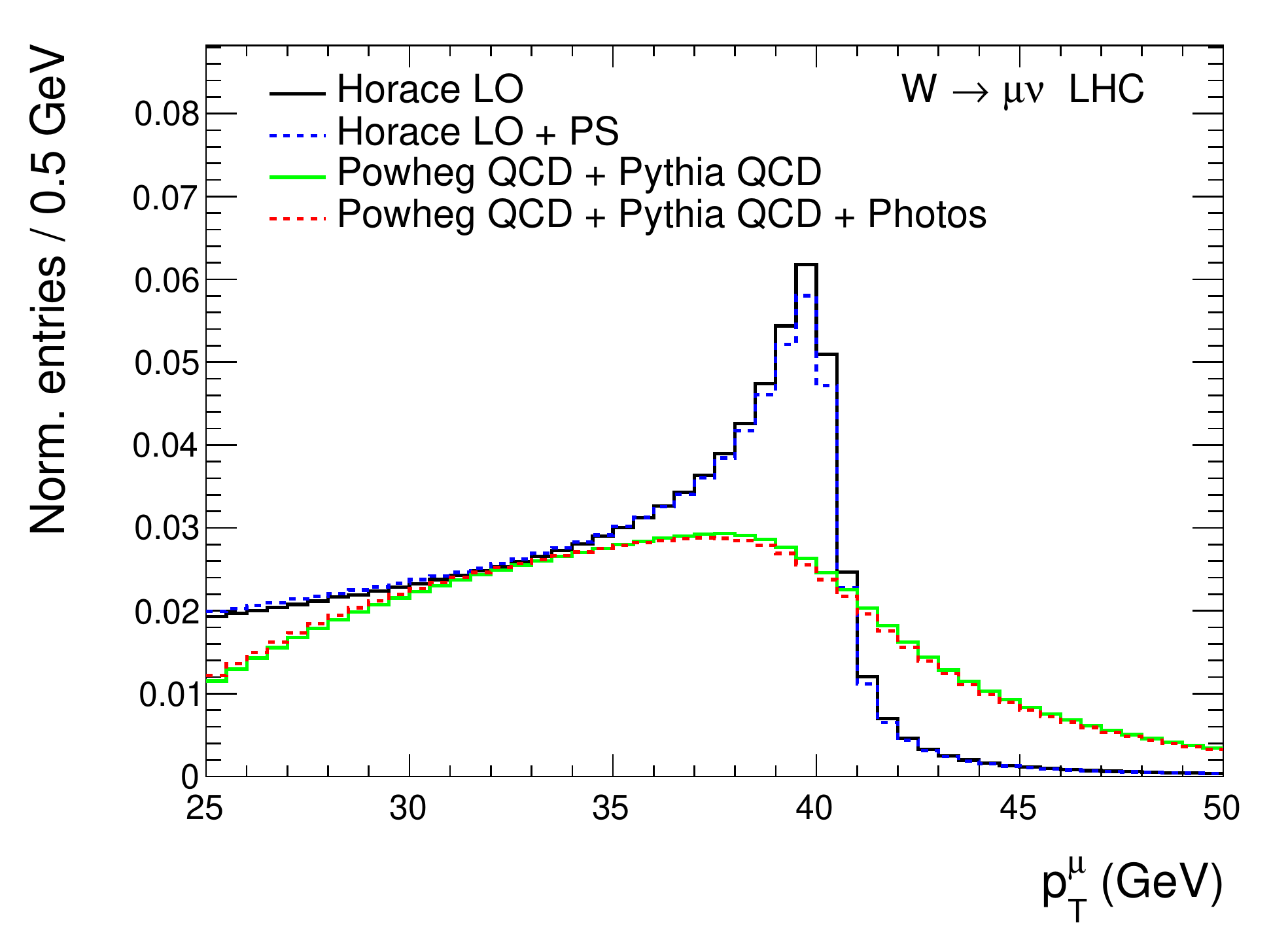} \\
\includegraphics[height=55mm]{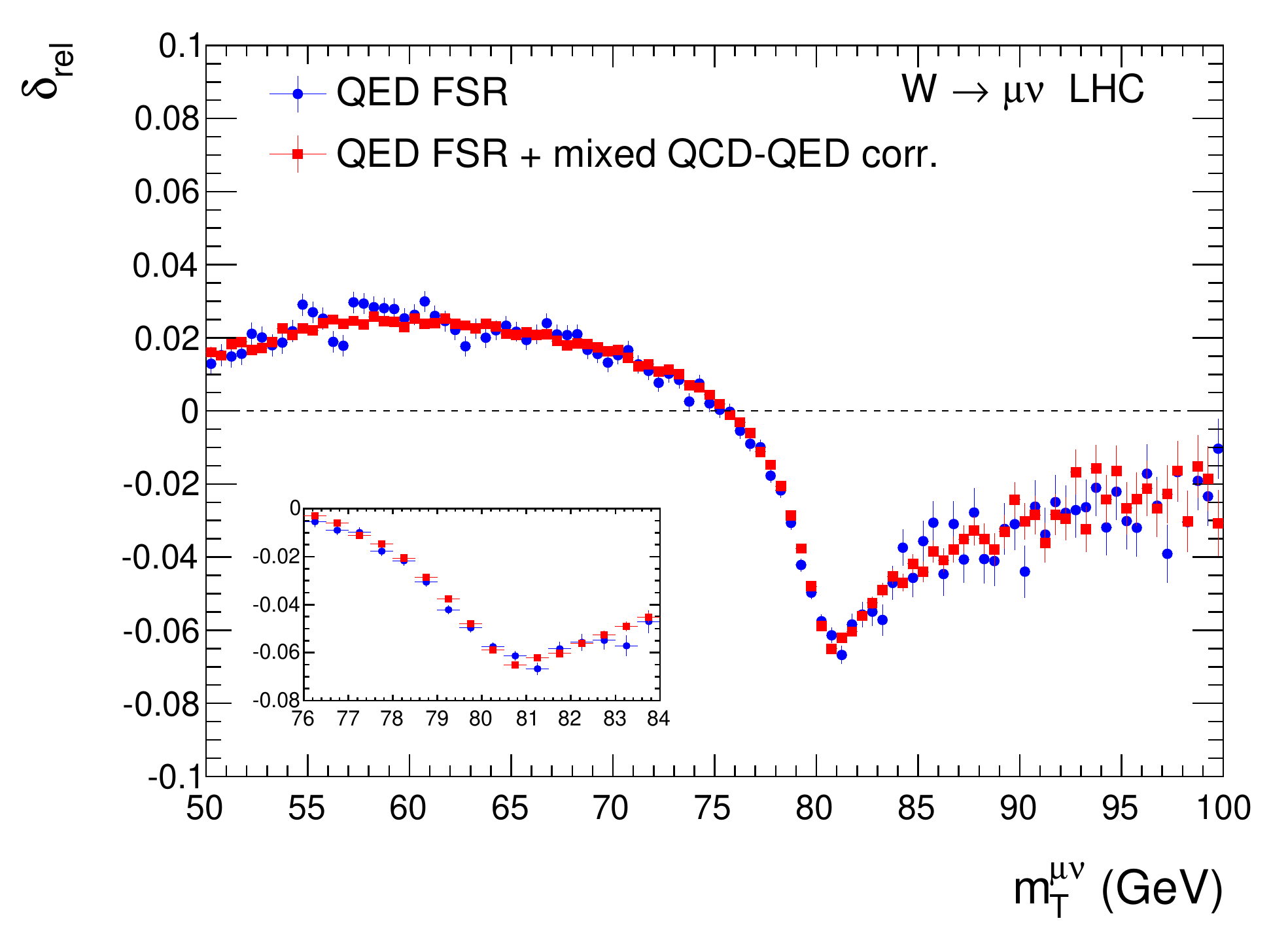}~~
\includegraphics[height=55mm]{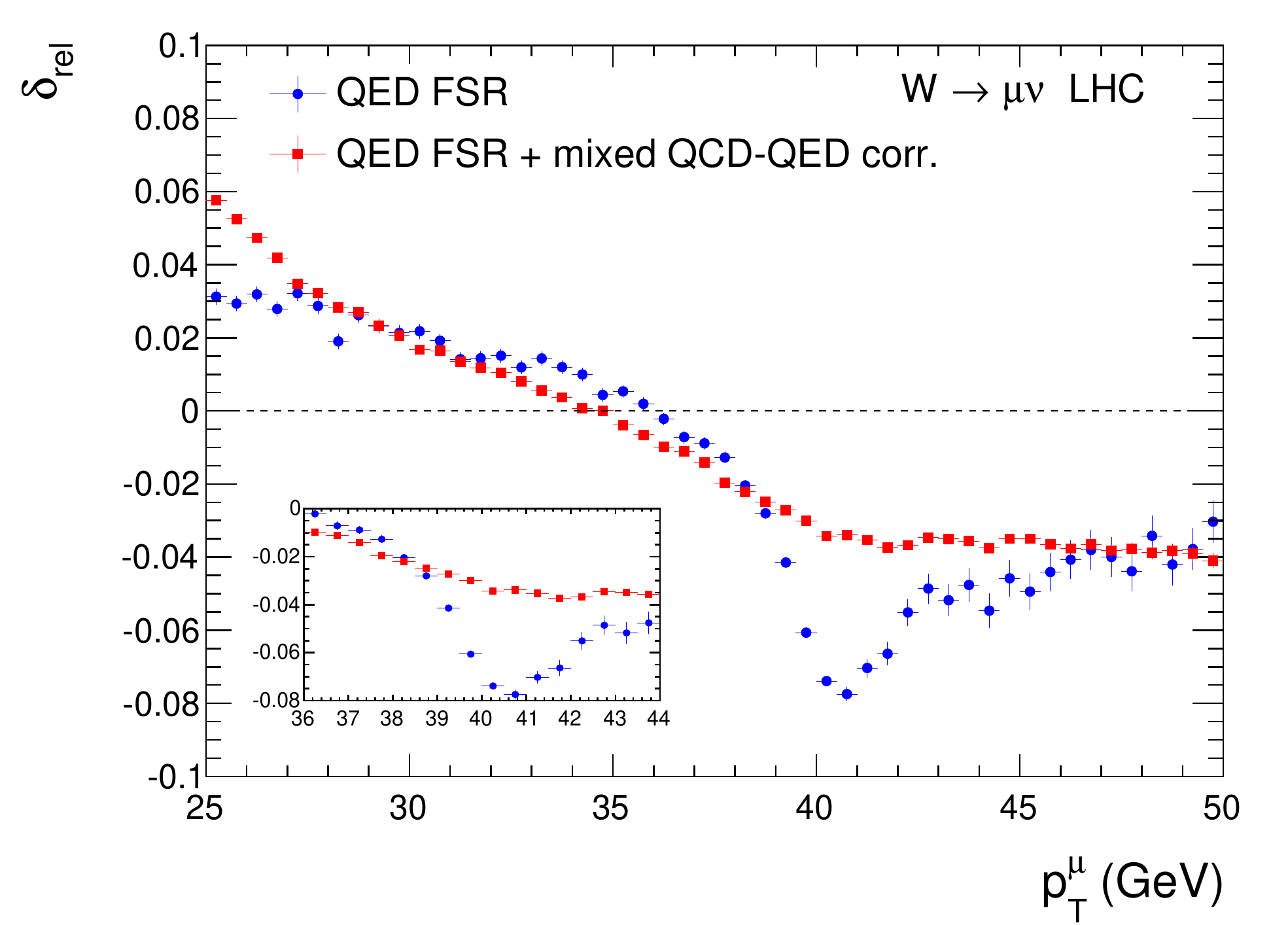}
\caption{\small Upper plots: lepton-pair transverse mass (left plots) 
and lepton transverse momentum (right plots) distributions
in different approximations:
without QCD corrections (\horaceo\, LO and \horaceo\, with QED FSR PS) 
and with QCD corrections (\powhegvtwo\, NLO QCD + QCD PS and \powhegvtwo\, NLO QCD + QCD PS interfaced to \photos) 
for the decay $W^+ \to \mu^+ \nu$ at the LHC 14 TeV, with acceptance cuts as in table~\ref{tab:parameters_lhc}. 
Lower plots: relative contribution of 
QED FSR normalized to the LO predictions
and of 
QED FSR + mixed QCD-QED corrections normalized to the \powhegvtwo\ NLO QCD + QCD PS predictions.
\label{fig:qcd}
} 
\end{center}
\end{figure}

The combination of two sets of corrections like QCD ISR and QED FSR,
which separately induce large changes of the distributions,
may also yield important effects at the level of mixed QCD-QED terms,
which are present in simulation tools based on a factorized ansatz for the inclusion of QCD and QED terms.
We show in the upper plots of figure \ref{fig:qcd}
the lepton-pair transverse mass (left plot) and the lepton transverse momentum (right plot) distributions in different perturbative approximations:
at LO (black), at LO convoluted with a QED PS (blue), at NLO QCD matched with a QCD PS (green)
and finally at NLO QCD matched with a QCD PS and convoluted with a tool describing QED FSR to all orders (red).
We dub the last approximation \qcdqed.
In figure \ref{fig:qcd} we use \photos\ as tool to describe QED FSR.
The comparison of the first two approximations (blue {\it vs} black lines) shows the negative effect of QED FSR at the jacobian peak of both observables.
The comparison of the first and of the third approximations (green {\it vs} black lines) shows the impact of QCD corrections with respect to the LO predictions: the lepton-pair transverse mass distribution is mildly modified by QCD effects, whereas the lepton transverse momentum distribution has a much broader and smeared shape compared to the LO one.
The comparison of the third and of the fourth approximations (red {\it vs} green lines) shows the effect of the convolution with a tool for the simulation of QED FSR to all orders on top of the results obtained with the full set of available QCD corrections.
In the lower plots of figure~\ref{fig:qcd} we focus on the relative size of QED FSR effects, evaluated  in terms of the LO predictions (blue dots);
we then consider the predictions 
in \qcdqed\ approximation
and take the ratio with purely QCD corrected distributions (red dots). 
With this ratio we express the impact of QED FSR corrections 
together with the one of mixed QCD-QED terms present in a tool based on a factorized ansatz for the combination of QCD and QED terms,
removing exactly the effect of pure QCD corrections.
The QED FSR corrections are common to the blue and red dots and the difference between the two sets of points is induced by the mixed QCD-QED corrections.
As it can be seen from figure~\ref{fig:qcd}, 
the shape and size of the QED FSR corrections to the transverse mass distribution is largely maintained after the inclusion of QCD corrections;
the mixed QCD-QED contributions 
are moderate but not negligible, with an effect at the few per mille level.
On the contrary, the lepton $p_T$ distribution is strongly modified by mixed QCD-QED effects,
which amount to some per cent and, more importantly, smear the varying shape of the QED FSR contribution, especially around the Jacobian peak, 
as emphasized in previous studies~\cite{Balossini:2009sa,Barze:2012tt,Barze':2013yca}.

%%%%%%%%%%%%%%%%%%%%%%%%%%%%%%%%%%%%%%%%%%%%%%%%%%%%%%%%%%%%%%%%%%%%%%%%%%
\subsection{Mixed QCD-EW corrections}
\label{sec:distr:qcdew}
As described in sections \ref{sec:powheg} and \ref{sec:powheg-improved},
there are two further sets of factorizable \oaas cor\-rections, beyond the ones obtained with the convolution of QED FSR effects on top of events with NLO QCD + QCD PS accuracy.
The first set comes from the inclusion 
of the full set of EW corrections at NLO accuracy in association to QCD radiation described by a QCD PS;
they are given by all the terms absent in pure QED FSR approximation of the ${\cal O}(\alpha)$ corrections, multiplied by the relevant QCD terms. 
The second set 
is due to the interplay of QCD effects with
subleading QED terms present in different ways in the tools that model
QED FSR to all orders.
These two classes of factorizable QCD-EW corrections
constitute a source of theoretical uncertainty, if a theoretical model like,
for instance, the standard QCD \powheg\, interfaced to a QED tool is adopted.
\begin{figure}[hbt]
\begin{center}
\includegraphics[height=55mm]{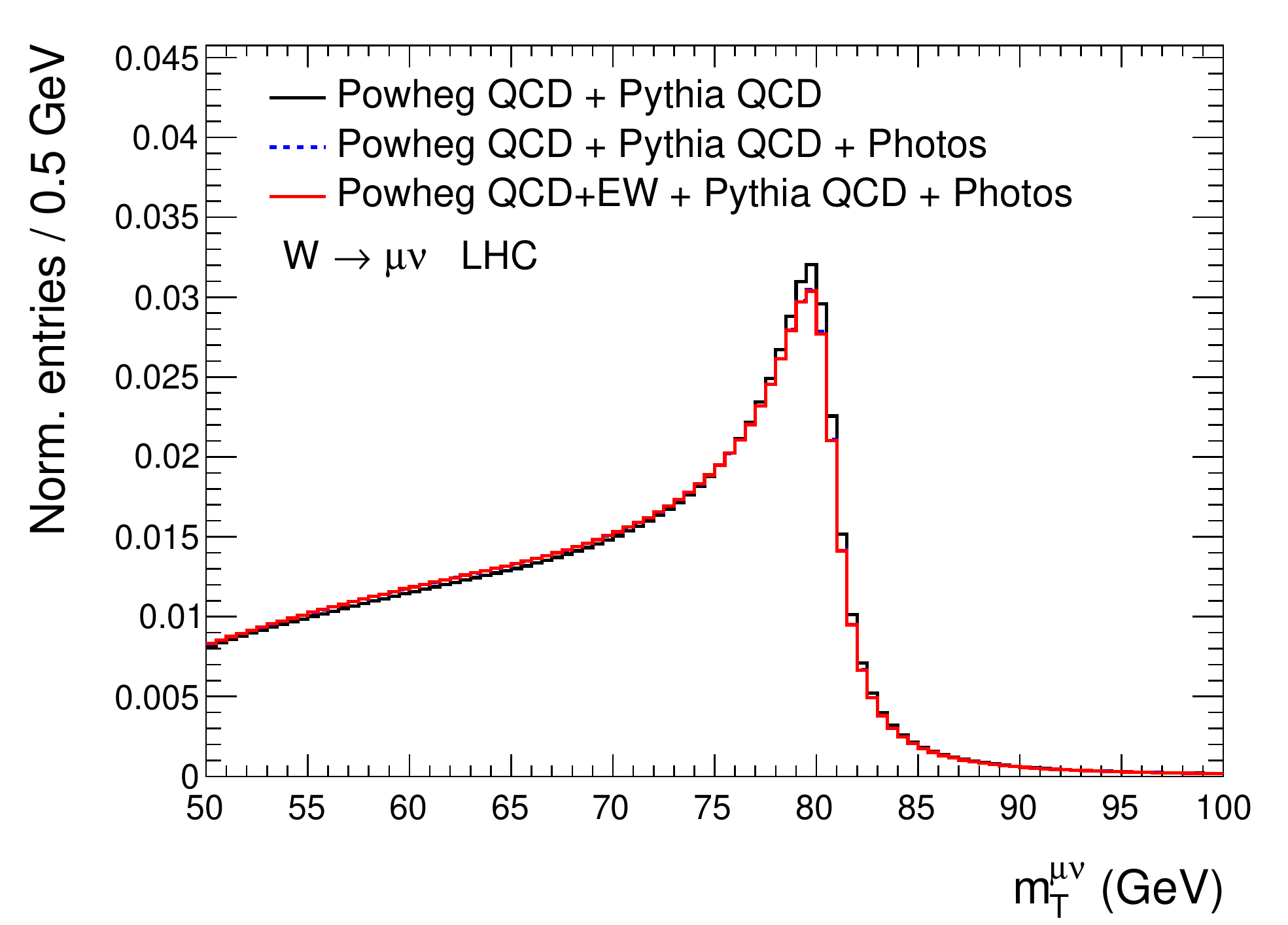}~~
\includegraphics[height=55mm]{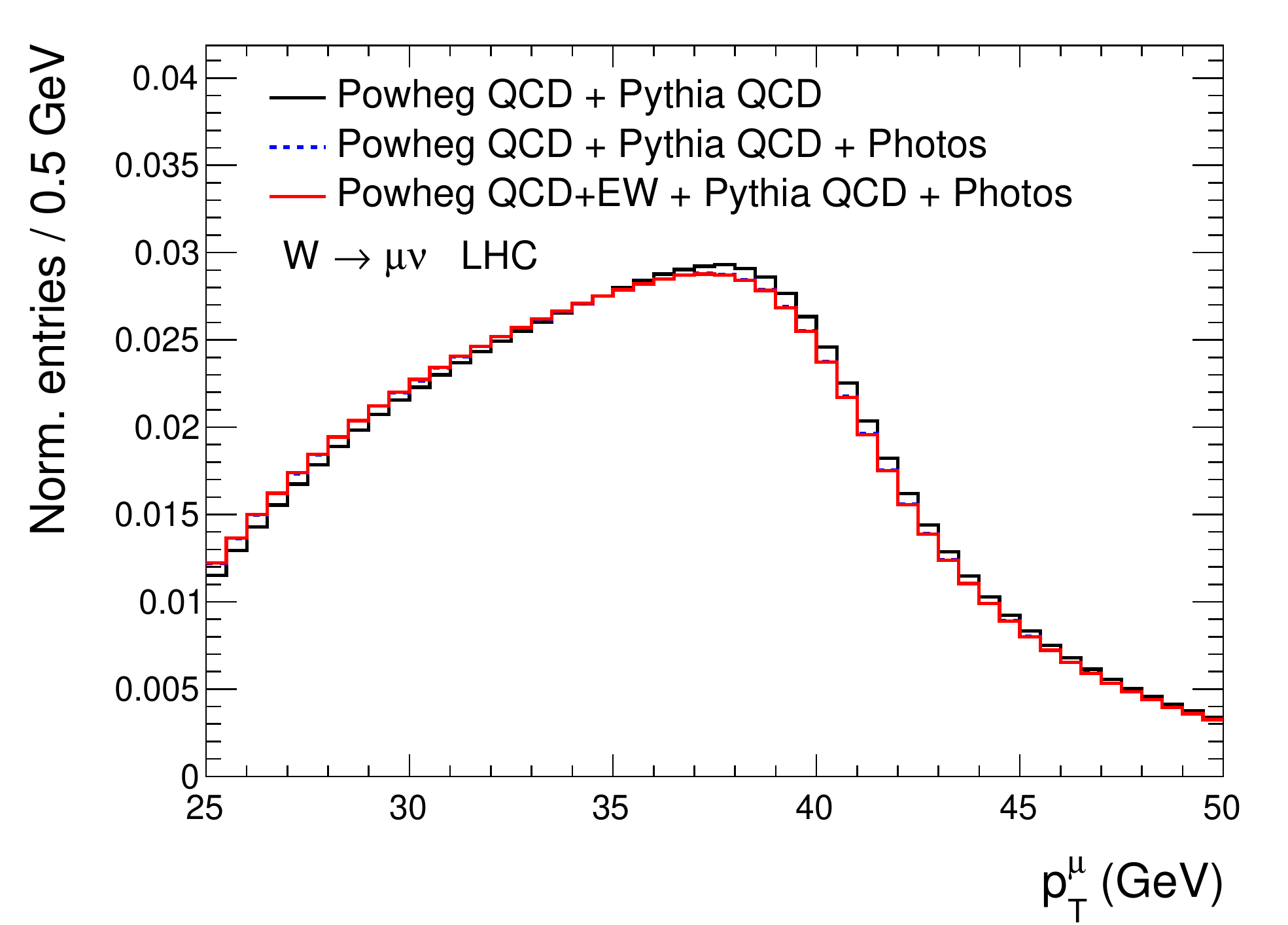} \\
\includegraphics[height=55mm]{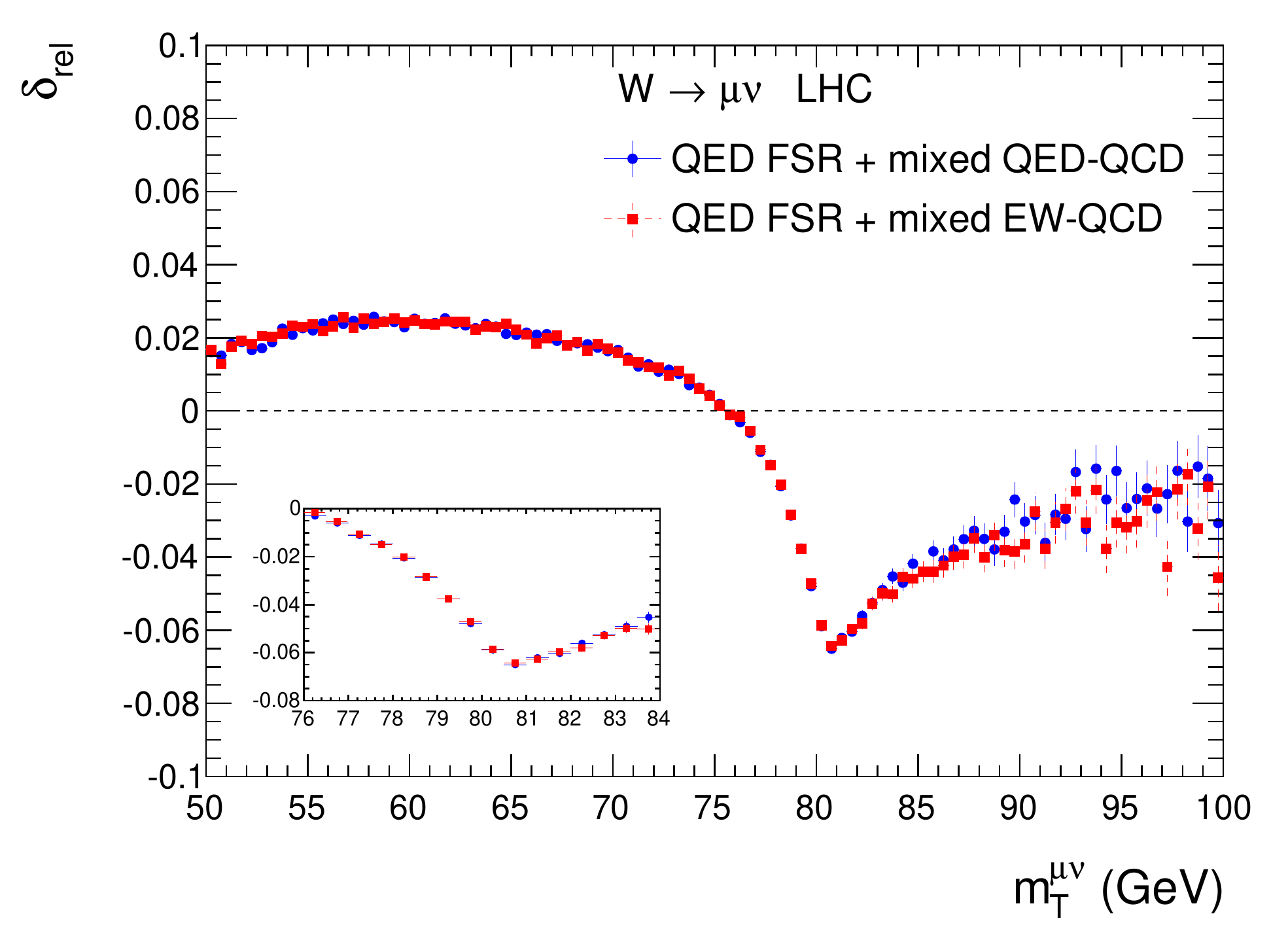}~~
\includegraphics[height=55mm]{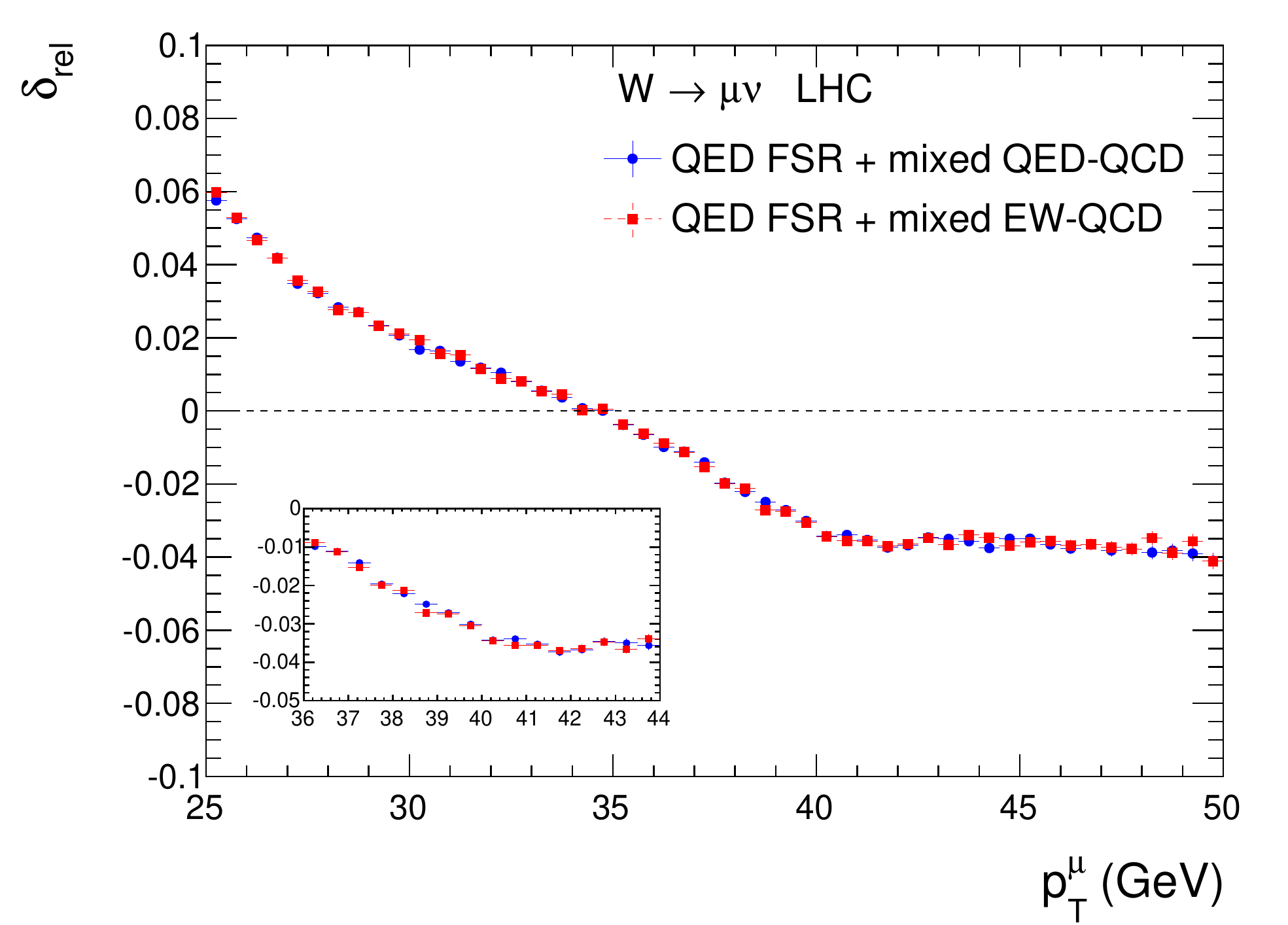}
\caption{\small Upper plots: lepton-pair transverse mass (left plots) and lepton transverse momentum (right plots) distributions,
for the decay $W^+ \to \mu^+ \nu$ at the LHC 14 TeV,
with acceptance cuts as in table~\ref{tab:parameters_lhc}, 
according to three different
approximations: \powhegvtwo\, with only QCD corrections, 
\powhegvtwo\, with only QCD corrections interfaced to \photos\, and 
\powhegvtwo\, {\tt two-rad} with NLO (QCD + EW) corrections
matched to  \photos.
Lower plots: relative contribution of 
QED FSR + mixed QCD-QED effects of the \qcdqed\, approximation
and of 
QED FSR + mixed QCD-EW corrections of the \qcdew\, approximation,
both normalized to the \powhegvtwo\ with only QCD corrections results.
\label{fig:mixed-lhc-mu}
} 
\end{center}
\end{figure}

The impact of these mixed QCD-EW contributions on the 
lepton-pair tranverse mass and lepton transverse momentum distributions 
is shown in figure~\ref{fig:mixed-lhc-mu} 
(for $W$ decays into muons at the LHC).
In the upper plots we present the results in three approximations:
{\it i)} the standard \powhegvtwo\, with only QCD corrections, with NLO QCD + QCD PS accuracy (black line), 
{\it ii)} the same events of the previous point convoluted with \photos\, to include QED FSR to all orders effects (blue line)
and
{\it iii)} events generated with \powhegvtwo\, {\tt two-rad}, with NLO (QCD+EW) + (QCD+QED) PS accuracy, the latter obtained using again \photos\, for the QED part (red line).
We dub the last approximation \qcdew.
In the lower plots of figure~\ref{fig:mixed-lhc-mu} we show the relative impact of approximations {\it ii)} and {\it iii)}
normalized to the pure QCD results of approximation {\it i)}.
The relative effect of QED FSR corrections with LL accuracy 
on top of QCD corrected events 
(approximation {\it ii)}\ ) is represented by the blue dots and
it has already been shown in figure \ref{fig:qcd}.
The results of {\it iii)} include, 
beyond QED FSR corrections in LL approximation convoluted with QCD effects, 
the additional \oaas corrections due to the NLO (QCD+EW) accuracy of \powhegvtwo\, {\tt two-rad}, matched with \photos;
their relative impact is expressed by the red dots.
The difference between the red and the blue dots shows the impact,
on the two observables under study,
of the mixed QCD-EW corrections beyond the ones obtained with the convolution of QED FSR and QCD corrections;
this difference is very small, at the per mille level or below, in both cases.

\begin{figure}[!hbt]
\begin{center}
\includegraphics[height=55mm]{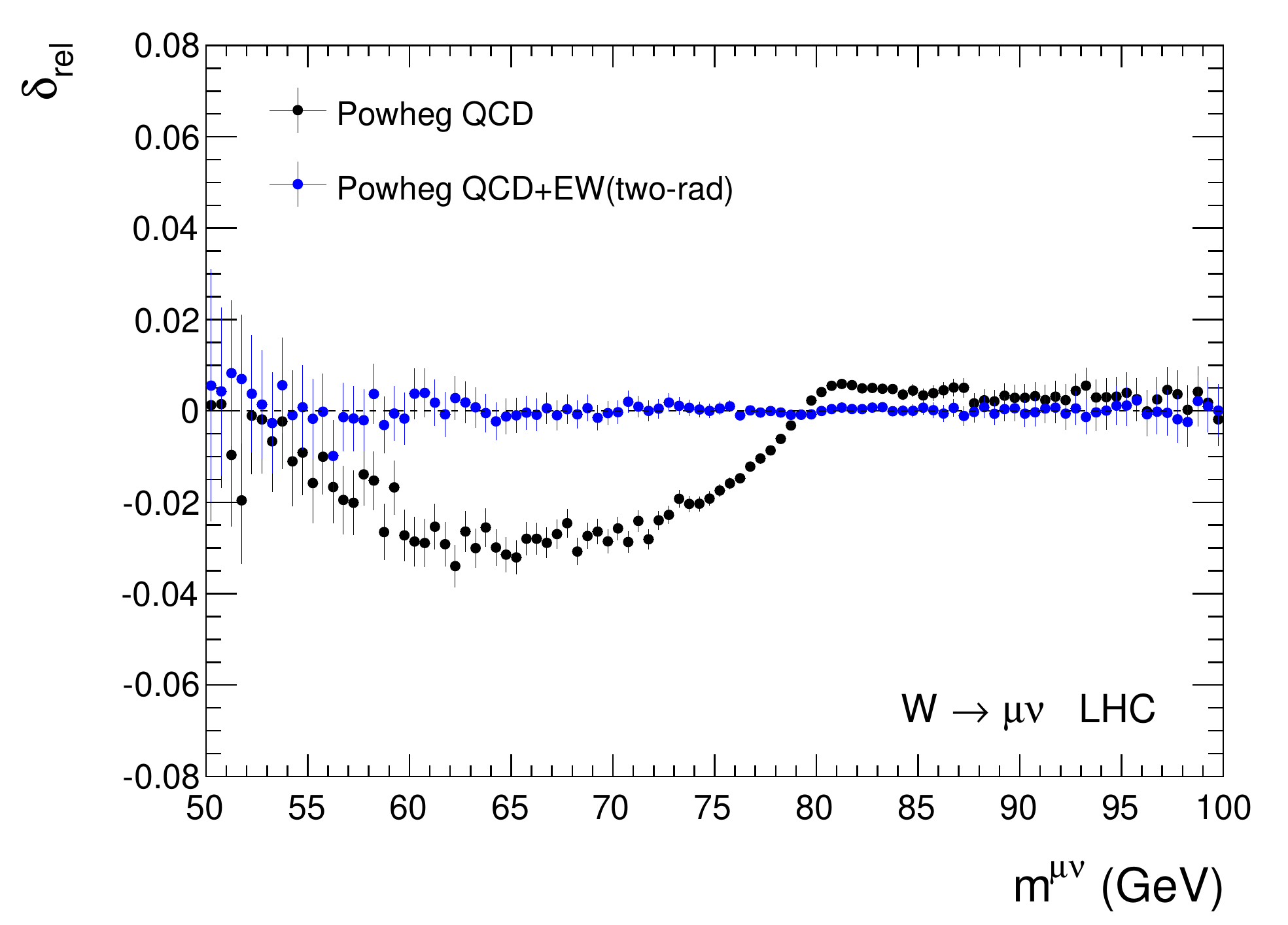}
\caption{\small 
Comparison of different approximations of the $\mu^+ \nu$ invariant mass distribution simulated at the LHC 14 TeV 
with acceptance cuts as in table~\ref{tab:parameters_lhc}.
Relative difference
of the predictions obtained with \photos\, compared to the ones with \pythiaqed\, as tools to simulate QED FSR effects.
The comparison is based on results computed with 
\powhegvtwo\, with only QCD corrections interfaced to \photos\,  or \pythiaqed\, (black dots)
and on results computed with
\powhegvtwo\, {\tt two-rad} with NLO (QCD+EW) corrections, matched to \photos\,
or \pythiaqed\, (blue dots).
\label{fig:pythia-photos-invm}
} 
\end{center}
\end{figure}

\begin{figure}[!hbt]
\begin{center}
\includegraphics[height=55mm]{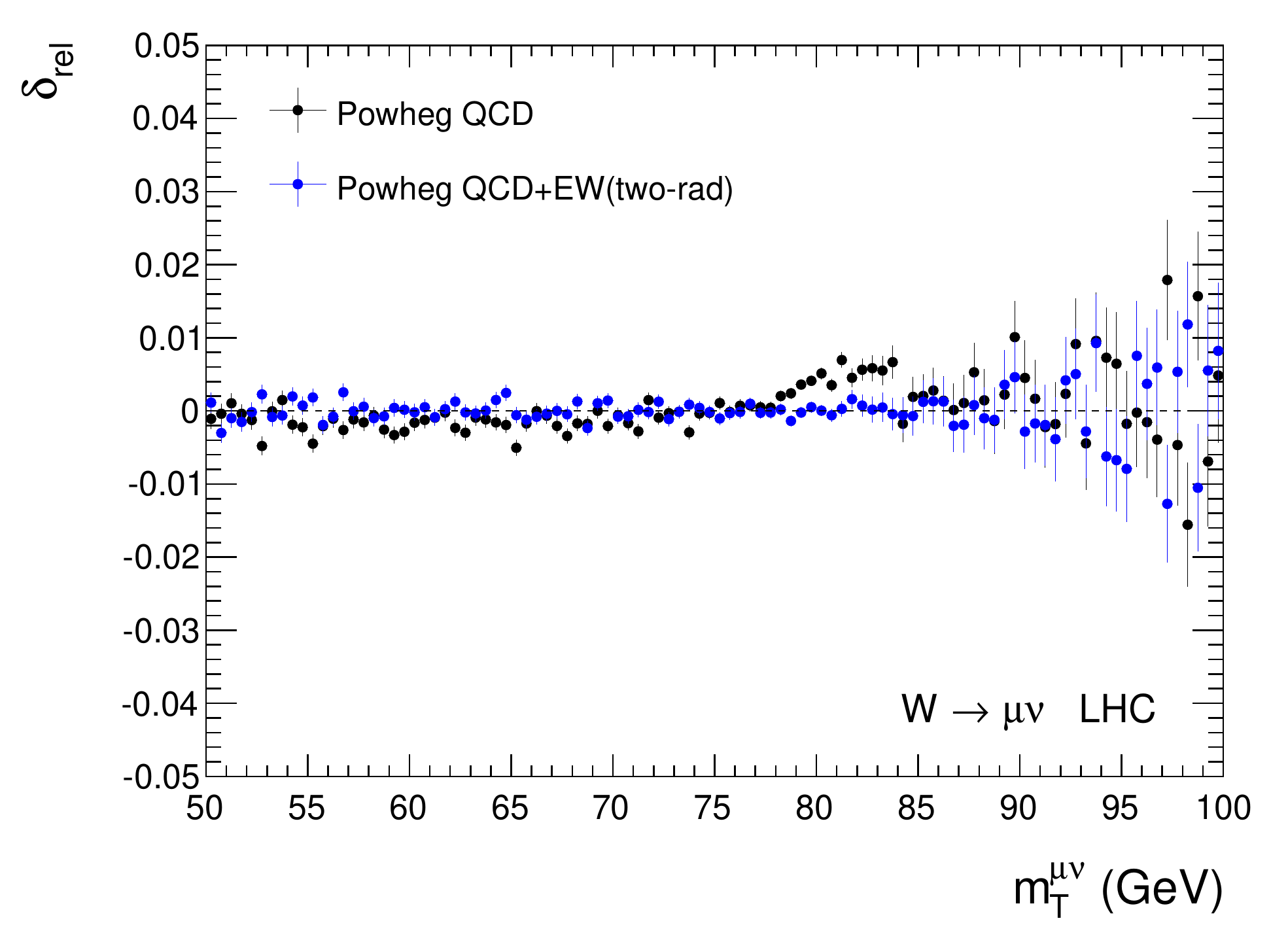}~~
\includegraphics[height=55mm]{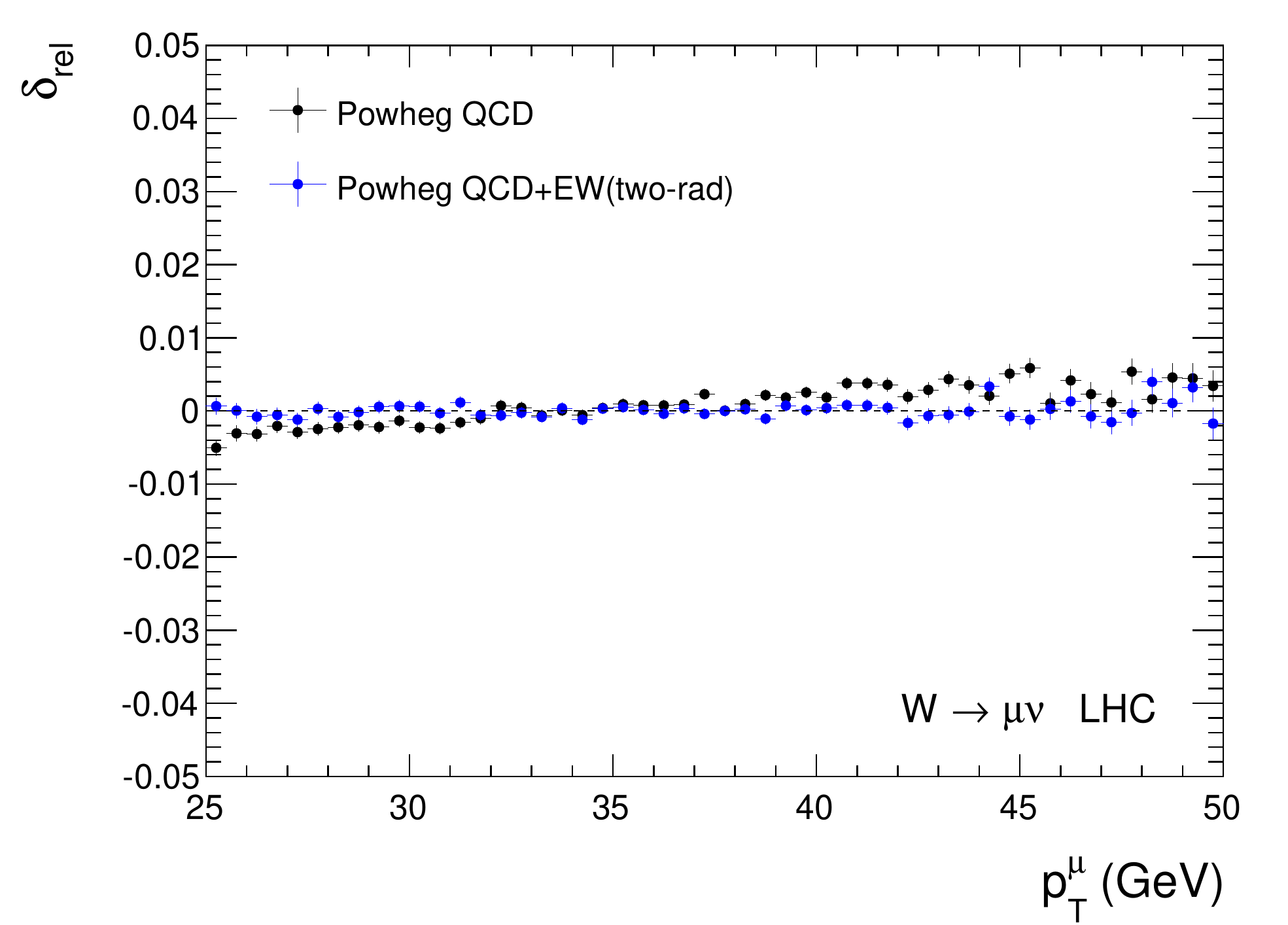}
\caption{\small Same as in Figure~\ref{fig:pythia-photos-invm} 
for the lepton-pair transverse mass (left plot) and
for the lepton transverse momentum (right plot) distributions. 
\label{fig:pythia-photos-mt-pt}
} 
\end{center}
\end{figure}

As a last point we investigate 
the effects of the different treatment of QED radiation between 
\photos\, and \pythiaqed.
In figures~\ref{fig:pythia-photos-invm} and \ref{fig:pythia-photos-mt-pt}  
we study  the lepton-pair invariant mass and transverse mass distributions, and the lepton transverse momentum distribution.
We show the ratio between the predictions obtained with \photos\, and the ones
with \pythiaqed\, in two different perturbative approximations:
the QED FSR effects convoluted on top of the pure QCD \powhegvtwo\, events (black dots) and these QED tools matched with the NLO (QCD+EW) accurate
\powhegvtwo\, {\tt two-rad} code (blue dots).
As it can be seen, the differences between \pythia\, and \photos\,
are not negligible for some specific observable, like e.g. the lepton-pair invariant mass, when the two tools are used stand-alone in convolution on top of the QCD events.
This deviation from one is due to subleading terms of \oa, which do not cancel in the ratio.
The differences instead become negligible when \photos\, and \pythia\,
are matched with
\powhegvtwo\, {\tt two-rad} with NLO (QCD+EW) accuracy.
This better agreement is expected, because the first photon emission is now described with the exact matrix elements in both cases and differences start at ${\cal O}(\alpha^2)$ and are subleading, i.e. without a $L_{\rm QED}$ logarithmic enhancement.
The same pattern as for the lepton-pair invariant mass can be observed in figure~\ref{fig:pythia-photos-mt-pt}
for the lepton-pair transverse mass and lepton transverse momentum, 
but all the effects are smaller in size.

\section{Impact of radiative corrections on the \mw\ determination}
\label{sec:numerics}

\subsection{Template fitting method and simulation details}
\label{sec:template}

At the Tevatron, the $W$-boson mass is determined by a template fit to the lepton-pair transverse mass ($M_T$), charged lepton transverse momentum (\ptl) or missing transverse energy (\met) distributions, and a similar strategy is being implemented at the LHC. In this procedure, distributions are generated assuming a given theoretical model, for different values of \mw, and compared to data.  The value of \mw\ is extracted as the one which gives the best agreement between the predicted and measured distributions. 

In order to propagate the effect of the different theoretical corrections to the \mw\ extraction, we use a procedure inspired by the experimental one described above. 
We work with large MC samples, generated with different theoretical options, and use one in order to generate ``templates" (predicted distributions for different input values of \mw), and the others as ``pseudo--data" (distribution that will be probed by the templates). 

All the samples are initially generated with the same input value for the 
$W$-boson mass, namely \mwzero=80.398~\gev\ and a fixed value of the width, $\gw=2.141~\gev$. In order to produce the templates, we perform a reweighting of the distributions, assuming that the only dependence on \mw\ comes from the relativistic Breit--Wigner shape of the $W$ resonance:
\begin{equation}
\frac{d\sigma}{d \shat} \propto \frac{1}{(\shat-M_W^2)^2 +  M_W^2 \Gamma_W^2}~~,
\label{eq-bw}
\end{equation}
where $\hat{s}$ is the reduced squared c.m. energy.
Notice that this is correct as long as the distributions are generated at LO accuracy in the treatment of EW corrections (further QCD corrections do not change this dependence), as we do in practice in our study. The reweighting avoids producing a large number of MC samples with different \mw\ input values, which would take a prohibitively large amount of CPU time, given that the target precision requires samples with a size of the order of $10^9$ events. 

For a particular pseudo--data distribution (and hence, for a particular theoretical assumption), we extract a measured value of \mw\ as the minimum of the parabola obtained by fitting the $\chi^2$ vs. \mw\ curve, whose points come from the comparison of each template with that pseudo--data distribution. The difference between this extracted value of \mw\ and \mwzero\ provides an estimate of the effect introduced in the pseudo--data with respect to the theoretical setup used in the generation of the templates. 
In other words it is an
estimate of the shift that one would obtain by fitting the real data with templates based on that specific theoretical option (instead of the one of the reference templates). The shift
uncertainty comes from the statistics of the MC samples and is estimated from the 
rule $\Delta\chi^2 \equiv \chi^2 - \chi^2_{\rm min} = 1$.

Both for templates and pseudo-data, we consider normalized 
differential cross sections, as we are interested to derive information on the impact of the different theoretical approximations 
just due to shape differences. 

We present results for $W$ decaying into muons and electrons, providing results for bare muons and recombined (or dressed) electrons (unless specified otherwise). For dressed
electrons, the lepton and photon three-momenta are recombined into an effective
momentum  $\mathbf{p}^{eff} = \mathbf{p}^e + \mathbf{p}^{\gamma}$ for each photon satisfying 
the condition $\Delta R_{e, \gamma} = \left( \Delta\eta_{e, \gamma}^2
+   \Delta\phi_{e, \gamma}^2 \right)^{1/2} \leq 0.1$,
where $\Delta\phi_{e, \gamma}$ is the lepton-photon separation angle in the 
transverse plane. We focus on $W^+$ production and decays, but for a particular case (LHC setup), we show also results for $W^-$, in order to discuss the differences 
between the two cases.

In analogy to the Tevatron strategy, we perform template fits on the $M_T$ and  \ptl\ distributions ($M_T \equiv \left( 2\, p_T^{\ell} \, p_T^{\nu} \, [1 - \cos\Delta\phi_{\ell,\nu}] \right) ^{1/2}$), using the following fit windows:
\begin{equation}
50~{\rm GeV} \leq M_T \leq 100~{\rm GeV}, \quad 27.5~{\rm GeV} \leq p_T^{\ell} \, \leq 47.5~{\rm GeV}
\end{equation}
The distributions are obtained with a bin separation of 0.5~GeV. Only for the comparison of the effect of mixed QCD-EW corrections with fixed order result (section \ref{sec:comparison_with_fixed_order}), we use a different setup, specified therein.

The template fitting procedure explained above is applied by taking into account typical acceptance cuts 
and using events at particle (or generator) level, in accordance with the Tevatron analyses and preliminary studies for the measurement of the $W$ mass at the LHC. 
The details of the event selection and parameters are given in the following subsections.

Notice that the specific choice of the PDFs sets and of the factorization/renormalization
scale is irrelevant for the study of purely EW effects on \mw, 
the theoretical contributions under scrutiny here being independent of those QCD details.
In the case of mixed QCD-EW effects, these choices enter as a higher-order correction.

The results of the template fit procedure
depend on the approximation used to compute the templates.
In the following we consider two different options: 
templates computed at LO, without any radiative effect either of QCD or EW origin, and templates computed using the standard \powhegvtwo\  with only QCD NLO corrections matched with the \pythia\  QCD PS.
The second option provides a sensible approximation of the shape of the observables and in turn allows the assessment of the impact of higher-order corrections on the $\mw$ determination.

%%%%%%%%%%%%%%%%%%%%%%%%%%%%%%%%%%%%%%%%%%%%%%%%%%%%%%%%%%%%%%%%%%%%%%%%%%%%%
\subsection{QED, EW and lepton-pair corrections}
\label{sec:qed}

The results of this section are obtained using \horaceo. 
The distributions are computed at LO accuracy in QCD, 
in accordance with the procedure adopted at the Tevatron for the assessment of the QED/EW uncertainties, and include different subsets of EW corrections. 
The generation of events has been performed using MRST2004QED~\cite{Martin:2009iq} as PDFs set, 
with factorization scale $\mu_F = M_{\ell \nu (\gamma)}$, 
where $M_{\ell \nu (\gamma)}$ is the invariant mass of the decaying $W$ boson. 
In table~\ref{tab:parameters_qed} we summarize the event selection used for this part of the analysis.
\begin{table}[!h]
\centering
\begin{tabular}{ll}
\toprule
Process & $pp \rightarrow W^+ \rightarrow \mu^+ \nu$, $\sqrt{s}=14$~TeV \\
PDF & MRST2004QED \\
Event selection & $|\eta^\ell| < 2.5, \, p_T^{\ell} > 20~{\rm GeV}, \, p_T^{\nu} > 20~{\rm GeV} $ \\
\bottomrule
\end{tabular}
\caption{Event selection used for the study of pure EW and QED effects.}
\label{tab:parameters_qed}
\end{table}
In table~\ref{tab:ewshifts} we show the $W$ mass shifts induced by QED and EW contributions at different accuracy levels, for bare muons and electrons. 
The last line contains results derived from \photos\  as QED tool on top of the LO events generated with \horaceo. 
The templates are at LO accuracy.
\begin{table}[!htb]
\centering
\begin{tabular}{clcccc}
\toprule
 \multicolumn{2}{c}{$pp \rightarrow W^+$, $\sqrt{s}=14$~TeV} & \multicolumn{4}{|c}{\mw\ shifts (MeV)} \\
 \multicolumn{2}{c}{Templates accuracy: LO} & \multicolumn{2}{|c}{$W^+\rightarrow\mu^+\nu$} & \multicolumn{2}{|c}{$W^+\rightarrow e^+\nu$} \\
 \multicolumn{2}{c}{Pseudo--data accuracy}   & \multicolumn{1}{|c}{$M_T$} & $p_T^{\ell}$ & \multicolumn{1}{|c}{$M_T$} & $p_T^{\ell}$ \\
\midrule
1 & \horaceo\ only FSR-LL at ${\cal O}(\alpha)$ & -94\rpm1 & -104\rpm1 & -204\rpm1 & -230\rpm2\\ 
2 & \horaceo\ FSR-LL & -89\rpm1 & -97\rpm1 & -179\rpm1 & -195\rpm1\\ 
3 & \horaceo\ NLO-EW with QED shower & -90\rpm1 & -94\rpm1 & -177\rpm1 & -190\rpm2\\ 
4 & \horaceo\ FSR-LL + Pairs & -94\rpm1 & -102\rpm1 & -182\rpm2 & -199\rpm1\\ 
5 & \photos\  FSR-LL & -92\rpm1 & -100\rpm2 & -182\rpm1 & -199\rpm2\\
\bottomrule
\end{tabular}
\caption{$W$ mass shifts (in MeV) due to different QED/EW contributions and
lepton-pair radiation, for muons and bare electrons at 14 TeV LHC. The templates are computed at LO without any shower correction, the pseudodata with the accuracy and the QED effects as indicated in the table.}
\label{tab:ewshifts}
\end{table}
In general, 
one can see that for the two most important observables, 
i.e. $M_T$ and $p_T^\ell$, 
the shifts are of similar size, 
of the order of 100~MeV for muons and 200~MeV for bare electrons.
This is just a direct consequence of the fact that the EW corrections, dominated by QED FSR, 
give to $M_T$ and $p_T^\ell$ a very similar relative effect, 
when normalized to the LO predictions,
as it can be observed in section \ref{sec:distr:qcd} in figure \ref{fig:qcd}
in the lower plots, with the blue dots.
Notice that these differences, obtained for LHC energies, are valid to a large extent for the Tevatron as well. 
Actually, the QED and lepton-pair corrections to the determination
of the $W$ mass are in practice independent of the nominal c.m. energy. 
This feature follows from the fact that 
these theoretical contributions are driven by logarithmic terms of the form 
$L_{\rm QED} = \ln (\hat{s} / m_\ell^2)$, where $m_\ell$ is the mass of the radiating particle. 
Independently of the accelerator energy, 
the configurations with $\hat{s} \simeq M_W^2$, the $W$ resonance, dominate the cross section and the kinematical distributions relevant for the determination of $\mw$.

Comparing the different lines of table~\ref{tab:ewshifts}, it can be noticed that:
\begin{itemize}
\item 1 vs. 2: 
the contribution due to multiple photon emission, beyond \oa, dominated by  two-photon radiation terms, 
amounts to some MeV for muons and to about 20 - 30 MeV for bare electrons, because of the very different impact of lepton-mass dependent collinear 
logarithms $L_{\rm QED}$. 
This is in agreement with previous studies at Tevatron energies, 
where the contribution of multiple FSR is taken into account using \photos.

\item 2 vs. 3: 
the contribution of non-logarithmic NLO EW corrections is a small effect, 
at a few MeV level, for both muons and electrons, and independent of the considered observable. 
This result emphasizes the dominant r\^ole played by QED FSR at LL level 
within the full set of NLO EW corrections.

\item 2 vs. 4: 
the \oalphasq contribution due to lepton-pair radiation 
induces a shift of $\mw$
of about 5$\pm$1~MeV for muons and 3$\pm$1~MeV for electrons, 
when considering the fits to the transverse mass distribution. 
It is a not negligible effect 
given the present accuracy of the measurement at the Tevatron, 
where it is presently treated as a contribution to the QED uncertainty, because 
the \photos\  version included in the Tevatron analyses did not simulate pair radiation\footnote{At
  present a version of \photos\ including the effects of light-pair radiation is
  available, as described in ref.~\cite{Davidson:2010ew}.}.
For $W$ decays into muons, 
the shift is of the same order of the one induced by multiple photon emission, 
whereas for $W$ decays into bare electrons 
it is much smaller than multiple FSR.
The lepton-flavor dependence of the pair radiation correction 
is a direct consequence of the $\beta_f (s), f = e, \mu$, expressions 
given in section \ref{sec:horacepairs} and, in turn,
of the distortion of the kinematical distributions
shown in section~\ref{sec:distr:pair}.
As remarked in section~\ref{sec:distr:pair}, the recombination procedure 
reduces the impact of multiple photon radiation on dressed electrons,
but does not modify the contribution of additional soft light pairs which instead are not recombined and
constitute the largest fraction of the emitted pairs. Therefore,
the 3~MeV shift due to pairs applies to both bare and dressed electrons.

\item 2 vs. 5: the predictions of \horace\ and \photos\  for the shifts due to
 multiple QED FSR agree at $\simeq 3 \pm 1$ MeV both for $M_T$ and \ptl,
both in muon and electron final states.
This agreement is certainly satisfactory, 
given the LL approximation inherent in the two programs.

\end{itemize}
The comparison of \horace\ and \photos\  in lines 2 and 5 tests
the {\it technical} precision of the two codes,
which claim the same accuracy but differ by subleading terms and in the implementation of the generation of radiation.
An estimate of the {\it physical} precision of \photos, 
instead, can be obtained with a comparison of its results against
those of the a code that includes a more complete set of higher-order radiative corrections.
Since \photos\  does not include exact NLO EW corrections nor the effect of lepton-pair radiation,
we can derive the size of these effects within the \horace\ framework
from the comparison of line 2 with lines 3 and 4 of table~\ref{tab:ewshifts}.
By summing the two 
 sources in quadrature, we conclude that the uncertainty of a QED 
 modeling based on \photos\  is in the range $4-6 \pm 1~{\rm MeV}$
both for $M_T$ and $p_T^\ell$, 
and it is independent of the final state lepton flavor.
This conclusion is in agreement with the estimates of the 
QED uncertainty presently provided by \cdf\ and \dzero\ collaborations, 
which make use of a procedure very similar to the one here described.

%%%%%%%%%%%%%%%%%%%%%%%%%%%%%%%%%%%%%%%%%%%%%%%%%%%%%%%%%%%%%%%%%%%%%%%%%%%%%%
\subsection{EW input scheme variation}
\label{sec:ewkscheme}

We compare now the shifts due to
the choice of the input parameter scheme, 
to provide an estimate of the theoretical uncertainty of EW origin at NNLO accuracy. 
Indeed, as already emphasized, 
a complete calculation of the NNLO EW corrections to DY processes 
is presently unavailable.
The results in table~\ref{tab:wshifts-scheme} correspond to 
the three different input schemes introduced in section~\ref{sec:inputscheme} 
and come from simulations both at NLO and in the NLO+PS formulation of 
\horace\ at the Tevatron energy.
All the numbers in table~\ref{tab:wshifts-scheme} are computed 
using the same templates with LO accuracy without any QCD correction, 
neither fixed-order nor from Parton Shower. 
The details of the event generation and selection are given in table~\ref{tab:parameters_input_scheme}.
\begin{table}[!htb] 
\centering 
\begin{tabular}{clccc} 
\toprule 
\multicolumn{3}{c}{$p\bar{p} \rightarrow W^+$, $\sqrt{s}=1.96$~TeV} & \multicolumn{2}{|c}{\mw\ shifts (MeV)} \\ 
\multicolumn{3}{c}{Templates accuracy: LO} & \multicolumn{2}{|c}{$W^+\rightarrow\mu^+\nu$} \\ 
\multicolumn{2}{c}{Pseudodata accuracy}  & Input scheme & \multicolumn{1}{|c}{$M_T$} & $p_T^{\ell}$ \\ 
\midrule 
1 & \horaceo\ NLO-EW & $\alpha_0$ & -101\rpm1 & -117\rpm2 \\ 
2 &                 & $G_{\mu}-I$ & -112\rpm1 & -130\rpm1 \\ 
3 &                 & $G_{\mu}-II$ & -101\rpm1 & -117\rpm1 \\ 
4 & \horaceo\ NLO-EW+QED-PS & $\alpha_0$ & ~-70\rpm1 & ~-81\rpm1 \\ 
5 &                        & $G_{\mu}-I$ & ~-72\rpm2 & ~-83\rpm1 \\ 
6 &                        & $G_{\mu}-II$ & ~-72\rpm1 & ~-82\rpm2 \\ 
\bottomrule 
\end{tabular} 
\caption{$W$ mass shifts (in MeV) induced by different input scheme choices, 
at NLO-EW (lines 1, 2 and 3) and NLO-EW+ QED-PS (lines 4, 5 and 6) 
accuracy, for the muon channel, at the Tevatron with $\sqrt{s}=1.96$~TeV. The templates have been computed at LO without any shower correction.} 
\label{tab:wshifts-scheme} 
\end{table} 
\begin{table}[htb]
\centering
\begin{tabular}{ll}
\toprule
Process & $p\bar{p} \rightarrow W^+ \rightarrow \mu^+ \nu$, $\sqrt{s}=1.96$~TeV \\
PDF & MRST2004QED \\
Event selection & $|\eta^\ell| < 1.05, \, p_T^{\ell} > 25~{\rm GeV}, \, p_T^{\nu} > 25~{\rm GeV} $
 \\
\bottomrule
\end{tabular}
\caption{Event selection used for the study of the EW input scheme variation.}
\label{tab:parameters_input_scheme}
\end{table}

The main comments are the following.
\begin{itemize}
\item  When considering NLO predictions (lines 1, 2 and 3), 
the shifts are not negligible, reaching the 10 MeV level. They
are induced by the different ${\cal O} (\alpha^2)$ components 
present in the three schemes.

\item However, the shifts are 
considerably reduced, down to about 1--2 MeV, when NLO corrections are matched 
with higher--order contributions (lines 4, 5 and 6). This follows 
from the fact that the sharing of the different photon multiplicities is the same in the three schemes, 
as remarked in Section \ref{sec:inputscheme}.

\item The $\alpha_0$ and the $G_\mu-II$ schemes behave in a very similar
way, as it can be clearly noticed from the results of the \oa\ analysis (line 2).
This result is a consequence of the equality of the relative fraction of the 
0-- and 1--photon samples in the two cases.

\item Since both versions of the $G_\mu$ scheme are {\it a priori} 
acceptable in the absence of a complete NNLO EW calculation, it follows from 
the results shown in line 6 that there is an intrinsic input scheme arbitrariness that 
induces an uncertainty on the $W$ mass at the 1~MeV level.

\end{itemize}
In summary, the uncertainty due to missing NNLO EW corrections, 
as estimated through input scheme variation, is at the MeV level and could be further
reduced only with a complete ${\cal O}(\alpha^2)$ calculation. 

%%%%%%%%%%%%%%%%%%%%%%%%%%%%%%%%%%%%%%%%%%%%%%%%%%%%%%%%%%%%%%%%%%%%%%%%%
\subsection{Mixed QCD-EW corrections}
\label{sec:mix}
In this section we study the $\mw$ shift induced by mixed QCD-EW corrections, in different perturbative approximations.
In section~\ref{sec:comparison_with_fixed_order} we 
compare the size of the $\mw$ shift induced by mixed the \oaas 
corrections present in our theoretical formulation,
based on \powhegvtwo\  {\tt two-rad} with NLO (QCD+EW) accuracy,
with the results available in the literature, 
obtained with the fixed-order calculation in pole approximation~\cite{Dittmaier:2014qza,Dittmaier:2014koa,Dittmaier:2015rxo,Dittmaier:2016egk}.
In sections (\ref{sec:mixtev}-\ref{sec:mixlhc})
we systematically compare
two different approximations available in \powhegvtwo:
{\it i)} we consider distributions computed with
\qcdqed\, approximation
and use,
as a tool that simulates QED FSR effects, either \photos\  or \pythiaqed;
{\it ii)} we generate distributions in
\qcdew\ approximation,
using \powhegvtwo\  {\tt two-rad} with NLO (QCD+EW) accuracy and 
matched with \pythia\  as QCD PS and with \photos\ or \pythiaqed\  as QED PS.
The comparison of these two approximations offers some hints for a
critical assessment of  the theoretical uncertainty induced by mixed 
QCD-EW corrections. 
This contribution is presently neglected in the theoretical error estimate by the Tevatron collaborations, 
but it can be nonetheless evaluated, 
using the state of the art of theoretical tools.
Eventually, in section \ref{sec:v2vstworad} we comment on the impact on the $\mw$ determination of the upgrade {\tt two-rad} with respect to the previous public versions of \powhegvtwo\  with NLO (QCD+EW) accuracy.

Unless stated otherwise, 
the templates used in the fitting procedure have been computed with the standard \powhegvtwo\  version that includes only QCD corrections, namely with NLO QCD accuracy and matched with \pythia 8
for the simulation of multiple parton QCD ISR.
The simulations have been done using
MSTW2008 NLO~\cite{Martin:2009iq} as PDFs set, 
with factorization/renormalization scale $\mu_F = \mu_R = M_{\ell \nu (\gamma)}$.
For the separation between Btilde and Remnant contributions, 
we use the default setting, which means that a radiative contribution 
is considered remnant if the ratios between the full matrix element and its soft/collinear approximations are greater than five. 
For the simulation of QED FSR, we use both \pythia 8 and \photos\ version 3.56, the latter without QED matrix element corrections.

%%%%%%%%%%%%%%%%%%%%%%%%%%%%%%%%%%%%%%%%%%%%%%%%%%%%%%%%%%%%%%%%%%%%%%%%%
\subsubsection{Comparisons with fixed-order results}
\label{sec:comparison_with_fixed_order}

In this section we focus on the study of the lepton-pair transverse mass distribution and 
we compare the $W$ mass shifts induced 
by the mixed QCD-EW corrections 
contained in the \powhegvtwo\  {\tt two-rad} predictions 
with the ones based on the fixed-order results 
of refs.~\cite{Dittmaier:2015rxo,Dittmaier:2016egk}, where results for both 
bare muons and calorimetric electrons are reported. Here, we limit ourselves 
to consider bare muons only, since this channel displays the largest effects. 
As already discussed in Section~\ref{sec:oalphaalphascorrections}, 
the factorized approach implemented in \powhegvtwo\  {\tt two-rad}
contains part of the \oaas terms. 
Actually, through the use of QCD and QED parton showers, 
terms of ${\cal O}(\alpha_s^m \alpha^n)$, with $m,n \geq 1$, are included to all orders as well. 
This means, strictly speaking, that a comparison with the fixed-order results 
of refs.~\cite{Dittmaier:2015rxo,Dittmaier:2016egk} is not possible. 
However, for the $M_T$ distribution, QCD corrections have a mild impact on the distribution, as remarked in section \ref{sec:distr:qcd};
we can thus assume that the dominant part of the mixed corrections 
is given by the lowest order term with $m=n=1$, 
with negligible effects from higher-order terms. 
In order to comply with refs.~\cite{Dittmaier:2015rxo,Dittmaier:2016egk}, 
we adopt the event selection and fit setup used there, 
as detailed in table \ref{tab:parameters_comparison_with_fixed_order}.
The templates used in this comparison, at variance with the rest
of section~\ref{sec:mix}, are generated with LO accuracy. 
\begin{table}[htb]
\centering
\begin{tabular}{ll}
\toprule
Process & $pp \rightarrow W^+ \rightarrow \mu^+ \nu$, $\sqrt{s}=14$~TeV \\
PDF & MSTW2008 NLO \\
Event selection & $|\eta^\ell| < 2.5, \, p_T^{\ell} > 25~{\rm GeV}, \, p_T^{\nu} > 25~{\rm GeV}$ \\
Fit window &  $64~{\rm GeV} \leq M_T \leq 91~{\rm GeV}$ \\
Bin width of $M_T$ distribution & 1 GeV \\
\bottomrule
\end{tabular}
\caption{Event selection and fit setup used for the comparison with the fixed order results of refs.\cite{Dittmaier:2014qza,Dittmaier:2014koa,Dittmaier:2015rxo,Dittmaier:2016egk}.}
\label{tab:parameters_comparison_with_fixed_order}
\end{table}

In order to isolate a contribution that can be compared to the one obtained in refs.~\cite{Dittmaier:2015rxo,Dittmaier:2016egk}, we observe that
the perturbative content, with respect to the LO prediction, of the fully differential
\powheg\ result, including showering effects, can be cast in the following form: 
\begin{equation}
d\sigma_{\rm POWHEG} = d\sigma_0\,
\left[
1~+~
\delta_{\alpha_s} + \delta_{\alpha}
+ \sum_{m=1,n=1}^{\infty}\delta^\prime_{\alpha_s^m \alpha^n} + \sum_{m=2}^{\infty}\delta^\prime_{\alpha_s^m} 
+ \sum_{n=2}^{\infty}\delta^\prime_{\alpha^n}
\right], 
\label{eq:powheg-pert}
\end{equation}
where the factors $\delta$ represent the correction, normalized to the LO result, induced by different subsets of higher-order terms, the latter labelled by the indices.
We add a prime to those $\delta$ factors where the corresponding correction is not known exactly but it is only approximated.
We extract the contribution to the $W$ mass shift given by 
$\sum_{m=1,n=1}^{\infty} \delta^\prime_{\alpha_s^m \alpha^n}$ in eq.~\ref{eq:powheg-pert}, 
subtracting from the full result 
the shift induced by the NLO contribution $(\delta_{\alpha_s} + \delta_{\alpha})$ and the one of
the higher-order contributions 
$\sum_{m=2}^{\infty}\delta^\prime_{\alpha_s^m}$ and $\sum_{n=2}^{\infty}\delta^\prime_{\alpha^n}$. 

\begin{table}
\begin{tabular}{>{\small}c>{\small}c>{\small}l>{\small}r} 
\toprule 
& Templates & Pseudodata & $M_W$ shifts (MeV) \\ 
\midrule 
1 & LO & POWHEG(QCD) NLO &  56.0 $\pm$ 1.0  \\
2 & LO & POWHEG(QCD)+PYTHIA(QCD) & 74.4 $\pm$ 2.0 \\
3 & LO & HORACE(EW) NLO  & -94.0 $\pm$ 1.0 \\ 
4 & LO & HORACE (EW,QEDPS) & -88.0 $\pm$ 1.0 \\
5 & LO & POWHEG(QCD,EW) NLO & -14.0 $\pm$ 1.0 \\
6 & LO & POWHEG(QCD,EW) {\tt two-rad}+PYTHIA(QCD)+PHOTOS & -5.6 $\pm$ 1.0 \\
\bottomrule 
\end{tabular}
\caption{$W$ mass shift (in MeV) induced by different sets of perturbative corrections and evaluated with templates computed at LO, at the LHC 14 TeV for $\mu^+ \nu$
production.  }
\label{tab:fixed-comp}
\end{table}

%%%%%%%%%%%%%%%%%%%%%%%
\begin{table}
\begin{center}
\begin{tabular}{lcr}
\hline
correction factor in eq.~\ref{eq:powheg-pert} & samples in table \ref{tab:fixed-comp} & $\mw$ shift (MeV)\\
\hline
$\sum_{m=1,n=1}^{\infty}\delta^\prime_{\alpha_s^m \alpha^n} + \sum_{m=2}^{\infty}\delta^\prime_{\alpha_s^m} 
+ \sum_{n=2}^{\infty}\delta^\prime_{\alpha^n}$ & [6]-[5]  & 8.4 $\pm 1.4~{\rm MeV}$\\
$\sum_{m=2}^{\infty}\delta^\prime_{\alpha_s^m}$ & [2]-[1] & 18.4 $\pm 2.2~{\rm MeV}$\\
$\sum_{n=2}^{\infty}\delta^\prime_{\alpha^n}$ & [4]-[3] & 6.0 $\pm 1.4~{\rm MeV}$\\
\hline
\end{tabular}
\caption{\label{tab:cfr-dhs}
Impact in terms of $\mw$ shifts 
of the correction factors present in eq.~\ref{eq:powheg-pert},
contributing to the \powhegvtwo\  {\tt two-rad} simulations with NLO (QCD+EW) accuracy, derived from the results of table \ref{tab:fixed-comp}.
}
\end{center}
\end{table}
For this analysis we generate
pseudodata samples with different perturbative accuracies, including: 
1) only fixed-order NLO QCD;
2) NLO QCD matched with QCD PS; 
3) only fixed-order NLO EW corrections; 
4) NLO EW matched with QED PS;
5) only fixed-order NLO (QCD+EW); 
6) NLO (QCD+EW) matched with (QCD+QED) PS.
In table~\ref{tab:fixed-comp} we present the shifts associated to these 6 samples, extracted with templates computed at LO,
while in table~\ref{tab:cfr-dhs} we show the combinations relevant for the determination of $\sum_{m=1,n=1}^{\infty} \delta^\prime_{\alpha_s^m \alpha^n}$.
Subtracting the second and third lines from the first line of table~\ref{tab:cfr-dhs} 
we obtain our estimate for the shift $\Delta\mw^{\alpha_s\alpha}$ induced by the correction factor
$\sum_{m=1,n=1}^{\infty}\delta^\prime_{\alpha_s^m \alpha^n}$,
which turns out to be 
$$
\Delta\mw^{\alpha_s\alpha}
=
-16.0 \pm 3.0~{\rm MeV},
$$
in nice agreement with $\delta_{\rm NNLO} = -14$~MeV of refs.~\cite{Dittmaier:2015rxo,Dittmaier:2016egk}. 

From table~\ref{tab:fixed-comp} we can obtain additional information. 
We remark that
the shift induced by NLO QCD corrections is 
positive and sizeable, $56 \pm 1$~MeV. 
Given the large cancellation of the NLO QCD and NLO EW corrections
at the jacobian peak of the $M_T$ distribution,
which is illustrated in figure \ref{fig:nlo},
and given also the non-linear behaviour of the $\chi^2$ function in the fitting procedure,
we observe that the shift extracted from sample 5, namely with the simultaneous presence of NLO QCD and NLO EW effects, is different from
the sum of the two shifts obtained with one set of corrections at a time
(samples 1 and 3 ).

%%%%%%%%%%%%%%%%%%%%%%%%%%%%%%%%%%%%%%%%%%%%%%%%%%%%%%%%%%%%%%%%%%%
\subsubsection{Results for the Tevatron}
\label{sec:mixtev}

In this section we focus on $W$ production at the Tevatron. 
The details of the event selection are shown in table~\ref{tab:parameters_tevatron} and
we notice the introduction of a cut in the transverse momentum of the $W$ boson ($p_T^W$), defined as $ p_T^W \equiv | \mathbf{p}_T^{\ell} +  \mathbf{p}_T^{\nu} + \sum \mathbf{p}_T^{\gamma}|$, where the sum runs on all the photons emitted by the charged lepton.
\begin{table}[htb]
\centering
\begin{tabular}{ll}
\toprule
Process & $p\bar{p} \rightarrow W^+ \rightarrow \mu^+ \nu$, $\sqrt{s}=1.96$~TeV \\
PDF & MSTW2008 NLO \\
Event selection & $|\eta^\ell| < 1.05, \, p_T^{\ell} > 25~{\rm GeV}, \, p_T^{\nu} > 25~{\rm GeV}, \, p_T^{W} < 15~{\rm GeV}$\\
\bottomrule
\end{tabular}
\caption{Event selection used for the study of QED and mixed QCD-EW effects at Tevatron.}
\label{tab:parameters_tevatron}
\end{table}
As already anticipated, here and in the following sections 
we consider the following two approximations:
{\it i)} \qcdqed\ using either \photos\ or \pythiaqed\  to simulate QED FSR effects,
{\it ii)} \qcdew\ using again either \photos\ or \pythiaqed\  to simulate QED FSR effects.
In table~\ref{tab:qedqcdshifts-tev} we present the corresponding shifts
(lines 1 and 2 for approximation {\it i)}, lines 3 and 4 for approximation {\it ii)}). 
\begin{table}[htb]
\centering
\begin{tabular}{>{\small}c>{\footnotesize}l>{\small}c>{\small}c>{\small}c>{\small}c>{\small}c} 
\toprule 
\multicolumn{3}{c}{$p\bar{p} \rightarrow W^+$, $\sqrt{s}=1.96$~TeV} & \multicolumn{4}{|c}{\mw\ shifts (MeV)} \\ 
\multicolumn{3}{c}{Templates accuracy: NLO-QCD+QCD$_{\rm PS}$} & \multicolumn{2}{|c}{$W^+\rightarrow\mu^+\nu$} & \multicolumn{2}{|c}{$W^+\rightarrow e^+\nu$(dres)} \\ 
\multicolumn{2}{c}{\normalsize Pseudodata accuracy} & QED FSR & \multicolumn{1}{|c}{$M_T$} & $p_T^{\ell}$ & \multicolumn{1}{|c}{$M_T$} & $p_T^{\ell}$ \\ 
\midrule 
1 & NLO-QCD+(QCD+QED)$_{\rm PS}$& \pythia\  & -91\rpm1 & -308\rpm4 & -37\rpm1 & -116\rpm4\\ 
2 & NLO-QCD+(QCD+QED)$_{\rm PS}$& \photos\ & -83\rpm1 & -282\rpm4 & -36\rpm1 & -114\rpm3\\ 
3 & NLO-(QCD+EW)-{\tt two-rad}+(QCD+QED)$_{\rm PS}$ & \pythia\  & -86\rpm1 & -291\rpm3 & -38\rpm1 &
 -115\rpm3\\ 
4 & NLO-(QCD+EW)-{\tt two-rad}+(QCD+QED)$_{\rm PS}$ & \photos\ & -85\rpm1 & -290\rpm4 & -37\rpm2 &
 -113\rpm3\\ 
\bottomrule 
\end{tabular} 
\caption{
$W$ mass determination for muons and dressed electrons at the Tevatron.
$\mw$ shifts (in MeV) due to multiple QED FSR and mixed QCD-EW corrections,
computed with \pythiaqed\  and \photos\ as tools for the simulation of QED FSR effects.
\pythiaqed\  and \photos\ have been interfaced to 
\powhegvtwo\  with only QCD corrections (lines 1 and 2)
or matched to
\powhegvtwo\  {\tt two-rad} with NLO (QCD+EW) accuracy (lines 3 and 4).
The templates have been computed with \powhegvtwo\  with only QCD corrections.
The results are based on MC samples with 1$\times 10^8$ events.
}
\label{tab:qedqcdshifts-tev}
\end{table}
We can notice that:
\begin{itemize}

\item 1 vs. 2: 
there is a not negligible difference between the predictions of  
 \pythiaqed\  and \photos\ for the QED FSR contribution. 
These differences amount to about $8 \pm 1$~MeV for the lepton-pair 
transverse mass 
and to about $26 \pm 5$~MeV for the lepton \ptl\  for muons and disappear for dressed electrons. 
The origin of the difference in size for the two observables 
has been discussed in section \ref{sec:distr:fsr} and
derives from the different modeling of QED radiation in the two programs.
The impact of this difference on the observables relevant for the $\mw$ determination has been shown in figure \ref{fig:pythia-photos-mt-pt} (black dots). 
Notice that this difference is robust, 
as we carefully checked that the parameters and theoretical ingredients used in our \pythiaqed\  simulations 
are fully consistent with  those of \photos\ 
(same value of the electromagnetic coupling constant given by $\alpha(0)$, 
no pair radiation and negligible effect of QED ISR in \pythiaqed).

\item 3 vs. 4:
the shifts induced by mixed \oaas~corrections are independent of the QED radiation model, or, in other words, the effect of QED terms subleading in an expansions in powers of $L_{\rm QED}$ is negligible. 
In fact the shifts of lines 3 and 4 agree
at the level of 1~MeV, within the statistical error, both for $M_T$ and \ptl\ in the case of muons and dressed electrons. This can be understood by the fact that the hardest QED final state photon 
is described, in both approaches, with NLO matrix element accuracy
and the QED LL shower simulates only higher-order effects. 
As a consequence, the differences stemming from
different QED simulations between \pythiaqed\  and \photos\ start from ${\cal O}(\alpha^2)$.
The differences for both lepton-pair transverse mass and lepton transverse momentum distributions are at the 0.1\% level, as shown in figure \ref{fig:pythia-photos-mt-pt} (blue dots) and flat around the jacobian peak,
yielding differences in the $\mw$ shifts below the 1~MeV target uncertainty.

\item 1 vs. 3 and 2 vs. 4: 
the difference between these theoretical options 
provides an estimate of the contribution of mixed \oaas~corrections, 
that are not included in the stand-alone tools that simulate QED FSR
and that become available only after matching these tools with an exact NLO EW calculation.

We note that the estimate of the mixed \oaas~corrections depends on the tool used to simulate QED FSR. 
In particular, 
the estimate of these effects with FSR simulated with \pythiaqed\  amounts to 
a $\sim 5 \pm 1$~MeV shift for the lepton-pair transverse mass and
to a shift of the order of $\sim 17 \pm 5$~MeV for the lepton transverse momentum, in the case of muons; 
for recombined electrons the shifts are of the size of $\sim 1 \pm 1$~MeV and $\sim 1 \pm 5$~MeV for $M_T$ and \ptl, respectively. 
When simulating QED FSR with \photos\ the effects amount to 
a $\sim 2 \pm 1$~MeV shift for the transverse mass and
to a shift of the order of $\sim 8 \pm 5$~MeV for the lepton transverse momentum, in the case of muons; 
for recombined electrons the shifts are of the size of $\sim 1 \pm 2$~MeV and $\sim 1 \pm 4$~MeV for $M_T$ and \ptl, respectively. 

These results show that a QED-LL approach without matching is more accurate, 
at the  level of precision required for the $M_W$ determination, 
when QED FSR is simulated with \photos\ (line 2). 
The small difference between the shifts obtained with \photos\ with and without matching with the NLO EW results can also be understood from figure \ref{fig:mixed-lhc-mu}, where the relative impact of the EW effects
in the two cases is almost identical.

These comparisons can be considered as a measure of the accuracy inherent in the  use of a generator given 
by a tandem of tools  like {\resbos+\photos} 
(like in the present Tevatron measurements)  
in the sector of mixed QCD-EW corrections.

\end{itemize}

The assessment of the uncertainty for the Tevatron as explained in the third item above, is, in our opinion, one of the most important and original aspects of our study.

%%%%%%%%%%%%%%%%%%%%%%%
\subsubsection{\label{sec:mixlhc}Results for the LHC}

In this section we present the results for a similar analysis to the one addressed in Section~\ref{sec:mixtev}, but under LHC conditions. The details of the event selection are shown in table~\ref{tab:parameters_lhc}, and the corresponding mass shifts in table~\ref{tab:qedqcdshifts-lhc}. 

\begin{table}[htb]
\centering
\begin{tabular}{ll}
\toprule
Process & $pp \rightarrow W^+ \rightarrow \mu^+ \nu$, $\sqrt{s}=14$~TeV \\
PDF & MSTW2008 NLO \\
Event selection & $|\eta^\ell| < 2.5, \, p_T^{\ell} > 20~{\rm GeV}, \, p_T^{\nu} > 20~{\rm GeV}, \, p_T^{W} < 30~{\rm GeV}$ \\
\bottomrule
\end{tabular}
\caption{Event selection used for the study of QED and mixed QCD-EW effects at LHC.}
\label{tab:parameters_lhc}
\end{table}

\begin{table}[htb] 
\centering
\begin{tabular}{>{\small}c>{\footnotesize}l>{\small}c>{\small}c>{\small}c>{\small}c>{\small}c} 
\toprule 
\multicolumn{3}{c}{$pp \rightarrow W^+$, $\sqrt{s}=14$~TeV} & \multicolumn{4}{|c}{\mw\ shifts (MeV)} \\ 
\multicolumn{3}{c}{Templates accuracy: NLO-QCD+QCD$_{\rm PS}$} & \multicolumn{2}{|c}{$W^+\rightarrow\mu^+\nu$} & \multicolumn{2}{|c}{$W^+\rightarrow e^+\nu$(dres)} \\ 
\multicolumn{2}{c}{\normalsize Pseudodata accuracy} & QED FSR & \multicolumn{1}{|c}{$M_T$} & $p_T^{\ell}$ & \multicolumn{1}{|c}{$M_T$} & $p_T^{\ell}$ \\ 
\midrule 
1 & NLO-QCD+(QCD+QED)$_{\rm PS}$& \pythia\  & -95.2\rpm0.6 & -400\rpm3 & -38.0\rpm0.6 & -149\rpm2\\ 
2 & NLO-QCD+(QCD+QED)$_{\rm PS}$& \photos\ & -88.0\rpm0.6 & -368\rpm2 & -38.4\rpm0.6 & -150\rpm3\\ 
3 & NLO-(QCD+EW)+(QCD+QED)$_{\rm PS}${\tt two-rad} & \pythia\  & -89.0\rpm0.6 & -371\rpm3 & -38.8\rpm0.6 &
 -157\rpm3\\ 
4 & NLO-(QCD+EW)+(QCD+QED)$_{\rm PS}${\tt two-rad} & \photos\ & -88.6\rpm0.6 & -370\rpm3 & -39.2\rpm0.6 &
 -159\rpm2\\ 
\bottomrule 
\end{tabular} 
\caption{$W$ mass determination for muons and dressed electrons at the LHC 14 TeV in the case of $W^+$ production.
$\mw$ shifts (in MeV) due to multiple QED FSR and mixed QCD-EW corrections,
computed with \pythiaqed\  and \photos\ as tools for the simulation of QED FSR effects.
\pythiaqed\  and \photos\ have been interfaced to 
\powhegvtwo\  with only QCD corrections (lines 1 and 2)
or matched to
\powhegvtwo\  {\tt two-rad} with NLO (QCD+EW) accuracy (lines 3 and 4).
The templates have been computed with \powhegvtwo\  with only QCD corrections.
The results are based on MC samples with 4$\times 10^8$ events.}
\label{tab:qedqcdshifts-lhc}
\end{table}

Similar remarks on the comparison between \pythiaqed\  and \photos, 
as well as on mixed QCD-EW corrections, apply in this case. 
However, further considerations can be drawn by comparing the results of
table~\ref{tab:qedqcdshifts-lhc}, 
where QCD corrections are included with NLO QCD + QCD PS accuracy, 
with those in table~\ref{tab:ewshifts},
which correspond to LHC simulations in the same setup but at LO accuracy in QCD.
In particular, this comparison is meaningful for the $W$ mass shifts
obtained with \photos\ for the modeling of QED FSR in table~\ref{tab:qedqcdshifts-lhc}. 
One can notice that:
\begin{itemize}

\item By comparing the results with muons,
in the last line of table~\ref{tab:ewshifts} with those in the second line of table~\ref{tab:qedqcdshifts-lhc}, 
the shift is largely independent of the presence of QCD corrections
for fits to the lepton-pair transverse mass, 
whereas the shift extracted from the lepton transverse momentum distribution is strongly influenced by the inclusion of QCD contributions.

As it has been discussed in section~\ref{sec:distr:qcd}
and shown in figure~\ref{fig:qcd}, 
QCD corrections preserve to a large extent the LO shape of the lepton-pair transverse mass and strongly modify the LO shape of the lepton transverse momentum distributions.
In the latter case, the broader shape enhances the impact of radiative corrections
in the template fit procedure. 
The large corrections induce in turn large mixed QCD-QED effects, which contribute to the differences between tables \ref{tab:ewshifts} and \ref{tab:qedqcdshifts-lhc}.

\item 1 vs. 3 and 2 vs. 4: as a consequence of the above point, 
the effect due to mixed QCD-EW corrections, beyond those due to QED FSR,
is of the same order at Tevatron and the LHC for fits to $M_T$,
while they are enhanced at the LHC for fits to $p_T^\ell$.
As already noticed in the analysis for the Tevatron,
in the case of the lepton transverse momentum these effects depend on the model of QED FSR.
If we consider \pythiaqed, the $\mw$ shift is in the range $\sim 29 \pm 5$~MeV for $p_T^\ell$ with muons
(even if, considering the associated numerical uncertainty, 
LHC and Tevatron results could be compatible).
If QED FSR is instead simulated with \photos,
the mixed QCD-EW corrections are already well accounted for 
by the convolution of \photos\ stand-alone
with the events generated with NLO QCD + QCD PS accuracy, 
with an uncertainty at the MeV level. 

\item 1 vs. 2: as at the Tevatron, there are differences between the predictions of 
\pythiaqed\  and \photos\ for QED FSR for muons under a bare event selection, 
which however disappear when considering dressed electrons or,
in full generality, disappear after matching with an exact NLO EW calculation (lines 3 and 4).

\end{itemize}

In order to summarize the results shown in section~\ref{sec:mixtev} and in the present section,
we collect in table~\ref{tab:finalshifts-lhc} the shifts induced on $\mw$ 
(as extracted from the
$M_T$ or $p_T^{\ell}$ distributions in the case of $W^+$ production,
with bare muon event selections, at Tevatron and LHC)  
by the subset of mixed QCD-EW corrections present in \qcdew\ but not included in \qcdqed\ approximation.
\begin{table}[htb] 
\centering 
\hskip-1cm
\begin{tabular}{>{\small}c>{\small}c>{\small}c>{\small}c} 
\toprule
\multicolumn{2}{c}{
} &
\multicolumn{2}{c}{$\Delta M_W({\rm MeV})$}\\
\midrule 
 & QED FSR model & \multicolumn{1}{c}{$M_T$} & $p_T^{\ell}$ \\
\midrule 
Tevatron & \pythia\  & +5 $\pm$ 2 & +17 $\pm$ 5 \\
                                 & \photos\ & ~-2 $\pm$ 1 & ~~-8 $\pm$  5 \\
LHC      & \pythia\  & +6.2 $\pm$ 0.8 & +29 $\pm$ 4 \\
                                 & \photos\ & ~-0.6 $\pm$ 0.8 & ~~-2 $\pm$ 4 \\
\bottomrule 
\end{tabular} 
\caption{Residual $\mw$ shifts
computed as the difference of the results of the simulations with \qcdew\ 
and \qcdqed\  accuracy, with \pythiaqed\  and \photos,
for Tevatron and LHC 14 TeV energies, in the case of $W^+$ production and bare muons.
}
\label{tab:finalshifts-lhc}
\end{table}
As a reference, we remind that, 
by inspection of the shifts as in the last line of table~\ref{tab:ewshifts}
(QED FSR effects)
in comparison with those in the second line of table~\ref{tab:qedqcdshifts-lhc} 
(QED FSR and mixed QCD-QED effects), 
the $\mw$ shift due to mixed QCD-QED factorized effects 
amounts to $+4 \pm 1$~MeV for fits to the transverse mass and QED FSR simulated with \photos.

We show in table~\ref{tab:qedqcdshifts-lhc-wm} results for the process 
$pp \to W^- \to \ell^- \nu$ at LHC energies, 
to be compared with the results for $W^+$ production 
in table~\ref{tab:qedqcdshifts-lhc}. 
In this case, the parameters and event selection are the same as those shown in table~\ref{tab:parameters_lhc}. 
From lines 3 and 4 of table~\ref{tab:qedqcdshifts-lhc-wm}
we remark that in \qcdew\ approximation the mixed QCD-EW effects due to a different modeling of QED FSR are at the 1~MeV level, as in the $W^+$ case.
The comparison between lines 1 and 3 and between lines 2 and 4
shows the impact of mixed QCD-EW corrections present in \qcdew\ but absent in \qcdqed: while \photos\ stand-alone provides already a good approximation of the NLO results (lines 2 and 4), the results of \pythiaqed\ stand-alone differ with respect to those with \qcdew\ accuracy (lines 1 and 3);
we also remark that in the \pythiaqed\ case the difference of the shifts is larger than the corresponding one in the $W^+$ case and reaches 34 MeV 
for the muon transverse momentum distribution.
Although  the statistics is not sufficient to draw a firm conclusion about a different behaviour of $W^+$ and $W^-$,
our results suggest that 
the evaluation of the theoretical shifts and related uncertainties requires 
at the LHC a separate study of $W^-$ and $W^+$ production and decay, at least for the lepton \ptl.
\begin{table}[htb] 
\centering
\begin{tabular}{>{\small}c>{\footnotesize}l>{\small}c>{\small}c>{\small}c>{\small}c>{\small}c} 
\toprule 
\multicolumn{3}{c}{$pp \rightarrow W^-$, $\sqrt{s}=14$~TeV} & \multicolumn{4}{|c}{\mw\ shifts (MeV)} \\ 
\multicolumn{3}{c}{Templates accuracy: NLO-QCD+QCD$_{\rm PS}$} & \multicolumn{2}{|c}{$W^-\rightarrow\mu^-{\bar{\nu}}$} 
& \multicolumn{2}{|c}{$W^-\rightarrow e^- {\bar{\nu}}$(dres)} \\ 
\multicolumn{2}{c}{\normalsize Pseudodata accuracy} & QED FSR & \multicolumn{1}{|c}{$M_T$} & $p_T^{\ell}$ & \multicolumn{1}{|c}{$M_T$} & $p_T^{\ell}$ \\ 
\midrule 
1 & NLO-QCD+(QCD+QED)$_{\rm PS}$& \pythia\  & -97\rpm1 & -413\rpm4 & -39\rpm1 & -155\rpm4\\ 
2 & NLO-QCD+(QCD+QED)$_{\rm PS}$& \photos\ & -90\rpm1 & -379\rpm5 & -40\rpm1 & -154\rpm5\\ 
3 & NLO-(QCD+EW)+(QCD+QED)$_{\rm PS}${\tt two-rad} & \pythia\  & -90\rpm1 & -379\rpm5 & -40\rpm1 &
 -166\rpm5\\ 
4 & NLO-(QCD+EW)+(QCD+QED)$_{\rm PS}${\tt two-rad} & \photos\ & -89\rpm1 & -377\rpm4 & -40\rpm1 &
 -164\rpm4\\ 
\bottomrule 
\end{tabular} 
\caption{$W$ mass determination for muons and dressed electrons at the LHC 14 TeV in the case of $W^-$ production.
$\mw$ shifts (in MeV) due to multiple QED FSR and mixed QCD-EW corrections,
computed with \pythiaqed\  and \photos\ as tools for the simulation of QED FSR effects.
\pythiaqed\  and \photos\ have been interfaced to 
\powhegvtwo\  with only QCD corrections (lines 1 and 2)
or matched to
\powhegvtwo\  {\tt two-rad} with NLO (QCD+EW) accuracy (lines 3 and 4).
The templates have been computed with \powhegvtwo\  with only QCD corrections.
The results are based on MC samples with 1$\times 10^8$ events.}
\label{tab:qedqcdshifts-lhc-wm}
\end{table}

As a last remark, we stress that, aiming at an $\mw$ determination with $\sim$MeV uncertainty,
we have validated the description of mixed \oaas contributions 
obtained in a \qcdqed\, approximation with \photos. 
If, on the contrary, QED FSR is described with \pythiaqed, 
the matching with NLO EW results in a \qcdew\, approximation is needed.

%% %%%%%%%%%%%%%%%%%%%%%%%%%%%%%%%%%%%%%%%%%%%%%%%%%%%%%%%%%%%%%%%%%%%%%%%%

\subsubsection{\label{sec:v2vstworad}Impact on $\mw$ determination of the \powhegvtwo\ {\tt two-rad} improvement}
In this section we discuss the impact on the $\mw$ determination
of the upgrade dubbed {\tt two-rad}
of the code \powhegvtwo\  with NLO (QCD+EW) accuracy matched to (QCD+QED) PS,
described in section \ref{sec:powheg-improved}, compared to the previous public versions of the code.
\begin{table}[htb] 
\centering
\begin{tabular}{>{\small}c>{\footnotesize}l>{\small}c>{\small}c>{\small}c>{\small}c>{\small}c} 
\toprule 
\multicolumn{3}{c}{$pp \rightarrow W^+$, $\sqrt{s}=14$~TeV} & \multicolumn{4}{|c}{\mw\ shifts (MeV)} \\ 
\multicolumn{3}{c}{Templates accuracy: NLO-QCD+QCD$_{\rm PS}$} & \multicolumn{2}{|c}{$W^+\rightarrow\mu^+\nu$} & \multicolumn{2}{|c}{$W^+\rightarrow e^+\nu$(dres)} \\ 
\multicolumn{2}{c}{\normalsize Pseudodata accuracy} & QED FSR & \multicolumn{1}{|c}{$M_T$} & $p_T^{\ell}$ & \multicolumn{1}{|c}{$M_T$} & $p_T^{\ell}$ \\ 
\midrule 
1 & NLO-QCD+(QCD+QED)$_{\rm PS}$& \pythia\  & -95.2\rpm0.6 & -400\rpm3 & -38.0\rpm0.6 & -149\rpm2\\ 
2 & NLO-QCD+(QCD+QED)$_{\rm PS}$& \photos\ & -88.0\rpm0.6 & -368\rpm2 & -38.4\rpm0.6 & -150\rpm3\\ 
3 & NLO-(QCD+EW)+(QCD+QED)$_{\rm PS}${\tt two-rad} & \pythia\  & -89.0\rpm0.6 & -371\rpm3 & -38.8\rpm0.6 &
 -157\rpm3\\ 
4 & NLO-(QCD+EW)+(QCD+QED)$_{\rm PS}${\tt two-rad} & \photos\ & -88.6\rpm0.6 & -370\rpm3 & -39.2\rpm0.6 &
 -159\rpm2\\ 
5 & NLO-(QCD+EW)+(QCD+QED)$_{\rm PS}$ & \pythia\   & -101.8\rpm0.4 & -423\rpm2 & -45.0\rpm0.6 & -179\rpm2\\ 
6 & NLO-(QCD+EW)+(QCD+QED)$_{\rm PS}$ & \photos\ & -94.2\rpm0.6 & -392\rpm2 & -45.2\rpm0.6 & -181\rpm2\\ 
\bottomrule 
\end{tabular} 

\caption{
Analysis of the impact on the $\mw$ determination 
of the {\tt two-rad} upgrade of
\powhegvtwo\  with NLO (QCD+EW) accuracy (lines 3 and 4), 
compared to the older public versions (lines 5 and 6).
All the parameters and specifications as in table \ref{tab:qedqcdshifts-lhc}.
}
\label{tab:qedqcdshifts-lhc-bis}
\end{table}
\begin{table}[htb]
\centering
\begin{tabular}{>{\small}c>{\footnotesize}l>{\small}c>{\small}c>{\small}c>{\small}c>{\small}c} 
\toprule 
\multicolumn{3}{c}{$p\bar{p} \rightarrow W^+$, $\sqrt{s}=1.96$~TeV} & \multicolumn{4}{|c}{\mw\ shifts (MeV)} \\ 
\multicolumn{3}{c}{Templates accuracy: NLO-QCD+QCD$_{\rm PS}$} & \multicolumn{2}{|c}{$W^+\rightarrow\mu^+\nu$} & \multicolumn{2}{|c}{$W^+\rightarrow e^+\nu$(dres)} \\ 
\multicolumn{2}{c}{\normalsize Pseudodata accuracy} & QED FSR & \multicolumn{1}{|c}{$M_T$} & $p_T^{\ell}$ & \multicolumn{1}{|c}{$M_T$} & $p_T^{\ell}$ \\ 
\midrule 
1 & NLO-QCD+(QCD+QED)$_{\rm PS}$& \pythia\  & -91\rpm1 & -308\rpm4 & -37\rpm1 & -116\rpm4\\ 
2 & NLO-QCD+(QCD+QED)$_{\rm PS}$& \photos\ & -83\rpm1 & -282\rpm4 & -36\rpm1 & -114\rpm3\\ 
3 & NLO-(QCD+EW)+(QCD+QED)$_{\rm PS}${\tt two-rad} & \pythia\  & -86\rpm1 & -291\rpm3 & -38\rpm1 &
 -115\rpm3\\ 
4 & NLO-(QCD+EW)+(QCD+QED)$_{\rm PS}${\tt two-rad} & \photos\ & -85\rpm1 & -290\rpm4 & -37\rpm2 &
 -113\rpm3\\ 
5 & NLO-(QCD+EW)+(QCD+QED)$_{\rm PS}$ & \pythia\   & -96\rpm1 & -323\rpm3 & -45\rpm1 & -129\rpm3\\ 
6 & NLO-(QCD+EW)+(QCD+QED)$_{\rm PS}$ & \photos\ & -89\rpm1 & -300\rpm3 & -44\rpm2 & -134\rpm3\\ 
\bottomrule 
\end{tabular} 
\caption{
Analysis of the impact on the $\mw$ determination 
of the {\tt two-rad} upgrade of
\powhegvtwo\  with NLO (QCD+EW) accuracy (lines 3 and 4), 
compared to the older public versions (lines 5 and 6).
All the parameters and specifications as in table \ref{tab:qedqcdshifts-tev}.
\label{tab:qedqcdshifts-tev-bis}
}
\end{table}

We stress that the new version {\tt two-rad} superseeds the previous public versions and is the only version that must be used for physical analyses.
As illustrated in figure \ref{fig:mixed-mt-pt-lhc-mu},
the {\tt two-rad} version removes a double counting of QED FSR contributions
and of mixed QCD-QED contributions; 
the latter, present in the previous versions of the code, 
causes a distortion at the jacobian peak of the shape of the lepton-pair
transverse mass and lepton transverse momentum distributions.
The distortion, visible by comparing the blue dots with respect to the red dots
in figure \ref{fig:mixed-mt-pt-lhc-mu}, 
induces an additional shift of $\mw$, 
shown in table \ref{tab:qedqcdshifts-lhc-bis} where we repeat for convenience also the other values of table \ref{tab:qedqcdshifts-lhc},
computed with LHC conditions in the case of $W^+$ production.
We can observe, by comparing lines 3 and 5 or lines 4 and 6,
the shift of order -10 MeV induced on $\mw$ extracted from the lepton-pair transverse mass and the shift of about -25/-50~MeV induced on $\mw$ extracted from the lepton transverse momentum distributions.
A clear sign of the problems in the previous implementations
emerges in the comparison of lines 5 and 6, because the dependence on the model that describes QED FSR, i.e. \pythiaqed\  vs. \photos,
is not reduced after matching with an exact NLO EW calculation
and remains at the level of the comparison between lines 1 and 2, 
where, instead, a discrepancy is justified.

A similar behaviour of the previous versions of \powhegvtwo\  with NLO (QCD+EW) accuracy has been observed also with the Tevatron setup and is illustrated in table~\ref{tab:qedqcdshifts-tev-bis}.

\section{\label{sec:conclusions}Conclusions}

In this paper, we have presented a comprehensive study of electroweak, QED 
and mixed factorizable QCD-electroweak corrections underlying 
the theoretical modeling necessary in the 
precise measurement of the $W$-boson mass at hadron colliders. 

We have shown that particular attention must be paid in the treatment of QED FSR, 
as the models implemented in \pythia\ and \photos\ give rise to different \mw\ shifts;
the \photos\ simulation of FSR is more reliable because 
it better approximates a matrix element behaviour
and it has been validated with the \horace\ independent results.

We have shown that neglecting the contribution of light lepton pairs 
in the evaluation of the theoretical templates
introduces an uncertainty 
(both for Tevatron and LHC) of $3 - 5 \pm 1~{\rm MeV}$;
the first value is for final state electrons, the second one applies to muons 
and the conclusion holds for fits both to \mtr\ and \ptl. 

We have pointed out that mixed factorizable \oaas corrections, 
as calculated by means of an improved version of 
\powhegvtwo\ with QCD and EW corrections,\, 
i.e. \powhegvtwo\ {\tt two-rad},\, 
induce a shift on \mw\ of $-16 \pm 3~{\rm MeV}$, 
in good agreement with the estimate of Ref.~\cite{Dittmaier:2015rxo} 
based on a NNLO calculation in pole approximation. 
We have provided clear evidence that 
\mw\ shifts due to factorizable \oaas corrections 
are largely dominated by the interplay between QCD radiation and QED FSR, 
the residual mixed corrections beyond the approximation \qcdqed\
being very small, at the 1~MeV level. 
Again these results apply to \mtr\ and \ptl\
and are based on simulations generated with \powhegvtwo\ {\tt two-rad}.

The study of mixed QCD-EW corrections to DY processes
motivated the development of the code \powhegvtwo\ with NLO (QCD+EW)\ accuracy
and eventually lead to the upgraded \powhegvtwo\ {\tt two-rad} version for
DY processes; the latter superseeds the previous
versions in the \powhegvtwo\ framework.

For the purpose of measuring \mw\ with high accuracy, 
the results of our study imply that, 
if tools such as \resbos\ or \powhegvtwo\ interfaced to ``standard" 
\photos\ (i.e. with no pair radiation) are used in the measurements, 
a theoretical systematic uncertainty of some MeV has to be accounted for.
This is in agreement with the completely different and independent 
Tevatron estimate of the theoretical errors of perturbative nature. 
On the other hand, 
if more refined and up-to-date generators like \powhegvtwo\ {\tt two-rad} interfaced to \photos\ including pairs are used, 
the uncertainty due to perturbative contributions 
is reduced to the $\sim 1-2~{\rm MeV}$ level. 
Advances in the field of high-precision calculations 
and MC generators for DY processes 
will allow to sustain this expectation.

In summary, our results can serve as a guideline 
for the assessment of the theoretical systematics at the Tevatron and LHC 
and allow a more robust precision measurement  of the $W$ mass at 
hadron colliders.

\vskip 12pt\noindent
{\bf Acknowledgments}\\
This work was supported in part by the Italian 
Ministry of University and Research 
under the PRIN project 2010YJ2NYW. 
AV is supported by the European Commission through the HiggsTools Initial Training Network PITN-GA2012-316704.
The work of HM has been supported by the Research Executive Agency (REA) of
the European Union under the Grant Agreement number PITN-2010- 264564 (LHCPhenoNet) 
and is presently supported by the University of Pavia under the grant ``Fondo Giovani".
\\
The authors thank the 
Galileo Galilei Institute for theoretical physics for giving
them the possibility of hosting the first meeting of the workshop 
``$W$ mass measurement at the LHC" (October 2014) and INFN for partial
support. They are also grateful to M. Boonekamp, L. Perrozzi and D.~Wackeroth
for their help in the organization of the meeting. \\
The authors wish to thank M.L. Mangano for his continuous support of the
present study within the CERN LPCC activities. \\
The authors acknowledge the participation of I. Bizijak to the early stage of this work.
\\
They are indebted to P.~Nason for his crucial 
collaboration in the development of \powheg v2 with EW corrections
and Z.~Was for his help in using \photos.
They are also grateful to M. Boonekamp, A. Kotwal, L. Perrozzi and
D.~Wackeroth for continuous encouragement and exchange of information.
\\
The authors wish to thank A.~Arbuzov, 
M.R.~D'Alfonso, S.~Dittmaier, D.~Froidevaux, C.~Hays, A.~Huss, 
T.~Kurca, A.~M\"uck, C.~Oleari, L.~Oymanns, R.~Lopes de Sa,  
C.~Schwinn, P.~Skands, J.~Stark and Z.~Was for useful discussions.
\\
FP is grateful to the Mainz Institute for Theoretical Physics 
(MITP) and to the CERN Theory Unit of the Physics Department
for their hospitality and partial support
during the development of this work. 
AV is grateful to the Kavli Institute for Theoretical Physics (KITP)
for hospitality and support during the
``LHC Run II and the Precision Frontier'' 
supported by NSF PHY11-25915,
and to the CERN Theory Unit of the Physics Department
for its hospitality and partial support
during the development of this work. 
\\
\vskip 12pt\noindent
{\bf Note added in proof}\\
During the completion of this work, 
another paper~\cite{Muck:2016pko} 
aiming at an improved treatment of vector-boson resonances with the \powheg\ method, 
including electroweak corrections, appeared in the literature. 
It can be considered as an independent method to perform parton-shower matching for Drell-Yan production of 
$W$ and $Z$ bosons at the LHC at NLO QCD and NLO electroweak accuracy, as provided by 
our code \powhegvtwo\ {\tt two-rad} developed in the context of the present study. For the future, it would be
worthwhile to compare the results of the two approaches.

\clearpage
\appendix
\section{\label{sec:photossetup} Appendix: \photos\, setup   }
In this Appendix we detail the input parameter setting
used to run \photos:\\
{\tt
  **  PHOTOS setup ** \\
 ->   Version: 3.56\\
 -> Output initializaton screen:\\
INTERF= 1 \\
ISEC= 0\\
ITRE= 0\\ 
IEXP= 1\\
IFTOP= 1\\ 
IFW= 1\\
ALPHA\_QED= 0.00729735\\
XPHCUT= 1e-07\\

\noindent
                    Option with interference is active\\
                    Option with exponentiation is active EPSEXP=0.0001\\
                    Emision in t tbar production is active\\
                    Correction wt in decay of W is active\\

\noindent
 -> Explicit values of flags\\
  phokey\_.interf = 1        (interference weight, on by default)\\
  phokey\_.isec = 0          (double photon, off by default)\\
  phokey\_.ifw = 1           (correction weight in decay of W, on by default)\\
  Photos::meCorrectionWtForW  = 0   (ME correction in decay of W, off by default)\\

\noindent
 ->  Setting random seed for each run, using:\\
     srand (time(NULL));\\
     int s1 = rand() \% 31327;   // (number between 0 and 31327)\\
     int s2 = rand() \% 30080;   // (number between 0 and 30080)\\
     Photos::setSeed(s1, s2); \\

\noindent
 ->  Setting infrared cutoff for each event, using:\\

\noindent
     kt2minqed  = 0.001d0**2\\
     xphcut = 2d0*sqrt( kt2minqed )/pup(5,3) // pup(5,3): Invariant mass of decaying W\\
     Photos::setInfraredCutOff(xphcut);

}

%%%%%%%%%%%%%%%%%%%%%%%%%%%%%%%%%%%%%%%%%%%%%%%%%%%%%%%%%%%%%%%%%%%%%%%%%%%%
\section{\label{sec:lpa} Appendix: Lepton-pair radiation}

In this appendix, we describe how the original PS algorithm implemented in 
\horaceo\ to simulate photon radiation has been generalized to account for
lepton-pair emission in \horace.

In QED the probability that a fermion evolves from a virtuality $s_i$ to $s_f$ emitting photons of  energy fraction below a
 threshold $\epsilon$ is given by the Sudakov form factor $\Pi (s_f,s_i)$, describing the so--called ``no emission"
 probability. It can be written as
\begin{eqnarray}
 \Pi (s_f,s_i) = \exp \left[  - \int_{s_i}^{s_f} \frac{\alpha(s' )}{2\pi} \frac{d s'}{s'}  I_+ \right]
\end{eqnarray}
where 
\begin{eqnarray}
\!\!\!\!\!\! I_+ \equiv \int_0^{1-\epsilon} dz \, P(z) = 
 - 2 \ln \epsilon - \frac{1}{2}(1-\epsilon)^2 - 1 + \epsilon
\end{eqnarray}
and $P(z) = (1+z^2)/(1-z)$ is the unregularized Altarelli--Parisi electron $\to$ electron + photon splitting function.
By definition, the Sudakov form factor includes the contribution of virtual and real soft photons to all orders of QED.
If we set $\alpha (s) = \alpha (0) \equiv \alpha$, which is the 
physical value to be used for the electromagnetic 
coupling constant to describe photon emission, the Sudakov form factor is 
simply given by
\begin{eqnarray}
 \Pi (s_f,s_i) = \exp \left[  - \frac{\alpha}{2\pi} \ln\frac{s_f}{s_i}  I_+ \right] 
  \label{eq:sg}
\end{eqnarray}
that is the formula used in the {\sc{Horace}} default version. If we 
consider photon radation from a lepton that evolves from $s_i = m^2$ to $s_f \equiv s$, the 
form factor can be rewritten as
\begin{eqnarray}
 \Pi (s, m^2) = \exp \left[  - \frac{\alpha}{2\pi} \ln\frac{s}{m^2}  I_+ \right] \xrightarrow[s \gg m^2]{} \exp \left[  - \frac{\beta}{4}  I_+ \right] 
 \label{eq:sgp}
\end{eqnarray}
where $\beta \equiv 2 {\alpha}/{\pi} \left( \ln {s}/{m^2} - 1 \right)$ is the QED collinear factor associated 
to photon emission from leptons.
On the other hand, if we introduce a running 
electromagnetic coupling constant $\alpha (s)$ as in eq.~(\ref{eq:alfas}) to 
describe photon emission accompanied by the conversion of photons into lepton pairs, the form factor becomes
\begin{eqnarray}
 && \Pi(s,m^2) \rightarrow \Pi_s (s, m^2) = \exp \left[  - \frac{\beta (s) }{4}  I_+ \right] 
\end{eqnarray}
An expansion of the Sudakov form factor $ \Pi_s = \epsilon^{- \beta(s) / 2} $ up to second order reads
\begin{eqnarray}
\Pi_s \simeq 1 - \frac{\beta (s)}{2} \ln \epsilon + \frac{\beta^2(s)}{8} \ln^2 \epsilon + \cdots
\label{eq:sudexp}
\end{eqnarray}
If we consider the dominant contribution of $e^+ e^-$ pair emission, 
one gets for radiating electrons ($f = e$)
\begin{eqnarray}
\beta (s) \simeq \beta_e + \frac{1}{12} \beta_e^2 + \cdots
\label{eq:betae}
\end{eqnarray}
 by definition of $\beta(s)$ as in the second formula of eq.~(\ref{eq-betarunning}) and expansion 
 of the logarithms entering eq.~(\ref{eq-betarunning}). 
 In eq.~(\ref{eq:betae}) $\beta_e  \equiv 2 {\alpha}/{\pi} \left( \ln {s}/{m_e^2} - 1 \right)$. Substituting eq.~(\ref{eq:betae}) in eq.~(\ref{eq:sudexp}) 
gives
\begin{eqnarray}
\Pi_s \simeq 1 - \frac{\beta_e}{2} \ln \epsilon + \frac{\beta^2_e}{8} \ln^2 \epsilon - \frac{1}{24} \beta_e^2 \ln \epsilon \cdots
\end{eqnarray}
where the second and third term correspond to one and two photon emission and the last one to pair radiation. 
Therefore, for electrons radiating electron pairs, the ratio between two photon and pair radiation is given by
\begin{equation}
2\gamma / {\rm pairs} ~\vert_{\rm electrons} \, \simeq \, 3 \, \ln \epsilon
\label{eq:re}
\end{equation}
On the other hand, by applying the same reasoning to muons ($f = \mu$) emitting $e^+ e^-$ pairs, one gets
\begin{eqnarray}
\beta (s) \simeq \beta_\mu + \frac{{\cal L}}{12} \, \beta_\mu^2 + \cdots
\label{eq:betamu}
\end{eqnarray}
where  $\beta_\mu  \equiv 2 {\alpha}/{\pi} \left( \ln {s}/{m_\mu^2} - 1 \right)$ and $\cal L$ is a ratio of big logarithms given by 
\begin{eqnarray}
{\cal L} \, = \, \ln \left( \frac{s m_\mu^2}{m_e^4} \right) / \ln \left( \frac{s}{m_\mu^2} \right)
\end{eqnarray}
Note that for $s \simeq M_W^2$ one has ${\cal L} \simeq 3$. 
Now, substitution of eq.~(\ref{eq:betamu}) in eq.~(\ref{eq:sudexp}) gives
\begin{eqnarray}
\Pi_s \simeq 1 - \frac{\beta_\mu}{2} \ln \epsilon + \frac{\beta^2_\mu}{8} \ln^2 \epsilon - \frac{1}{24} \beta_\mu^2 {\cal L} \ln \epsilon \cdots
\end{eqnarray}
This implies that for muons radiating electron pairs, the ratio between two photon and pair radiation is 
enhanced with respect to the electron case and is given by
\begin{equation}
2\gamma / {\rm pairs} ~\vert_{\rm muons} \, \simeq \, \frac{3}{\cal L} \, \ln \epsilon \simeq \ln \epsilon
\label{eq:rmu}
\end{equation}
Equation~(\ref{eq:re}) and eq.~(\ref{eq:rmu}) explain the results discussed in section \ref{sec:distr:pair} and 
section \ref{sec:qed}.

In \horace, in addition to the modification of the Sudakov form factor, the treatment of lepton-pair radiation 
is completed as follows. Because of the meaning of the Sudakov form factor, the number of emitted photons of energy fraction above $\epsilon$ inside a sample 
of $N$ events is given by
\begin{eqnarray}
N_{\gamma} = (1 - \Pi) \, N
\label{eq:ng}
\end{eqnarray}
Analogously, the number of emitted photons plus pairs above $\epsilon$ is
\begin{eqnarray}
N_{\gamma} +  N_{\rm pairs} = (1 - \Pi_s) \, N
\label{eq:ngp}
\end{eqnarray}
Therefore, the fraction of emitted pairs $\nu_{\rm pairs}$ is given by
\begin{eqnarray}
1 + \nu_{\rm pairs} = \frac{ N_{\gamma} + N_{\rm pairs}}{N_{\gamma} }
\label{eq:fpairs}
\end{eqnarray}
Substituting eq. (\ref{eq:ng}) and eq. (\ref{eq:ngp}) in eq. (\ref{eq:fpairs}) it follows that the 
fraction of pairs can be cast in the form
\begin{eqnarray}
\nu_{\rm pairs} = \frac{ N_{\gamma} + N_{\rm pairs}}{ N_{\gamma} } - 1 =  \frac{(1 - \Pi_s) N}{(1 - \Pi)  N} - 1
\end{eqnarray}
This formula is used in \horace\ to account for $e^+ e^-$ and $\mu^+ \mu^-$ pair radiation above $\epsilon$ 
according to the appropriate relative fractions.

\bibliographystyle{JHEP}
\bibliography{Wmass}

\end{document}